\documentclass[prd,onecolumn,nofootinbib,showpacs]{revtex4}%
\usepackage{epsfig,graphicx,bm}
\usepackage[latin2]{inputenc}
\usepackage{color}
\usepackage{t1enc}
\usepackage{amssymb}
\usepackage{amsmath}
\usepackage{graphicx}
\usepackage{bm}
\usepackage{amsfonts}%
\newcommand{\bea}{\begin{eqnarray}}
\newcommand{\eea}{\end{eqnarray}}
\newcommand{\be}{\begin{equation}}
\newcommand{\ee}{\end{equation}}

\begin{document}
\title{Meson vacuum phenomenology in a three-flavor linear sigma model with
(axial-)vector mesons}
\author{$\text{D.\ Parganlija}^{1,2}$}
\email{denisp@hep.itp.tuwien.ac.at}
\author{$\text{P.\ Kov{\'a}cs}^{2,3}$}
\email{kovacs.peter@wigner.mta.hu}
\author{$\text{Gy.\ Wolf}^{3}$}
\email{wolf.gyorgy@wigner.mta.hu}
\author{$\text{F.\ Giacosa}^{2}$}
\email{giacosa@th.physik.uni-frankfurt.de}
\author{$\text{D.H.\ Rischke}^{2,4}$}
\email{drischke@th.physik.uni-frankfurt.de}
\affiliation{$^{1}$Institute for Theoretical Physics, Vienna University of Technology,
Wiedner Hauptstr.\ 8-10, A-1040 Vienna, Austria}
\affiliation{$^{2}$Institute for Theoretical Physics, Johann Wolfgang Goethe University,
Max-von-Laue-Str.\ 1, D-60438 Frankfurt am Main, Germany}
\affiliation{$^{3}$Institute for Particle and Nuclear Physics, Wigner Research Center for
Physics, Hungarian Academy of Sciences, H-1525 Budapest, Hungary}
\affiliation{$^{4}$Frankfurt Institute for Advanced Studies, Ruth-Moufang-Str.\ 1 D-60438
Frankfurt am Main, Germany}

\begin{abstract}
We study scalar, pseudoscalar, vector, and axial-vector mesons with
non-strange and strange quantum numbers in the framework of a linear sigma
model with global chiral $U(N_{f})_{L} \times U(N_{f})_{R}$ symmetry. We
perform a global fit of meson masses, decay widths, as well as decay
amplitudes. The quality of the fit is, for a hadronic model that does not
consider isospin-breaking effects, surprisingly good. We also investigate the
question whether the scalar $\bar{q}q$ states lie below or above 1 GeV and
find the scalar states above 1 GeV to be preferred as $\bar{q}q$ states.
Additionally, we also describe the axial-vector resonances as $\bar{q}q$ states.

\end{abstract}

\maketitle

\allowdisplaybreaks

\section{INTRODUCTION}

\label{intro}

Understanding the meson mass spectrum in the region below 2 GeV is one of the
fundamental problems of QCD. While the quark model seems to work very well for
many resonances [see, for instance, the summary in Ref.\ \cite{PDG}], some
fundamental questions, such as the constituent quark content of scalar and
axial-vector resonances, are still unanswered. \newline

The quark content of the scalar mesons has been a matter of debate for many
decades \cite{amslerrev}. The Particle Data Group (PDG) \cite{PDG} suggests
the existence of five $I(J^{PC})=0(0^{++})$ states in the region below 1.8
GeV: $f_{0}(500)$, $f_{0}(980)$, $f_{0}(1370)$, $f_{0}(1500)$, and
$f_{0}(1710)$. Also the existence of a sixth state, $f_{0}(1790)$, close to,
but distinct from, $f_{0}(1710)$ has been claimed \cite{f0(1790)}.
Additionally, there are scalar resonances in the isotriplet and in the
isodoublet sector as well: the $I=1$ states $a_{0}(980)$ and $a_{0}(1450)$ are
well established; the existence of the $I=1/2$ state $K_{0}^{\star}(800)$ (or
$\kappa$) is confirmed by some \cite{KappaY} and disputed by other authors
\cite{KappaN}, whereas the $I=1/2$ scalar kaon, $K_{0}^{\star}(1430)$, is an
established resonance.\newline

A description of all mentioned scalar states as $\bar{q}q$ states is not
possible for a simple reason: the number of physical resonances is much larger
than the number of resonances that can be constructed within a $\bar{q}q$
picture of mesons. Explicitly, there is only one $I(J^{PC})=0(0^{++})$ state
that can be constructed from the (non-strange) $u$ and $d$ quarks (provided
they are degenerate). Denoting this state as $\sigma_{N}$ we obtain
$\sigma_{N}\equiv(\bar{u}u+\bar{d}d)/\sqrt{2}$. An additional $I(J^{PC}%
)=0(0^{++})$ state can be constructed, if the strange quark is considered as
well: the pure strange state $\sigma_{S}\equiv\bar{s}s$. Since $\sigma_{N}$
and $\sigma_{S}$ have the same $I(J^{PC})$ quantum numbers, we expect the
physical spectrum to consist of mixed, rather than pure, states of $\sigma
_{N}$ and $\sigma_{S}$; however, such mixing will, of course, produce exactly
two new states. These states will correspond to at most two of the mentioned
five (six) $f_{0}$ states, and the natural question is then: \textit{which
two}?

Similarly, in both the $I=1$ and $I=1/2$ sectors one can construct only one
quark-antiquark resonance. Restricting for example to electric charge $+1$,
one has the states $a_{0}^{+}=\bar{d}u$ and $K_{0}^{\ast+}=\bar{s}u.$ The
state $a_{0}^{+}=\bar{d}u$ can be assigned to $a_{0}(980)$ or $a_{0}(1450)$,
and $K_{0}^{\ast+}=\bar{s}u$ to $K_{0}^{\star}(800)$ or $K_{0}^{\star}(1430).$
The question is: \textit{which is the correct assignment?}\newline

An answer to these questions is inevitably complicated for several reasons.
Firstly, as already indicated, states with the same $I(J^{PC})$ quantum
numbers are expected to mix -- this needs to be considered in particular in
the scalar sector due to the large number of physical resonances. Secondly,
basic features (pole mass and decay width) of some scalar mesons
[e.g.\ $f_{0}(500)$] are notoriously difficult to determine experimentally,
which makes it non-trivial to determine the structure of these states [see,
e.g.\ Ref.\ \cite{Leutwyler} for an example regarding the $f_{0}(500)$
resonance and Ref.\ \cite{buggf0} for an example regarding $f_{0}%
(1370)$].\newline

The question of how to correctly determine the quark structure of the mesons
is not only restricted to the scalar sector. More recently, also the nature of
the axial-vector mesons, most notably $a_{1}(1260)$, but also that of the
isoscalar states $f_{1}(1285)$, $f_{1}(1525)$, and the kaonic state
$K_{1}(1270)$ [or $K_{1}(1400)$, see the discussion in Ref.~\cite{av} and
refs.\ therein] have been investigated. Should one interpret the isotriplet
resonance $a_{1}(1260)$ as a quark-antiquark state, as the quark model
suggests (e.g.\ $a_{1}^{+}=\bar{d}u$), or is this state a broad $\rho\pi$
molecular state?\newline

Understanding these issues is not only crucial for hadron vacuum spectroscopy
but is also important at nonzero temperatures and densities, because the
correct identification of the chiral partner of the pion and of the $\rho$
meson is necessary for a proper description of the in-medium properties of
hadrons \cite{Achim}. In fact, if the $a_{1}(1260)$ is (predominantly) a
quark-antiquark state, it is the chiral partner of the $\rho$ meson, with
which it becomes degenerate at large $T$ and $\mu$ \cite{RS}. The chiral
partner of the pion is the scalar-isoscalar state $\sigma_{N}\equiv(\bar
{u}u+\bar{d}d)/\sqrt{2}$ which has been identified with the lightest scalar
state $f_{0}(500)$ in many works. However, as various theoretical results for
the vacuum have shown, and as we shall present in detail in this work, this
assignment is not correct: it turns out that $f_{0}(1370)$ emerges as the
chiral partner of the pion.\newline

Last but not least, the question of the origin of hadron masses is important.
One aims to understand to which extent the quark condensate, $\langle\bar{q} q
\rangle$, and the gluon condensate, $\langle G^{\mu\nu} G_{\mu\nu} \rangle$,
generate the hadron masses. To give an example, which condensate is
predominantly responsible for the mass of the $\rho$ meson? And related to
this, how will the mass of the $\rho$ (and that of other resonances) change in
the medium?\newline

The answers to all these fundamental questions are in principle contained in
the QCD Lagrangian. Unfortunately, QCD cannot be solved by analytic means from
first principles in the low-energy domain. For this reason, effective theories
have been developed which share some of the underlying symmetries of QCD. The
QCD Lagrangian exhibits, in addition to the local $SU(3)_{c}$ color symmetry
and the discrete $C$, $P$, and $T$ symmetries, a global chiral $U(N_{f}%
)_{L}\times U(N_{f})_{R}\equiv U(1)_{V}\times U(1)_{A}\times SU(N_{f}%
)_{V}\times SU(N_{f})_{A}$ symmetry which is broken in several ways:
spontaneously [due to the chiral condensate $\langle\bar{q}q\rangle
=\langle\bar{q}_{R}q_{L}+\bar{q}_{L}q_{R}\rangle\neq0$\ \cite{SSB}],
explicitly (due to non-vanishing quark masses), as well as at the quantum
level [the $U(1)_{A}$ anomaly \cite{Hooft}].\newline

In the framework of effective theories the chiral symmetry of QCD can be
realized along two lines: linearly \cite{gellmanlevy} and non-linearly
\cite{weinberg}. In the former case one obtains the so-called linear sigma
model: both scalar and pseudoscalar degrees of freedom are present and vectors
as well as axial-vectors can be included into the model in a straightforward
manner \cite{geffen,urban,UBW,Paper1,gallas}. Note that an important advantage
of linear sigma models is the possibility to investigate the state of matter
at large values of temperature and chemical potential [e.g.\ in the region of
the chiral transition, i.e., where chiral symmetry is restored
\cite{Lenaghan:2000ey, Szep}]. Every linear sigma model contains so-called
chiral partners -- states that mix with each other under axial transformations
-- that become degenerate when chiral symmetry is restored. In the non-linear
realization (i.e., the non-linear sigma models), the scalar and the
(axial-)vector states are integrated out and the pseudoscalar states are the
only degrees of the freedom: one obtains the Lagrangian of chiral perturbation
theory \cite{chpt}. If the vector mesons are not integrated out, one is left
with chiral perturbation theory with vector mesons, see
e.g.\ Ref.\ \cite{chptvm} and refs.\ therein.\newline

In this paper, we present a linear sigma model containing scalar,
pseudoscalar, vector, and axial-vector mesons with both non-strange and
strange quantum numbers. Although this project represents a straightforward
implementation of the principles of the linear realization of chiral symmetry
as already outlined in Ref.\ \cite{geffen}, this is -- to our knowledge -- the
first time that all these degrees of freedom are considered within a single
linear chiral framework. In view of the large number of the fields involved,
our model shall be referred to as \textquotedblleft extended Linear Sigma
Model\textquotedblright, or \textquotedblleft eLSM\textquotedblright.
Moreover, we also exploit a ''classical'' symmetry of the QCD
Lagrangian in the chiral limit: the dilatation symmetry. This symmetry is
broken by quantum effects (trace anomaly) and generates, through
dimensional transmutation, the QCD low-energy scale $\Lambda_{\rm QCD}$.
We describe this phenomenon by including a dilaton field in our model.
The associated potential for the dilaton field encodes the trace
anomaly by an explicit
breaking of dilatation symmetry. We assume that, except for terms
associated with the $U(1)_A$ anomaly and nonzero quark masses, all other
interactions are dilatation invariant. Assuming in addition
analyticity in the fields (i.e., absence of divergences in the 
Lagrangian in the limit of vanishing fields), 
the number of terms appearing
in the Lagrangian is finite. The
fluctuations of the dilaton field correspond to the glueball degree of
freedom. In this work, we neglect the coupling of the glueball
with the other mesonic degrees of freedom. Formally, this can be justified
by taking the large-$N_c$ limit. In a future work, we also plan to
study the coupling of the glueball to mesons in our framework, similar
to the $N_f=2$ study of Ref.\ \cite{Stani}.

In Refs.\ \cite{Paper1, References}, we have already presented a linear sigma
model with vector and axial-vector mesons for two flavors. Comparing our model
with experimental data for meson vacuum phenomenology led us to conclude that
the scalar $\bar{q}q$ states are located in the energy region above 1 GeV in
the hadron mass spectrum. Explicitly, we concluded that the resonances
$f_{0}(1370)$ and $a_{0}(1450)$ are strongly favoured to be scalar quarkonia.
The present work is more general: we now consider three flavors both in the
(pseudo)scalar and (axial-)vector channels in order to ascertain whether the
conclusion of Ref.\ \cite{Paper1} is still valid once strange mesons are
considered. We emphasize that, as discussed in Ref.\ \cite{Paper1}, all fields
entering our model describe pure quark-antiquark states. The reason is that
masses and decay widths of our theoretical states scale as $N_{c}^{0}$ and
$N_{c}^{-1}$, respectively, where $N_{c}$ is the number of colors; thus all
decay widths vanish in the large-$N_{c}$ limit. Note that the inclusion of
strange mesons provides us with a large number of very precise data [in
general, decisively more precise than in the case of non-strange mesons
\cite{PDG}]. As a consequence, our model parameters are much more constrained
than in Ref.\ \cite{Paper1}. Preliminary results of this work have already
been presented in Ref.\ \cite{References2}.

In order to (\textit{i}) test the performance of a linear sigma model in
describing the overall phenomenology of (pseudo)scalar and (axial-)vector
mesons and (\textit{ii}) ascertain which scalar states are predominantly
quarkonia, we perform a global fit in which 21 experimentally known quantities
(decay constants, masses, and decay widths as well as amplitudes) are
included. The situation in the scalar-isoscalar sector is extremely uncertain
because five (six) resonances, $f_{0}(500)$, $f_{0}(980)$, $f_{0}(1370)$,
$f_{0}(1500)$, and $f_{0}(1710)$ [and, possibly in the future, $f_{0}(1790)$]
are listed in the PDG. In addition, the $f_{0}(500)$ and $f_{0}(1370)$ decay
widths and branching ratios are poorly known and $f_{0}(980)$ suffers from a
distortion by the $K\bar{K}$ threshold. Moreover, a scalar glueball
state with a bare mass of
about $1.5-1.7$ GeV as predicted by lattice QCD \cite{Morningstar} can sizably
affect the masses and the branching ratios of the scalar states above 1 GeV,
that is $f_{0}(1370)$, $f_{0}(1500)$, and $f_{0}(1710)$ \cite{glueball}. In
view of all these reasons, we do \emph{not} include the scalar-isoscalar
states into the fit. We note that, although the coupling of the glueball state to the
other mesons is not considered in this paper, it does not affect our global fit.
We do, however, include the isotriplet and isodoublet
quark-antiquark scalar states and we test all four combinations $a_{0}%
(1450)~/~K_{0}^{\star}(1430)$, $a_{0}(980)~/~K_{0}^{\star}(800)$,
$a_{0}(980)~/~K_{0}^{\star}(1430)$, and $a_{0}(1450)~/~K_{0}^{\star}(800)$.
Quite remarkably, the outcome of the fit is univocal: only the pair
$a_{0}(1450)~/~K_{0}^{\star}(1430)$ yields a good fit, while the other
combinations do not. We thus conclude that the $I=1$ and $I=1/2$
quark-antiquark scalar resonances lie above 1 GeV. In fact, the quality of our
fit is surprisingly good. We describe \textit{all} experimental quantities
with an average error of 5\%, and most of them even to much better precision.
We perceive this to be a remarkable achievement within an (in principle, quite
simple) effective model for the strong interaction.

We then study the two scalar-isoscalar quark-antiquark states in our
model: a consequence of our fit
is their mass in the large-$N_{c}$ limit. These masses turn out to be about
$1.36$ GeV and $1.53$ GeV, respectively. Varying large-$N_{c}$ suppressed
parameters which cannot be determined by our fit, one can also study their
decays: the non-strange quark-antiquark state is well described by
$f_{0}(1370)$ and the heavier $\bar{s}s$ state might be $f_{0}(1710).$ Turning
to the axial-vector channel, both masses and decays are well described by
assuming the axial-vector resonances $a_{1}(1260)$, $f_{1}(1285)$,
$f_{1}(1525)$, and $K_{1}(1270)$ as (predominantly) quark-antiquark states.
Detailed calculations of all formulas used in this paper as well as further
discussion are presented in Ref.\ \cite{Doktorarbeit}.\newline

The outline of the paper is as follows. In Sec.\ \ref{model} we discuss the
$U(3)_{L}\times U(3)_{R}$ linear sigma model with vector and axial-vector
mesons. In Sec.\ \ref{fitprocedure} we discuss the global fit and its
consequences and in Sec.\ \ref{sec:conclusion} we present our conclusions. We
defer detailed calculations to Appendices~\ref{App:mass} and \ref{App:decay}.
Our units are $\hbar= c=1$; the metric tensor is $g_{\mu\nu}= \mathrm{diag}
(+,-,-,-)$.

\section{THE MODEL}

\label{model}

\subsection{Lagrangian}

In this section we present our model: a linear sigma model with (axial-)vector
mesons and global chiral $U(3)_{L}\times U(3)_{R}$ symmetry. To this end, we
discuss some important criteria for the construction of its
Lagrangian.\newline The aim of our work is to emulate the properties of the
QCD Lagrangian in our effective approach. This implies the necessity to
consider as many symmetries of QCD as possible. The QCD Lagrangian possesses
various symmetries: local (gauge) invariance with respect to the color group $SU(3)_c$,
discrete $C$, $P$, and $T$ symmetries, global chiral symmetry $U(3)_{L}\times
U(3)_{R}$ (which is exact in the chiral limit), and also the classical
dilatation (scale) symmetry. The local color symmetry is automatically
fulfilled when working with colorless hadronic degrees of freedom. The
discrete and chiral symmetries impose severe constraints on the terms which
are kept in the Lagrangian; still, infinitely many are allowed. Indeed, in
some older versions of the linear sigma model, terms with dimension larger
than four were considered, see for example Refs.\ \cite{geffen,
urban,Meissner}. In these approaches chiral symmetry was promoted to a
\textit{local symmetry}, up to the vector meson mass term which breaks
this symmetry explicitly. Sigma models with local chiral symmetry require the
inclusion of terms of order larger than four in the fields in order to correctly describe
experimental data. This procedure obviously breaks renormalizability, but this
is not an issue because a hadronic theory is obviously not fundamental and is
supposed to be valid only up to a mass scale of $1$--$2$ GeV. Still, the
problem of constraining the number of terms affects these effective approaches
of QCD.

We then turn our attention to the last of the mentioned symmetries of QCD: the
dilatation symmetry. It plays a key role to justify why we retain only a
\emph{finite} number of terms \cite{dynrec}. Let us
recall some of the basic features of the dilatation symmetry in the pure
gauge sector of QCD: The Yang-Mills (YM) Lagrangian (QCD without quarks) is
classically invariant under dilatations, but this symmetry is broken at the
quantum level. The divergence of the corresponding current is the trace of the
energy-momentum tensor $T_{\text{YM}}^{\mu\nu}$ of the YM Lagrangian $\left(
T_{\text{YM}}\right)  _{\mu}^{\mu}=\frac{\beta(g)}{4g}
G_{\mu\nu}^{a}G^{a,\mu\nu} \neq0$, where $G_{\mu\nu}^{a}$ is the gluon
field-strength tensor and $\beta(g)$ the beta function of
YM theory. At the composite level, one can parametrize this situation by
introducing the dilaton field $G$, which is described by the Lagrangian
\cite{schechter}
\begin{equation}
\mathcal{L}_{dil}=\frac{1}{2}(\partial_{\mu}G)^{2}-\frac{1}{4}\frac{m_{G}^{2}%
}{\Lambda^{2}}\left(  G^{4}\ln\frac{G^{2}}{\Lambda^{2}}-\frac{G^{4}}%
{4}\right)  \text{ .} \label{ldil}%
\end{equation}
The dilatation symmetry is explicitly broken by the scale $\Lambda$ under
the logarithm. The minimum of the dilaton potential is realized for
$G_{0}=\Lambda$; upon shifting $G\rightarrow G_{0}+G,$ a particle with mass
$m_{G}$ emerges, which is interpreted as the scalar glueball. The numerical
value has been evaluated in Lattice QCD and reads $m_{G}\sim1.6$ GeV
\cite{Morningstar}. In the large-$N_c$ limit, $m_G \sim N_c^0$, while
the $n$-glueball vertex $\sim N_c^{2-n}$ \cite{wittennc}. Applying this
to Eq.\ (\ref{ldil}), we observe that $\Lambda \sim N_c$.
In the large-$N_c$ limit, the glueball self-interaction terms vanishes
and the glueball becomes a free field.

We are now ready to present our Lagrangian. It follows from
requiring symmetry under $C,P,T$, (global) chiral 
\cite{UBW,Paper1,Doktorarbeit}, as well as dilatation
transformations. In accordance with QCD, we take the latter two symmetries 
to be explicitly broken only by nonzero quark masses, and dilatation
symmetry to be explicitly broken only
by the dilaton potential in Eq.\ (\ref{ldil}) as well as the 
$U_{A}(1)$ anomaly. Therefore, all other terms in the Lagrangian must 
be dilatation invariant and thus carry 
mass dimension equal to four. This would in principle still allow for
an infinite number of terms. However, assuming in addition that there are no
terms with non-analytic powers of the field variables makes the
number of possible terms finite.

Explicitly, the Lagrangian of the model has the form
\begin{align}
\mathcal{L}  &  =\mathcal{L}_{dil}%
+\mathop{\mathrm{Tr}}[(D_{\mu}\Phi)^{\dagger}(D_{\mu}\Phi)]-m_{0}^{2}\left(
\frac{G}{G_{0}}\right)  ^{2}\mathop{\mathrm{Tr}}(\Phi^{\dagger}\Phi
)-\lambda_{1}[\mathop{\mathrm{Tr}}(\Phi^{\dagger}\Phi)]^{2}-\lambda
_{2}\mathop{\mathrm{Tr}}(\Phi^{\dagger}\Phi)^{2}{\nonumber}\\
&  -\frac{1}{4}\mathop{\mathrm{Tr}}(L_{\mu\nu}^{2}+R_{\mu\nu}^{2}%
)+\mathop{\mathrm{Tr}}\left[  \left(  \left(  \frac{G}{G_{0}}\right)
^{2}\frac{m_{1}^{2}}{2}+\Delta\right)  (L_{\mu}^{2}+R_{\mu}^{2})\right]
+\mathop{\mathrm{Tr}}[H(\Phi+\Phi^{\dagger})]{\nonumber}\\
&  +c_{1}(\det\Phi-\det\Phi^{\dagger})^{2}+i\frac{g_{2}}{2}%
(\mathop{\mathrm{Tr}}\{L_{\mu\nu}[L^{\mu},L^{\nu}%
]\}+\mathop{\mathrm{Tr}}\{R_{\mu\nu}[R^{\mu},R^{\nu}]\}){\nonumber}\\
&  +\frac{h_{1}}{2}\mathop{\mathrm{Tr}}(\Phi^{\dagger}\Phi
)\mathop{\mathrm{Tr}}(L_{\mu}^{2}+R_{\mu}^{2})+h_{2}%
\mathop{\mathrm{Tr}}[\vert L_{\mu}\Phi \vert ^{2}+\vert \Phi R_{\mu} \vert ^{2}]+2h_{3}%
\mathop{\mathrm{Tr}}(L_{\mu}\Phi R^{\mu}\Phi^{\dagger}){\nonumber}\\
&  +g_{3}[\mathop{\mathrm{Tr}}(L_{\mu}L_{\nu}L^{\mu}L^{\nu}%
)+\mathop{\mathrm{Tr}}(R_{\mu}R_{\nu}R^{\mu}R^{\nu})]+g_{4}%
[\mathop{\mathrm{Tr}}\left(  L_{\mu}L^{\mu}L_{\nu}L^{\nu}\right)
+\mathop{\mathrm{Tr}}\left(  R_{\mu}R^{\mu}R_{\nu}R^{\nu}\right)
]{\nonumber}\\
&  +g_{5}\mathop{\mathrm{Tr}}\left(  L_{\mu}L^{\mu}\right)
\,\mathop{\mathrm{Tr}}\left(  R_{\nu}R^{\nu}\right)  +g_{6}%
[\mathop{\mathrm{Tr}}(L_{\mu}L^{\mu})\,\mathop{\mathrm{Tr}}(L_{\nu}L^{\nu
})+\mathop{\mathrm{Tr}}(R_{\mu}R^{\mu})\,\mathop{\mathrm{Tr}}(R_{\nu}R^{\nu
})]\text{ ,} \label{eq:Lagrangian}%
\end{align}
where $\mathcal{L}_{dil}$ is the dilaton term (\ref{ldil}) and
\begin{align}
D^{\mu}\Phi &  \equiv\partial^{\mu}\Phi-ig_{1}(L^{\mu}\Phi-\Phi R^{\mu
})-ieA^{\mu}[T_{3},\Phi]\;,\nonumber\\
L^{\mu\nu}  &  \equiv\partial^{\mu}L^{\nu}-ieA^{\mu}[T_{3},L^{\nu}]-\left\{
\partial^{\nu}L^{\mu}-ieA^{\nu}[T_{3},L^{\mu}]\right\}  \;\text{,}\nonumber\\
R^{\mu\nu}  &  \equiv\partial^{\mu}R^{\nu}-ieA^{\mu}[T_{3},R^{\nu}]-\left\{
\partial^{\nu}R^{\mu}-ieA^{\nu}[T_{3},R^{\mu}]\right\}  \;\text{.}\nonumber
\end{align}
The quantities $\Phi$, $R^{\mu}$, and $L^{\mu}$ represent the scalar and
vector nonets:
\begin{align}
\Phi &  =\sum_{i=0}^{8}(S_{i}+iP_{i})T_{i}=\frac{1}{\sqrt{2}}\left(
\begin{array}
[c]{ccc}%
\frac{(\sigma_{N}+a_{0}^{0})+i(\eta_{N}+\pi^{0})}{\sqrt{2}} & a_{0}^{+}%
+i\pi^{+} & K_{0}^{\star+}+iK^{+}\\
a_{0}^{-}+i\pi^{-} & \frac{(\sigma_{N}-a_{0}^{0})+i(\eta_{N}-\pi^{0})}%
{\sqrt{2}} & K_{0}^{\star0}+iK^{0}\\
K_{0}^{\star-}+iK^{-} & {\bar{K}_{0}^{\star0}}+i{\bar{K}^{0}} & \sigma
_{S}+i\eta_{S}%
\end{array}
\right)  \text{ ,}\label{eq:matrix_field_Phi}\\
L^{\mu}  &  =\sum_{i=0}^{8}(V_{i}^{\mu}+A_{i}^{\mu})T_{i}=\frac{1}{\sqrt{2}%
}\left(
\begin{array}
[c]{ccc}%
\frac{\omega_{N}+\rho^{0}}{\sqrt{2}}+\frac{f_{1N}+a_{1}^{0}}{\sqrt{2}} &
\rho^{+}+a_{1}^{+} & K^{\star+}+K_{1}^{+}\\
\rho^{-}+a_{1}^{-} & \frac{\omega_{N}-\rho^{0}}{\sqrt{2}}+\frac{f_{1N}%
-a_{1}^{0}}{\sqrt{2}} & K^{\star0}+K_{1}^{0}\\
K^{\star-}+K_{1}^{-} & {\bar{K}}^{\star0}+{\bar{K}}_{1}^{0} & \omega
_{S}+f_{1S}%
\end{array}
\right)  ^{\mu}\text{,}\label{eq:matrix_field_L}\\
R^{\mu}  &  =\sum_{i=0}^{8}(V_{i}^{\mu}-A_{i}^{\mu})T_{i}=\frac{1}{\sqrt{2}%
}\left(
\begin{array}
[c]{ccc}%
\frac{\omega_{N}+\rho^{0}}{\sqrt{2}}-\frac{f_{1N}+a_{1}^{0}}{\sqrt{2}} &
\rho^{+}-a_{1}^{+} & K^{\star+}-K_{1}^{+}\\
\rho^{-}-a_{1}^{-} & \frac{\omega_{N}-\rho^{0}}{\sqrt{2}}-\frac{f_{1N}%
-a_{1}^{0}}{\sqrt{2}} & K^{\star0}-K_{1}^{0}\\
K^{\star-}-K_{1}^{-} & {\bar{K}}^{\star0}-{\bar{K}}_{1}^{0} & \omega
_{S}-f_{1S}%
\end{array}
\right)  ^{\mu}\text{,} \label{eq:matrix_field_R}%
\end{align}
where the assignment to physical particles is shown as well\footnote{With the
exception of the $(0-8)$ sector where particle mixing takes place (see
below).}. Here, $T_{i}\,(i=0,\ldots,8)$ denote the generators of $U(3)$, while
$S_{i}$ represents the scalar, $P_{i}$ the pseudoscalar, $V_{i}^{\mu}$ the
vector, $A_{i}^{\mu}$ the axial-vector meson fields, and $A^{\mu}$ is the
electromagnetic field. It should be noted that here and below we use the
so-called non strange -- strange basis in the $(0-8)$ sector, defined as
\begin{align}
\varphi_{N}  &  =\frac{1}{\sqrt{3}}\left(  \sqrt{2}\;\varphi_{0}+\varphi
_{8}\right)  \;,\nonumber\\
\varphi_{S}  &  =\frac{1}{\sqrt{3}}\left(  \varphi_{0}-\sqrt{2}\;\varphi
_{8}\right)  \;,\quad\quad\varphi\in(S_{i},P_{i},V_{i}^{\mu},A_{i}^{\mu})\;,
\label{eq:nsbase}%
\end{align}
which is more suitable for our calculations. Moreover, $H$ and $\Delta$ are
constant external fields defined as
\begin{align}
H  &  =H_{0}T_{0}+H_{8}T_{8}=\left(
\begin{array}
[c]{ccc}%
\frac{h_{0N}}{2} & 0 & 0\\
0 & \frac{h_{0N}}{2} & 0\\
0 & 0 & \frac{h_{0S}}{\sqrt{2}}%
\end{array}
\right)  \;,\label{eq:expl_sym_br_epsilon}\\
\Delta &  =\Delta_{0}T_{0}+\Delta_{8}T_{8}=\left(
\begin{array}
[c]{ccc}%
\frac{\tilde{\delta}_{N}}{2} & 0 & 0\\
0 & \frac{\tilde{\delta}_{N}}{2} & 0\\
0 & 0 & \frac{\tilde{\delta}_{S}}{\sqrt{2}}%
\end{array}
\right)  \equiv\left(
\begin{array}
[c]{ccc}%
\delta_{N} & 0 & 0\\
0 & \delta_{N} & 0\\
0 & 0 & \delta_{S}%
\end{array}
\right)  \text{ .} \label{eq:expl_sym_br_delta}%
\end{align}
These terms describe the effect of nonzero quark masses in the (pseudo)scalar
and (axial-)vector sectors, respectively: $h_{N}\sim m_{u}$, $h_{S}\sim m_{s}%
$, $\delta_{N}\sim m_{u}^{2}$, $\delta_{S}\sim m_{s}^{2}$. Throughout this
work we assume exact isospin symmetry for $u$ and $d$ quarks, such that the
first two diagonal elements in Eq.\ \eqref{eq:expl_sym_br_epsilon} and
Eq.~\eqref{eq:expl_sym_br_delta} are identical. Thus, only the
scalar-isoscalar fields
$\sigma_{N},\, \sigma_{S}$ and $G$, carrying the same quantum numbers as the
vacuum, can have nonzero vacuum expectation values (vev's)\footnote{In case of
isospin breaking, also $\sigma_{3}$ could have a nonzero vev.}.

In the Lagrangian (\ref{eq:Lagrangian}) the following terms break the
original $U(3)_{L}\times U(3)_{R}$ [$=U(3)_{V}\times U(3)_{A}$]
symmetry: {\it (i)} the terms
proportional to the matrix $H$ and $\Delta$ of
Eqs.\ \eqref{eq:expl_sym_br_epsilon} and (\ref{eq:expl_sym_br_delta}), which
describes the explicit symmetry breaking due to the nonzero values of the
quark masses in the (pseudo)scalar and (axial-)vector sectors and break
$U(3)_{A}$ if $H_{0},\Delta_{0}\neq0$ and $U(3)_{V}\rightarrow SU(2)_{V}\times
U(1)_{V}$ if $H_{8},\Delta_{8}\neq0$ [for more details see,
e.g.\ Ref.\ \cite{Lenaghan:2000ey}], and {\it (ii)} 
the term proportional to the determinant
and parametrized by $c_{1}$, which breaks the $U(1)_{A}$ symmetry and
describes the axial anomaly, responsible for large mass of the
$\eta^{\prime}$ meson.

These terms also explicitly break dilatation symmetry because they
involve dimensionful coupling constants. This is expected: the first two terms
describe the bare quark masses which generate an explicit breaking of
dilatation symmetry at the level of the QCD Lagrangian, and the determinant
term describes an anomalous breaking of the dilatation symmetry arising from
the YM sector of the theory.

The interaction of the meson fields with the dilaton field $G$ enters
only in two terms. Upon condensation of the dilaton field, these
terms correspond to meson mass terms. In addition, there are interaction terms proportional
to one or two powers of the glueball field. Since
$G_0 \sim \Lambda \sim N_c$, an $(m+g)$-point vertex involving $m$
meson lines and $g$ glueball lines scales as $\sim N_c^{-(m/2+g-1)}$, while an
$m$-point vertex involving $m$ meson lines scales as
$N_c^{1-m/2}$. Thus, vertices with $n$ external lines involving glueballs vanish
faster than the corresponding $n$-point vertices involving only
mesons. As a first approximation, we assume that the glueball field
completely decouples, so that we neglect it in the following. Effects
from coupling the glueball to the other meson fields will be studied
in a subsequent work.

Let us now discuss in detail the assignment of fields in
Eqs.\ (\ref{eq:matrix_field_Phi}) -- (\ref{eq:matrix_field_R}). If we consider
isospin multiplets as single degrees of freedom, then there are 16 resonances
that can be described by the model: $\sigma_{N}$, $\sigma_{S}$, $\vec{a}_{0}$,
$K_{0}^{\star}$ (scalar); $\eta_{N}$, $\eta_{S}$, $\vec{\pi}$, $K$
(pseudoscalar); $\omega_{N}^{\mu}$, $\omega_{S}^{\mu}$, $\vec{\rho}^{\;\mu}$,
$K^{\star\mu}$ (vector), and $f_{1N}^{\mu}$, $f_{1S}^{\mu}$, $\vec{a}%
_{1}^{\;\mu}$, $K_{1}$ (axial-vector). All fields in our model represent
$\bar{q}q$ states, as discussed in Ref.\ \cite{Paper1}. If we assign a state
from our model to a physical resonance we, therefore, implicitly assume that
this resonance is a ${\bar{q}}q$ state. This assumption can be tested for a
multitude of physical resonances in the scalar and axial-vector sectors, as
discussed below (in the pseudoscalar and the vector channels, there are no ambiguities).

In the non-strange sector, we assign the fields $\vec{\pi}$ and $\eta_{N}$ to
the pion and the non-strange part of the $\eta$ and $\eta^{\prime}$ mesons,
$\eta_{N}\equiv(\bar{u}u+\bar{d}d)/\sqrt{2}$. The fields $\omega_{N}^{\mu}$
and $\vec{\rho}^{\;\mu}$ represent the $\omega(782)$ and $\rho(770)$ vector
mesons, respectively, and the fields $f_{1N}^{\mu}$ and $\vec{a}_{1}^{\;\mu}$
represent the $f_{1}(1285)$ and $a_{1}(1260)$ mesons, respectively. In the
strange sector, we assign the $K$ fields to the kaons; the $\eta_{S}$ field is
the strange contribution to the physical $\eta$ and $\eta^{\prime}$ fields
[$\eta_{S}\equiv\bar{s}s$]; the $\omega_{S}$, $f_{1S}$, $K^{\star}$, and
$K_{1}$ fields correspond to the $\phi(1020)$, $f_{1}(1420)$, $K^{\star}%
(892)$, and $K_{1}(1270)$ [or $K_{1}(1400)$] mesons, respectively.

Unfortunately, the assignment of the scalar fields is substantially less
clear. Experimental data suggest existence of five (six) scalar-isoscalar
states below $1.8$ GeV: $f_{0}(500)$, $f_{0}(980)$, $f_{0}(1370)$,
$f_{0}(1500)$, and $f_{0}(1710)$ [as well as $f_{0}(1790)$]. Note that the
existence of these five states is acknowledged by the Particle Data Group --
PDG \cite{PDG}. [The sixth state, $f_{0}(1790)$, will not be of importance for
the rest of our work because its predominant $\pi\pi$ decay renders it a
putative radial excitation of $f_{0}(1370)$, and our model describes
ground-state quarkonia only.]\newline

Our model contains a pure non-strange isoscalar $\sigma_{N}$ and a pure
strange isoscalar $\sigma_{S}$. We will demonstrate below that our model
yields mixing of $\sigma_{N}$ and $\sigma_{S}$, producing a predominantly
non-strange state labeled as $f_{0}^{L}$,\ and a predominantly strange state
labeled as $f_{0}^{H}$. Assignment of the mixed states to physical
resonances is ambiguous because, as already discussed, there are five physical
states all of which could, in principle, be candidates for $f_{0}^{L}$ and
$f_{0}^{H}$.

Similarly, the isospin triplet ${a}_{0}$ can be assigned to different physical
resonances -- although, in this case, there are only two candidate states:
$a_{0}(980)$ and $a_{0}(1450)$. A preliminary examination of the assignment
of the ${a}_{0}$\ field has been performed in Ref.\ \cite{Paper1} where it was
concluded that it most likely corresponds to the $a_{0}(1450)$ resonance [or,
equivalently, $a_{0}(1450)$ rather than $a_{0}(980)$ was favored to represent
a ${\bar{q}}q$ state]. The discussion in Ref.\ \cite{Paper1} was limited to
non-strange mesons. In this work, besides the assignment of ${a}_{0}$, we also
have to consider possible assignments for the strange scalar field
$K_{0}^{\star}$; there are also two possibilities: either $K_{0}^{\star}(800)$
or $K_{0}^{\star}(1430)$.\newline

\subsection{Tree-level masses and mixing terms}

\label{ssec:SSB_and_mass}

After spontaneous symmetry breaking, the fields with nonzero vev's are shifted
by their expectation values, namely, $\sigma_{N/S}\rightarrow\sigma_{N/S}%
+\phi_{N/S}$, where we have introduced $\phi_{N/S}\equiv\langle\sigma_{N/S}
\rangle$. After substituting the shifted fields into the Lagrangian
\eqref{eq:Lagrangian}, one obtains the tree-level masses by selecting all
terms quadratic in the fields,
\begin{align}
\mathcal{L}^{\mathrm{quad}}  &  =-\frac{1}{2}S_{i}\left[  \delta_{ij}\,
\Box+(m_{S}^{2})_{ij}\right]  S_{j}\nonumber\\
&  -\frac{1}{2}P_{i}\left[  \delta_{ij}\, \Box+(m_{P}^{2})_{ij}\right]
P_{j}\nonumber\\
&  -\frac{1}{2}V_{i\mu}\left[  (-g^{\mu\nu}\,\Box+\partial^{\mu}\partial^{\nu
})\delta_{ij}-g^{\mu\nu}(m_{V}^{2})_{ij}\right]  V_{j\nu}\nonumber\\
&  -\frac{1}{2}A_{i\mu}\left[  (-g^{\mu\nu}\,\Box+\partial^{\mu}\partial^{\nu
})\delta_{ij}-g^{\mu\nu}(m_{A}^{2})_{ij}\right]  A_{j\nu}\nonumber\\
&  -\frac{1}{2}V_{i\mu}\left(  ig_{1}f_{ijk}\phi_{k}\partial^{\mu}\right)
S_{j}-\frac{1}{2}S_{i}\left(  ig_{1}f_{ijk}\phi_{k}\partial^{\nu}\right)
V_{j\nu}\nonumber\\
&  +\frac{1}{2}A_{i\mu}\left(  g_{1}d_{ijk}\phi_{k}\partial^{\mu}\right)
P_{j}-\frac{1}{2}P_{i}\left(  g_{1}d_{ijk}\phi_{k}\partial^{\nu}\right)
A_{j\nu}, \label{eq:quad}%
\end{align}
where $(m_{S}^{2})_{ij},(m_{P}^{2})_{ij},(m_{V}^{2})_{ij}$, and $(m_{A}%
^{2})_{ij}$ are the scalar, pseudoscalar, vector, and axial-vector (squared)
mass matrices, respectively, see their explicit expressions in
Appendix~\ref{App:mass}. Moreover, $f_{ijk}$ and $d_{ijk}$ are the
antisymmetric and symmetric structure constants of $U(3)$. The (squared) mass
matrices are in general non-diagonal due to the mixing among particles sitting
in the center of a given nonet, and they can be diagonalized by appropriate
orthogonal transformations (for details see Appendix~\ref{App:mass}). Besides
the mixing inside the nonets there are other mixing terms, namely the last
four terms of Eq.~\eqref{eq:quad}, which mix different nonets.

In order to eliminate the latter, one performs the following shifts of the
(axial-)vector fields:
\begin{align}
f_{1N/S}^{\mu}  &  \longrightarrow f_{1N/S}^{\mu}+Z_{\eta_{N/S}}w_{f_{1N/S}%
}\partial^{\mu}\eta_{N/S}\text{ },\text{ }{a_{1}^{\mu}}^{\pm,0}\longrightarrow
{a_{1}^{\mu}}^{\pm,0}+Z_{\pi}w_{a_{1}}\partial^{\mu}\pi^{\pm,0},\nonumber\\
{K_{1}^{\mu}}^{\pm,0,\bar{0}}  &  \longrightarrow{K_{1}^{\mu}}^{\pm,0,\bar{0}%
}+Z_{K}w_{K_{1}}\partial^{\mu}K^{\pm0,\bar{0}}\text{ },\text{ }{K^{\star\mu}%
}^{\pm,0,\bar{0}}\longrightarrow{K^{\star\mu}}^{\pm,0,\bar{0}}+Z_{K^{\star}%
}w_{K^{\star}}\partial^{\mu}K_{0}^{\star\pm,0,\bar{0}}\text{ }.\text{ }
\label{eq:shifts}%
\end{align}
These shifts produce additional kinetic terms for the pseudoscalar fields. In
order to retain the canonical normalization for the latter, one has to
introduce wavefunction renormalization constants,
\begin{equation}
\pi^{\pm,0}\rightarrow Z_{\pi}\pi^{\pm,0},\text{ }K^{\pm,0,\bar{0}}\rightarrow
Z_{K}K^{\pm,0,\bar{0}}\text{ },\text{ }\eta_{N/S}\rightarrow Z_{\eta_{N}%
/\eta_{S}}\eta_{N/S}\text{ , }{K^{\star\mu}}^{\pm,0,\bar{0}}\rightarrow
Z_{K^{\star}}{K^{\star\mu}}^{\pm,0,\bar{0}}\;.
\end{equation}
For the sake of simplicity we have grouped together the isotriplet states with
the notation $\pi^{\pm,0},{a_{1}^{\mu}}^{\pm,0}$ and the isodoublet states
with the notation $K^{\pm,0,\bar{0}},{K^{\star\mu}}^{\pm,0,\bar{0}}$, where
$\bar{0}$ refers to $\bar{K}^{0}$. The coefficients $w_{i}$ and $Z_{i}$ are
determined in order to eliminate the last four mixing terms in
Eq.\ (\ref{eq:quad}) and to obtain the canonical normalization of the $\pi$,
$\eta_{N}$, $\eta_{S}$, $K$, and $K_{0}^{\star}$ fields. After some
straightforward calculation one finds the explicit expressions:
\begin{equation}
w_{f_{1N}}=w_{a_{1}}=\frac{g_{1}\phi_{N}}{m_{a_{1}}^{2}}\text{, }w_{f_{1S}%
}=\frac{\sqrt{2}g_{1}\phi_{S}}{m_{f_{1S}}^{2}}\text{, }w_{K^{\star}}%
=\frac{ig_{1}(\phi_{N}-\sqrt{2}\phi_{S})}{2m_{K^{\star}}^{2}}\text{, }%
w_{K_{1}}=\frac{g_{1}(\phi_{N}+\sqrt{2}\phi_{S})}{2m_{K_{1}}^{2}}\text{,}%
\end{equation}%
\begin{align}
Z_{\pi}  &  =Z_{\eta_{N}}=\frac{m_{a_{1}}}{\sqrt{m_{a_{1}}^{2}-g_{1}^{2}%
\phi_{N}^{2}}}\text{ ,} & Z_{K} =\frac{2m_{K_{1}}}{\sqrt{4m_{K_{1}}^{2}%
-g_{1}^{2}(\phi_{N}+\sqrt{2}\phi_{S})^{2}}}\text{ ,}\label{Z_pi}\\
Z_{\eta_{S}}  &  =\frac{m_{f_{1S}}}{\sqrt{m_{f_{1S}}^{2}-2g_{1}^{2}\phi
_{S}^{2}}}\text{ ,} & Z_{K^{\star}} =\frac{2m_{K^{\star}}}{\sqrt{4m_{K^{\star
}}^{2}-g_{1}^{2}(\phi_{N}-\sqrt{2}\phi_{S})^{2}}}\;. \label{Z_K_S}%
\end{align}
It can be seen from the expressions of the wavefunction renormalization
constants that they are always larger than one. Finally, using the explicit
expressions found in Appendix~\ref{App:mass} the tree-level (squared) masses
for the different nonets are obtained as follows:
\begin{align}
m_{\pi}^{2}  &  =Z_{\pi}^{2}\left[  m_{0}^{2}+\left(  \lambda_{1}%
+\frac{\lambda_{2}}{2}\right)  \phi_{N}^{2}+\lambda_{1}\phi_{S}^{2}\right]
\equiv\frac{Z_{\pi}^{2}h_{0N}}{\phi_{N}}\text{ ,}\label{m_pi}\\
m_{K}^{2}  &  =Z_{K}^{2}\left[  m_{0}^{2}+\left(  \lambda_{1}+\frac
{\lambda_{2}}{2}\right)  \phi_{N}^{2}-\frac{\lambda_{2}}{\sqrt{2}}\phi_{N}%
\phi_{S}+\left(  \lambda_{1}+\lambda_{2}\right)  \phi_{S}^{2}\right]  \text{
,}\label{mkaon}\\
m_{\eta_{N}}^{2}  &  =Z_{\pi}^{2}\left[  m_{0}^{2}+\left(  \lambda_{1}%
+\frac{\lambda_{2}}{2}\right)  \phi_{N}^{2}+\lambda_{1}\phi_{S}^{2}%
+c_{1}\,\phi_{N}^{2}\phi_{S}^{2}\right]  \equiv Z_{\pi}^{2} \left(
\frac{h_{0N}}{\phi_{N}}+c_{1}\,\phi_{N}^{2}\phi_{S}^{2}\right)
\;,\label{eq:etaN}\\
m_{\eta_{S}}^{2}  &  =Z_{\eta_{S}}^{2}\left[  m_{0}^{2}+\lambda_{1}\phi
_{N}^{2}+\left(  \lambda_{1}+\lambda_{2}\right)  \phi_{S}^{2}+\frac{c_{1}}%
{4}\phi_{N}^{4}\right]  \equiv Z_{\eta_{S}}^{2} \left(  \frac{h_{0S}}{\phi
_{S}}+\frac{c_{1}}{4}\phi_{N}^{4}\right)  \text{ ,}\label{eq:etaS}\\
m_{\eta_{NS}}^{2}  &  =Z_{\pi}Z_{\pi_{S}}\frac{c_{1}}{2}\phi_{N}^{3}\phi
_{S}\text{ ,} \label{eq:etaNS}%
\end{align}
are the (squared) pseudoscalar masses, while
\begin{align}
m_{a_{0}}^{2}  &  =m_{0}^{2}+\left(  \lambda_{1}+\frac{3}{2}\lambda
_{2}\right)  \phi_{N}^{2}+\lambda_{1}\phi_{S}^{2}\text{ ,}\label{m_a_0}\\
m_{K_{0}^{\star}}^{2}  &  =Z_{K_{0}^{\star}}^{2}\left[  m_{0}^{2}+\left(
\lambda_{1}+\frac{\lambda_{2}}{2}\right)  \phi_{N}^{2}+\frac{\lambda_{2}%
}{\sqrt{2}}\phi_{N}\phi_{S}+\left(  \lambda_{1}+\lambda_{2}\right)  \phi
_{S}^{2}\right]  \text{ ,}\\
m_{\sigma_{N}}^{2}  &  =m_{0}^{2}+3\left(  \lambda_{1}+\frac{\lambda_{2}}%
{2}\right)  \phi_{N}^{2}+\lambda_{1}\phi_{S}^{2}\text{ ,}\label{eq:sigN}\\
m_{\sigma_{S}}^{2}  &  =m_{0}^{2}+\lambda_{1}\phi_{N}^{2}+3\left(  \lambda
_{1}+\lambda_{2}\right)  \phi_{S}^{2}\text{ ,}\label{eq:sigS}\\
m_{\sigma_{NS}}^{2}  &  =2\lambda_{1}\phi_{N}\phi_{S}\text{ ,}
\label{eq:sigNS}%
\end{align}
are the (squared) scalar masses. The quantities $m_{\pi_{NS}}^{2}$ and
$m_{\sigma_{NS}}^{2}$ are mixing terms in the non-strange--strange sector.
These mixings can be removed by orthogonal transformations, and the resulting
mass eigenstates are found to be
\begin{align}
m_{{f_{0}^{H}}/{f_{0}^{L}}}^{2}  &  =\frac{1}{2}\left[  m_{\sigma_{N}}%
^{2}+m_{\sigma_{S}}^{2}\pm\sqrt{(m_{\sigma_{N}}^{2}-m_{\sigma_{S}}^{2}%
)^{2}+4m_{\sigma_{NS}}^{4}}\right]  \;,\label{m_f0LH}\\
m_{\eta^{\prime}/\eta}^{2}  &  =\frac{1}{2}\left[  m_{\eta_{N}}^{2}%
+m_{\eta_{S}}^{2}\pm\sqrt{(m_{\eta_{N}}^{2}-m_{\eta_{S}}^{2})^{2}%
+4m_{\eta_{NS}}^{4}}\right]  \;.
\end{align}
Moreover, the (squared) vector masses are given by
\begin{align}
m_{\rho}^{2}  &  =m_{1}^{2}+\frac{1}{2}(h_{1}+h_{2}+h_{3})\phi_{N}^{2}%
+\frac{h_{1}}{2}\phi_{S}^{2}+2\delta_{N}\;,\label{m_rho}\\
m_{K^{\star}}^{2}  &  =m_{1}^{2}+\frac{1}{4}\left(  g_{1}^{2}+2h_{1}%
+h_{2}\right)  \phi_{N}^{2}+\frac{1}{\sqrt{2}}\phi_{N}\phi_{S}(h_{3}-g_{1}%
^{2})+\frac{1}{2}(g_{1}^{2}+h_{1}+h_{2})\phi_{S}^{2}+\delta_{N}+\delta
_{S}\;,\label{m_K_star}\\
m_{\omega_{N}}^{2}  &  =m_{\rho}^{2}\;,\\
m_{\omega_{S}}^{2}  &  =m_{1}^{2}+\frac{h_{1}}{2}\phi_{N}^{2}+\left(
\frac{h_{1}}{2}+h_{2}+h_{3}\right)  \phi_{S}^{2}+2\delta_{S}\;,
\end{align}
while the (squared) axial-vector meson masses are
\begin{align}
m_{a_{1}}^{2}  &  =m_{1}^{2}+\frac{1}{2}(2g_{1}^{2}+h_{1}+h_{2}-h_{3})\phi
_{N}^{2}+\frac{h_{1}}{2}\phi_{S}^{2}+2\delta_{N}\;,\label{m_a_1}\\
m_{K_{1}}^{2}  &  =m_{1}^{2}+\frac{1}{4}\left(  g_{1}^{2}+2h_{1}+h_{2}\right)
\phi_{N}^{2}-\frac{1}{\sqrt{2}}\phi_{N}\phi_{S}(h_{3}-g_{1}^{2})+\frac{1}%
{2}\left(  g_{1}^{2}+h_{1}+h_{2}\right)  \phi_{S}^{2}+\delta_{N}+\delta
_{S}\;,\label{m_K_1}\\
m_{f_{1N}}^{2}  &  =m_{a_{1}}^{2}\;,\\
m_{f_{1S}}^{2}  &  =m_{1}^{2}+\frac{h_{1}}{2}\phi_{N}^{2}+\left(  2g_{1}%
^{2}+\frac{h_{1}}{2}+h_{2}-h_{3}\right)  \phi_{S}^{2}+2\delta_{S}\;.
\label{m_f1_S}%
\end{align}
It is interesting to note that in case of vectors and axial-vectors there are
no mixings in the non-strange--strange sector.

\subsection{Parameters}

\label{param}

The Lagrangian (\ref{eq:Lagrangian}) contains 18 parameters (as
mentioned above, we neglect the coupling of the glueball with the
other mesons in the present work):
\begin{equation}
m_{0}^{2}\text{, }m_{1}^{2}\text{, }c_{1}\text{, }\delta_{N}\text{, }%
\delta_{S}\text{, }g_{1}\text{, }g_{2}\text{, }g_{3}\text{, }g_{4}\text{,
}g_{5}\text{, }g_{6}\text{, }h_{0N}\text{, }h_{0S}\text{, }h_{1}\text{, }%
h_{2}\text{, }h_{3}\text{, }\lambda_{1}\text{, }\lambda_{2}\text{.}%
\end{equation}

We make the following observations regarding the model parameters:

\begin{itemize}
\item The parameters $h_{0N}$ and $h_{0S}$ model the explicit breaking of the
chiral symmetry (ESB) in the (pseudo)scalar sector via the term $\mathrm{Tr}%
[H(\Phi+\Phi^{\dagger})]$; they are uniquely determined from the mass terms of
the pion, Eq.\ (\ref{m_pi}), and of $\eta_{S}$,\ the strange part of the
$\eta$ meson [see Eq.\ (\ref{eq:etaS})], implying that the masses of $\vec
{\pi}$ and $\eta_{S}$ are generated by ESB.

\item The parameters $\delta_{N}$ and $\delta_{S}$ model the explicit symmetry
breaking in the vector and axial-vector channels. The ESB arises from
non-vanishing quark masses and therefore we employ the correspondence
$\delta_{N}\propto m_{u,d}^{2}$ and $\delta_{S}\propto m_{s}^{2}$. However, in
the vector-meson mass term $\mathrm{Tr}[(m_{1}^{2}/2+\Delta)(L_{\mu}%
^{2}+R_{\mu}^{2})]$ only the linear combinations $m_{1}^{2}/2+\delta_{N/S}$
appear. Therefore, it is possible to redefine $m_{1}^{2}/2\rightarrow
m_{1}^{2}/2-\delta_{N}$. Then, only the combination $\delta_{S}-\delta_{N}$
appears in the mass formulas. Only this difference will be determined by the
fit of the (axial-)vector masses. Alternately, we may set $\delta_{N}=0$ from
the beginning, and determine $\delta_{S}$ from the fit.

\item The parameters $g_{3}$, $g_{4}$, $g_{5}$, and $g_{6}$ do not influence
any of the decays to be discussed in this work and are therefore not
considered in the fit.
\end{itemize}

Consequently, we are left with the following 13 parameters:
\begin{equation}
m_{0}^{2}\text{, }m_{1}^{2}\text{, }c_{1}\text{, }\delta_{S}\text{, }%
g_{1}\text{, }g_{2}\text{ , }h_{0N}\text{, }h_{0S}\text{, }h_{1}\text{, }%
h_{2}\text{, }h_{3}\text{, }\lambda_{1}\text{, }\lambda_{2}\text{.}
\label{13p}%
\end{equation}

Their large-$N_{c}$ dependence reads \cite{Paper1}:%
\begin{equation}
m_{0}^{2}\text{, }m_{1}^{2}\text{, }\delta_S\propto N_{c}^{0}\text{ ; }\;h_{0N}\text{, }%
h_{0S}\propto N_{c}^{1/2}\text{ ; }\;g_{1}\text{, }g_{2}\propto N_{c}%
^{-1/2}\text{ ; }\;\lambda_{2},\text{ }h_{2}\text{, }h_{3}\propto N_{c}%
^{-1}\text{ ; }\;\lambda_{1},h_{1}\propto N_{c}^{-2}\text{ ; }\;c_{1}\propto
N_{c}^{-3}\;. \label{ncdep}%
\end{equation}

We recall that a non-suppressed $n$-meson interaction vertex scales as
$N_{c}^{1-n/2}$ \cite{wittennc}. In this respect the parameters $h_{1}$ and
$\lambda_{1}$ are suppressed in the large-$N_{c}$ limit: in fact, they scale
as $N_{c}^{-2}$ and not as $N_{c}^{-1}$. Similarly, the axial-anomaly
parameter $c_{1}$ scales as $N_{c}^{-3}$ and not as $N_{c}^{-2}$. Note that
the large-$N_{c}$ behaviour of the model parameters (\ref{ncdep}) implies that
the decay widths (the formulas for which are given in Appendix \ref{App:decay}%
) decrease with increasing $N_{c}$. For this reason, as already mentioned, the
states in our model represent ${\bar{q}}q$ states.

\section{Fit: Results and Consequences}

\label{fitprocedure}

\subsection{Input and constraints for the fit}

Let us now turn to the fit procedure for the parameters discussed above. Our
fit aims to ascertain whether (\textit{i}) it is possible to find a fit
containing masses and decay widths for (pseudo)scalar and (axial-)vector
degrees of freedom present in our model and (\textit{ii}) which physical
scalar states are best described by the ${\bar{q}}q$ states of our model.

All decays considered are two-particle decays, thus they can be calculated
from the appropriate tree-level terms of the Lagrangian \eqref{eq:Lagrangian}
after applying the necessary field shifts \eqref{eq:shifts} and orthogonal
transformations \eqref{eq:orthog}. If the decaying particle is denoted by $A$
and the decay products by $B$ and $C$, respectively, the decay width reads
\begin{equation}
\Gamma_{A\rightarrow BC}=\mathcal{I}\frac{|\mathbf{k}|}{8\pi m_{A}^{2}%
}\left\vert \mathcal{M}_{A\rightarrow BC}\right\vert ^{2} \; , \label{Gamma}%
\end{equation}
where $\mathbf{k}$ is the three-momentum of one of the resulting particles in
the rest frame of $A$ and $\mathcal{M}_{A\rightarrow BC}$ is the transition
matrix element (decay amplitude). Moreover, $\mathcal{I}$ refers to the
so-called isospin factor which is the number of sub-channels in a given decay
channel (for instance, if $B=C=K$ the $A\rightarrow KK$ decay can have two
sub-channels, namely $K^{+}K^{-}$ and $\bar{K}^{0}K^{0}$, which results in
$\mathcal{I}=2$). Equation (\ref{Gamma}) will be used to calculate all decay
widths entering the fit; details of the calculations can be found in Appendix
\ref{App:decay}. Moreover, when identical particles emerge in the final state,
the usual symmetry factors are included.

A brief remark is necessary regarding the errors used in the fit. The fit will
contain input information from experimental data regarding both (axial-)vector
and (pseudo)scalar states. The data are very precise for some of the
resonances described by our model. For example, the mass of the $\phi(1020)$
resonance (our $\omega_{S}$ state) is known with $0.002\%$ accuracy. Our model
does not aim to describe hadron vacuum phenomenology with this extreme
precision. The reason is simple: already isospin-breaking effects in the
physical hadron mass spectrum are of the order of 5\% [for instance the
difference between the charged and neutral pion masses, or the masses of the
$a_{1}(1260)$ and the $f_{1}(1285)$], but are completely neglected in our
model. We therefore artificially increase the experimental errors to 5\%, if
the actual error is smaller, or we use the experimental values, if the error
is larger than 5\%.\newline

Let us now discuss the input information for our fit. We do this separately
for mesons of different spin [central values from PDG \cite{PDG} unless
otherwise stated, errors according to the above discussion]:

\begin{itemize}
\item \textit{Weak decay constants:} We use
\begin{equation}
\label{eq:weakdecayconstants}f_{\pi} = (92.2\pm4.6)~\text{MeV, } \;\; f_{K} =
(155.6/ \sqrt{2}\pm5.5)~\text{MeV}\;.
\end{equation}
The following formulas relate the decay constants to the vacuum condensates:
$f_{\pi}=\phi_{N}/Z_{\pi}$ and $f_{K}=$ $\left(  \sqrt{2} \phi_{S} + \phi
_{N}\right)  /(2 Z_{K}).$

\item \textit{Pseudoscalar mesons:} We use
\begin{align}
m_{\pi}  &  = (138\pm6.9)~\text{MeV, } \;\;\; m_{K}=(495.6\pm24.8)~\text{MeV,}%
\nonumber\\
m_{\eta}  &  = (547.9\pm27.4)~\text{MeV, } \;\;\; m_{\eta^{\prime}} = (957.8
\pm47.9)~\text{MeV}\;. \label{PI}%
\end{align}
with the following notes: (\textit{i}) $m_{\pi}$ and $m_{K}$ represent
isospin-averaged values; (\textit{ii}) the relatively large error values come
from the criterion $\max$(5\%, experimental error), discussed above.

\item \textit{Vector mesons:} We use
\begin{align}
m_{\rho}  &  = (775.5 \pm38.8)~\text{MeV, }\;\; m_{K^{\star}} = (893.8
\pm44.7)~\text{MeV, }\;\; m_{\phi}=(1019.5 \pm51)~\text{MeV,}\nonumber\\
\Gamma_{\rho\rightarrow\pi\pi}  &  = (149.1 \pm7.4)~\text{MeV, }\;\;
\Gamma_{K^{\star}\rightarrow K\pi} = (46.2 \pm2.3)~\text{MeV, } \;\;\Gamma
_{\phi\rightarrow KK} = (3.54 \pm0.178)~\text{MeV} \;, \label{VI}%
\end{align}
with the following notes: (\textit{i}) we use the isospin-averaged value for
$m_{K^{\star}}$; (\textit{ii}) in case of the $\phi$ decay we use the physical
mass values in the kinematic factor $k_{K}^{3} = (m_{\phi}^{2} - 4m_{K}%
^{2})^{3/2}$ due to the proximity of $m_{\phi(1020)}$ to the $\bar{K}K$
threshold (in order to eliminate phase-space effects).

\item \textit{Axial-vector mesons:} We use
\begin{align}
m_{a_{1}}  &  = (1230 \pm61.5)~\text{MeV, }\;\; m_{f_{1}(1420)} = (1426.4
\pm71.3)~\text{MeV}\;,\nonumber\\
\Gamma_{a_{1}\rightarrow\rho\pi}  &  = (425 \pm175)~\text{MeV, }
\;\;\Gamma_{a_{1}\rightarrow\pi\gamma} = (0.640 \pm0.250)~\text{MeV, }\;\;
\Gamma_{f_{1}(1420)\rightarrow K^{\star}K} = (43.9 \pm2.2)~\text{MeV}\;,
\label{AVI}%
\end{align}
with the following notes: (\textit{i}) the value of $\Gamma_{a_{1}
\rightarrow\rho\pi}$ is not precisely known; there are experimental data
suggesting this decay channel to be dominant for $a_{1}$ \cite{Asner:1999} and
thus we estimate the possible range for $\Gamma_{a_{1}\rightarrow\rho\pi}$
from the interval for the full $a_{1}$ decay width [$=(250-600)$ MeV];
(\textit{ii}) according to PDG the channel $f_{1}(1420)\rightarrow K^{\star}K$
is dominant within the channel $f_{1}(1420)\rightarrow KK\pi$, with the latter
being the overall dominant decay mode for $f_{1}(1420)$; we have assumed
$\Gamma_{f_{1}(1420)\rightarrow KK\pi}$ to be equal to the full decay width
$\Gamma_{f_{1}(1420)}=(54.9\pm2.6)$ MeV and determined $\Gamma_{f_{1}
(1420)\rightarrow K^{\star}K}$ using an averaged branching ratio
$\Gamma_{f_{1}(1420)\rightarrow K^{\star}K}/\Gamma_{f_{1}(1420)\rightarrow
KK\pi}=(0.8\pm0.09)$ from Refs.\ \cite{Bromberg} and \cite{Dionisi}.

\item \textit{Isotriplet and isodoublet scalar mesons:} The observables from
Eqs.\ (\ref{PI}) -- (\ref{AVI}) will be used with any of the ${a}_{0}$
-$K_{0}^{\star}$ combinations [$a_{0}(980)$/$K_{0}^{\star}(800)$, $a_{0}
(980)$/$K_{0}^{\star}(1430)$, $a_{0}(1450)$/$K_{0}^{\star}(800)$,
$a_{0}(1450)$/$K_{0}^{\star}(1430)$], where the data to be used are as
follows:
\begin{align}
&  m_{a_{0}(980)} = (980 \pm49)~\text{MeV, }\;\; m_{a_{0}(1450)} = (1474
\pm74)~\text{MeV}\;,\nonumber\\
&  m_{K_{0}^{\star}(800)} = (676 \pm40)~\text{MeV, }\;\; m_{K_{0}^{\star}
(1430)}=(1425\pm71)\text{ MeV}\;,\nonumber\\
&  \Gamma_{a_{0}(1450)} = (265 \pm13.3)~\text{MeV, }\;\; \Gamma_{K_{0}^{\star
}(800)\rightarrow K\pi} = (548 \pm27.4)~\text{MeV, } \;\;\Gamma_{K_{0}^{\star
}(1430)\rightarrow K\pi} = (270 \pm80)~\text{MeV}\;,\nonumber\\
&  |\mathcal{M}_{a_{0}(980)\rightarrow KK}|\, = (3590 \pm440)~\text{MeV
\cite{Giacosa:2009qh}, }\;\; |\mathcal{M}_{a_{0}(980) \rightarrow\eta\pi}| \,
= (3300 \pm166.5)~\text{MeV \cite{Giacosa:2009qh}}\;. \label{SI1}%
\end{align}
We note the following: (\textit{i}) the interpretation of $K_{0}^{\star}(800)$
as a particle is controversial; we will nonetheless include it into our fits
in accordance with the conclusions of Ref.\ \cite{KappaY}; (\textit{ii}) as in
the case of $\phi(1020)$, we will use the decay amplitudes rather than the
decay widths for the processes $a_{0}(980)\rightarrow KK$ and $a_{0}%
(980)\rightarrow\eta\pi$ due to the proximity of $a_{0}(980)$ to the $\bar
{K}K$ threshold. The reader may find it somewhat surprising that we are
considering pairs of states above [$a_{0}(1450)$/$K_{0}^{\star}(1430)$] 1 GeV,
below [$a_{0} (980)$/$K_{0}^{\star}(800)$] 1 GeV, as well as "mixed" pairs
$a_{0} (980)$/$K_{0}^{\star}(1430)$, $a_{0}(1450)$/$K_{0}^{\star}(800)$. The
reason to consider also mixed pairs is that we want to avoid any kind of
prejudice in the assignment of our $\bar{q}q$ states, and thus explore all
possibilities. Nonetheless, these mixed pairs as well as the pair below 1 GeV
will be disfavored by our analysis (see below). Mass formulas used in the fit
are presented in Eqs.\ (\ref{m_pi}) -- (\ref{eq:etaS}), (\ref{m_a_0}) --
(\ref{eq:sigS}), and (\ref{m_f0LH}) -- (\ref{m_f1_S}), whereas formulas for
the decay widths are given in Appendix \ref{App:decay}.
\end{itemize}

Moreover, the fit shall be constrained by the following conditions:

\begin{itemize}
\item $Z_{\pi}$, $Z_{K}$, $Z_{\eta_{S}}$, $Z_{K_{0}^{\star}}>1$, due to
Eqs.\ (\ref{Z_pi}) -- (\ref{Z_K_S}).

\item $m_{\eta_{N}}<m_{\eta_{S}}$ and $m_{\sigma_{N}}<m_{\sigma_{S}}$, i.e.,
pure non-strange states should be lighter than pure strange states.

\item $m_{0}^{2}<0$. This is a necessary condition for the spontaneous
breaking of chiral symmetry.

\item $\lambda_{2}>0$ and $\lambda_{1}>-\lambda_{2}/2$. This is necessary for
the potential in the Lagrangian (\ref{eq:Lagrangian}) to be bounded from below.

\item $m_{1}^{2}\geq0$. The reason is that otherwise (\textit{i}) an
instability of the vacuum in the physical $\rho$ direction would occur [see
the Lagrangian (\ref{eq:Lagrangian})] and (\textit{ii}) $m_{\rho}$ and
$m_{a_{1}}$ would become imaginary in the chiral transition, i.e., once the
condensates vanish [see Eqs.\ (\ref{m_rho}) and (\ref{m_a_1})].

\item $m_{1}\leq m_{\rho}$ as otherwise $(h_{1}+h_{2}+h_{3})\phi_{N}
^{2}/2+h_{1}\phi_{S}^{2}/2$ in the $\rho$ mass term (\ref{m_rho}) would be
negative; this would imply that spontaneous chiral symmetry breaking decreases
the $\rho$ mass. This is clearly unnatural because the breaking of chiral
symmetry generates a sizable constituent mass for the light quarks, which is
expected to positively contribute to the meson masses. This positive
contribution is a feature of all known models (such as the Nambu--Jona-Lasinio
model and constituent quark approaches). Indeed, in an important class of
hadronic models [see Ref.\ \cite{harada} and refs.\ therein] the only and
obviously positive contribution to the $\rho$ mass squared is proportional to
$\phi^{2}$ (i.e., $m_{1}=0$).
\end{itemize}

Before discussing the results of the fit, it is important to stress which
particles have not been included in the list above and why. Namely, we have
omitted experimental information about the scalar-isoscalar and the axial-kaon states.

(\textit{i}) Scalar-isoscalar states are not included in the fit
because in this first study we have neglected the coupling of the
glueball and of additional light scalar states, such as tetraquarks, to the other mesons. Thus, although there are three scalar states
\cite{glueball} above 1 GeV, at most two can be described within our model. Below 1 GeV
the resonance $f_{0}(500)$ is still poorly known and $f_{0}(980)$ is distorted
by the nearby $\bar{K}K$ threshold.

(\textit{ii}) The axial-kaon state $K_{1}$ is not included in the fit for a
similar reason: the kaonic states from our axial-vector nonet $1^{++}$ mix
with the kaonic states of the nonet $1^{+-}$ which is not part of our
Lagrangian. (This is possible, because charge conjugation is not a
well-defined quantum number for kaons.) Also in this case we cannot assign our
theoretical axial kaon field to a specific resonance [in PDG there are two
axial kaons with masses $K_{1}(1270)$ and $K_{1}(1400)$], because the mixing
is expected to be large, see e.g.\ Ref.\ \cite{burakovsky}.

\subsection{Results of the fit}

\label{predictions}

The experimental quantities discussed in the previous subsection do not depend
on all 13 parameters of Eq.\ (\ref{13p}), but on $m_{0}, \lambda_{1}, m_{1},
h_{1}$ only through the two combinations
\begin{equation}
C_{1} = m_{0}^{2} + \lambda_{1} \left(  \phi_{N}^{2} +\phi_{S}^{2}\right)
\;,\;\;\; C_{2} = m_{1}^{2} + \frac{h_{1}}{2} \left(  \phi_{N}^{2} + \phi
_{S}^{2}\right)  \;.
\end{equation}
Moreover, instead of the parameters $h_{0N}$ and $h_{0S}$ we use the
condensates $\phi_{N}$ and $\phi_{S}$: this is equivalent, as $h_{0N},\,
h_{0S}$ are completely determined by the masses of pion and $\eta_{S}$,
cf.\ Eqs.\ (\ref{m_pi}), (\ref{eq:etaS}). Summarizing, we have the following
eleven parameters entering the fit:
\begin{equation}
C_{1}, ~C_{2} , ~c_{1}, ~\delta_{S}, ~g_{1}, ~g_{2}, ~\phi_{N}, ~\phi_{S},
~h_{2}, ~h_{3}, ~\lambda_{2}. \label{11p}%
\end{equation}

We perform our fit using the experimental values for the 17 quantities given
in Eqs.\ (\ref{eq:weakdecayconstants}) -- (\ref{AVI}). In addition, we use the
experimental values for the pairs $a_{0}(980)$~/~$K_{0}^{\star}(800)$,
$a_{0}(980)$~/~$K_{0}^{\star}(1430)$, $a_{0}(1450)$~/~$K_{0}^{\star}(800)$,
$a_{0}(1450)$~/~$K_{0}^{\star}(1430)$ given in Eq.\ (\ref{SI1}). These are
four additional experimental quantities [except in the case of $a_{0}(980)$,
where we use the two values for the decay amplitudes instead of one value for
the decay width]. In total, we therefore fit $21$ (22) experimental quantities
to the eleven parameters given in Eq.\ (\ref{11p}). The corresponding values
of $\chi^{2}$ are listed in Table~\ref{Tab:khi}. We see that the combination
$a_{0}(1450)$/$K_{0}^{\star}(1430)$ gives the best value for $\chi^{2}$. In
fact, with a $\chi^{2}_{red}$ of 1.23 this fit is remarkably good, meaning
that all physical quantities are reproduced within an average error of $5\%$.
In the framework of an effective model for the strong interaction, which
neglects isospin-breaking effects and describes all mesonic resonances up to
about 1.8 GeV, we perceive this to be a remarkable
achievement. 

Note that also the fit with the pair $a_{0}(980)$~/~$K_{0}^{\star}(1430)$ has
a $\chi_{red}^{2}$ of the same order of magnitude as the one with the pair
$a_{0}(1450)$~/~$K_{0}^{\star}(1430)$. Nevertheless, we will disregard this
fit for two reasons. Firstly, the $\chi_{red}^{2}$ for the pair $a_{0}%
(980)$~/~$K_{0}^{\star}(1430)$ is, although rather small, still larger by a
factor of about two than the $\chi_{red}^{2}$ for the pair $a_{0}%
(1450)$~/~$K_{0}^{\star}(1430)$. Secondly, and more importantly, the fit with
the pair $a_{0}(980)$~/~$K_{0}^{\star}(1430)$ produces a scalar-kaon mass of
$m_{K_{0}^{\star}}=1146$ MeV, which cannot be assigned to any physical
resonance as it is much larger than $m_{K_{0}^{\star}(800)}$ and much smaller
than $m_{K_{0}^{\star}(1430)}$. Note that this problem is also present in NJL
models with mixing between scalar mesons below and above 1 GeV \cite{Hiller}.

\begin{table}[th]
\centering
\begin{tabular}
[c]{|c|c|c|}\hline
Pair & $\chi^{2}$ & $\chi_{red}^{2}$\\\hline
$a_{0}(1450)$/$K_{0}^{\star}(1430)$ & $12.33$ & $1.23$\\\hline
$a_{0}(980)$/$K_{0}^{\star}(800)$ & $129.36$ & $11.76$\\\hline
$a_{0}(980)$/$K_{0}^{\star}(1430)$ & $22.00$ & $2.00$\\\hline
$a_{0}(1450)$/$K_{0}^{\star}(800)$ & $242.27$ & $24.23$\\\hline
\end{tabular}
\caption{Isotriplet and isodoublet scalar pairs and the corresponding values
of the total $\chi^{2}$ and the reduced $\chi^{2}_{red}=\chi^{2}/N_{dof}$,
where $N_{dof}$ is the difference between the number of experimental
quantities and the number of fit parameters (10 for the first and fourth rows
and 11 for the second and third rows).}%
\label{Tab:khi}%
\end{table}

The detailed comparison of theory with data for the pair $a_{0}(1450)$%
~/~$K_{0}^{\star}(1430)$ is presented in Table~\ref{Comparison1} where the
theoretical errors are also shown. Errors for the model parameters ($\delta
p_{i}$) are calculated as the inverse square roots of the eigenvalues of the
Hessian matrix obtained from $\chi^{2}(p_{j})$, where $p_{j}$ denotes elements
of the parameter set (\ref{11p}) and $i=1,2,\ldots,11$. Theoretical errors
$\Delta O_{i}$ for each observable $O_{i}$ (mass, decay width) can be
calculated according to the following formula:
\begin{equation}
\Delta O_{i}=\sqrt{\sum_{j=1}^{n}\left(  \left.  \frac{\partial O_{i}%
}{\partial p_{j}}\right\vert _{\text{at fit value of }O_{i}}\delta
p_{j}\right)  ^{2}}.
\end{equation}
The remarkable agreement of our results with experimental data in the
(pseudo)scalar and (axial-)vector sectors can now be explicitly seen for the
various quantities shown in Table~\ref{Comparison1}. We thus conclude that the
quark-antiquark states ${a}_{0}$ and $K_{0}^{\ast}$ should be assigned to the
resonances $a_{0}(1450)$ and $K_{0}^{\ast}(1450)$.
Moreover, also the axial-vector resonances are well described by the fit: we
therefore also interpret them as predominantly quark-antiquark states.

\begin{table}[th]
\centering
\begin{tabular}
[c]{|c|c|c|}\hline
Observable & Fit [MeV] & Experiment [MeV]\\\hline
$f_{\pi}$ & $96.3 \pm0.7 $ & $92.2 \pm4.6$\\\hline
$f_{K}$ & $106.9 \pm0.6$ & $110.4 \pm5.5$\\\hline
$m_{\pi}$ & $141.0 \pm5.8$ & $137.3 \pm6.9$\\\hline
$m_{K}$ & $485.6 \pm3.0$ & $495.6 \pm24.8$\\\hline
$m_{\eta}$ & $509.4 \pm3.0$ & $547.9 \pm27.4$\\\hline
$m_{\eta^{\prime}}$ & $962.5 \pm5.6$ & $957.8 \pm47.9$\\\hline
$m_{\rho}$ & $783.1 \pm7.0$ & $775.5 \pm38.8$\\\hline
$m_{K^{\star}}$ & $885.1 \pm6.3$ & $893.8 \pm44.7$\\\hline
$m_{\phi}$ & $975.1 \pm6.4$ & $1019.5 \pm51.0$\\\hline
$m_{a_{1}}$ & $1186 \pm6$ & $1230 \pm62$\\\hline
$m_{f_{1}(1420)}$ & $1372.5 \pm5.3$ & $1426.4 \pm71.3$\\\hline
$m_{a_{0}}$ & $1363 \pm1$ & $1474 \pm74$\\\hline
$m_{K_{0}^{\star}}$ & $1450 \pm1$ & $1425 \pm71$\\\hline
$\Gamma_{\rho\rightarrow\pi\pi}$ & $160.9 \pm4.4$ & $149.1 \pm7.4$\\\hline
$\Gamma_{K^{\star}\rightarrow K\pi}$ & $44.6 \pm1.9$ & $46.2 \pm2.3$\\\hline
$\Gamma_{\phi\rightarrow\bar{K}K}$ & $3.34 \pm0.14$ & $3.54 \pm0.18$\\\hline
$\Gamma_{a_{1}\rightarrow\rho\pi}$ & $549 \pm43$ & $425 \pm175$\\\hline
$\Gamma_{a_{1}\rightarrow\pi\gamma}$ & $0.66 \pm0.01$ & $0.64 \pm0.25$\\\hline
$\Gamma_{f_{1}(1420)\rightarrow K^{\star}K}$ & $44.6 \pm39.9$ & $43.9 \pm
2.2$\\\hline
$\Gamma_{a_{0}}$ & $266 \pm12$ & $265 \pm13$\\\hline
$\Gamma_{K_{0}^{\star}\rightarrow K\pi}$ & $285 \pm12$ & $270 \pm80$\\\hline
\end{tabular}
\caption{Best-fit results for masses and decay widths compared with
experiment.}%
\label{Comparison1}%
\end{table}

\begin{figure}[h]
\begin{center}%
\begin{tabular}
[c]{cc}%
\resizebox{88mm}{!}{\includegraphics{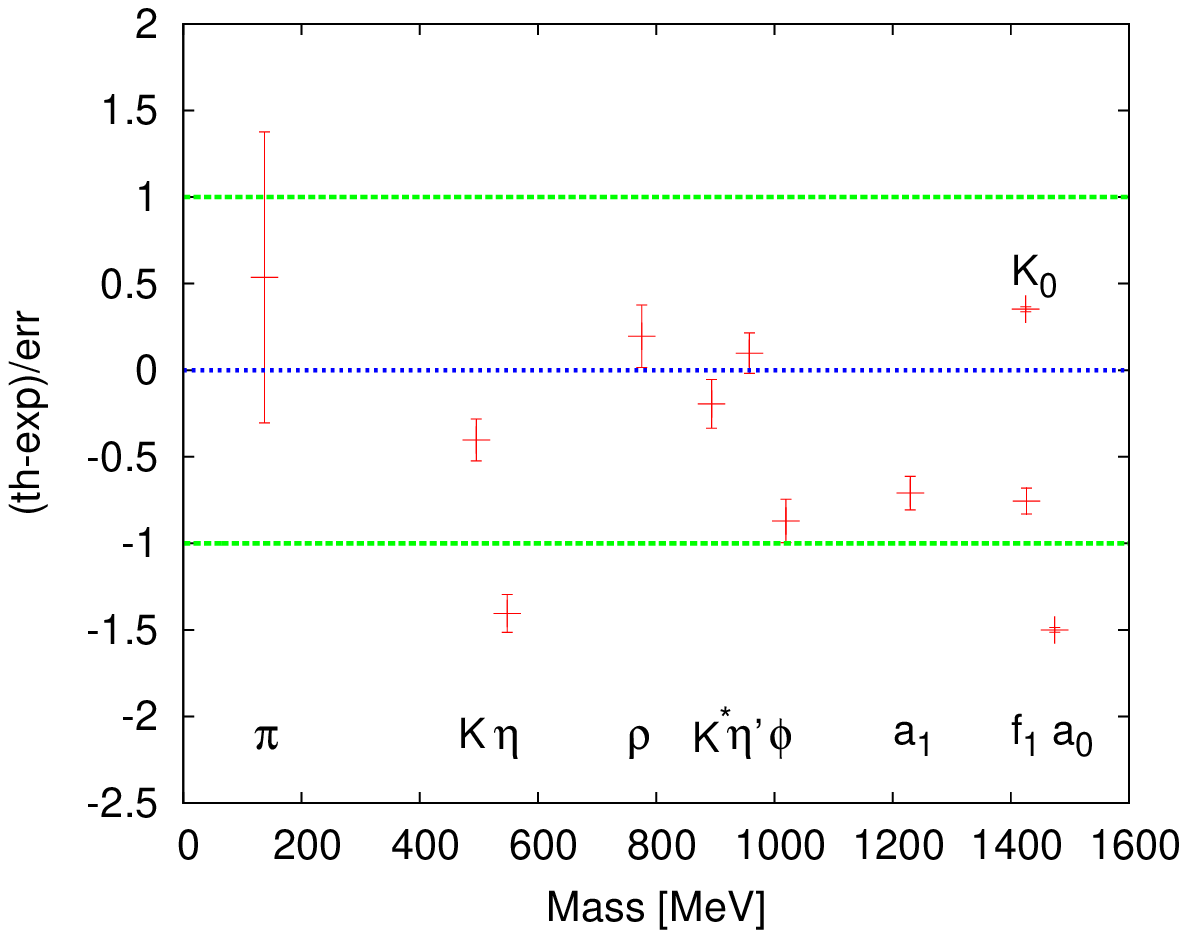}} &
\resizebox{88mm}{!}{\includegraphics{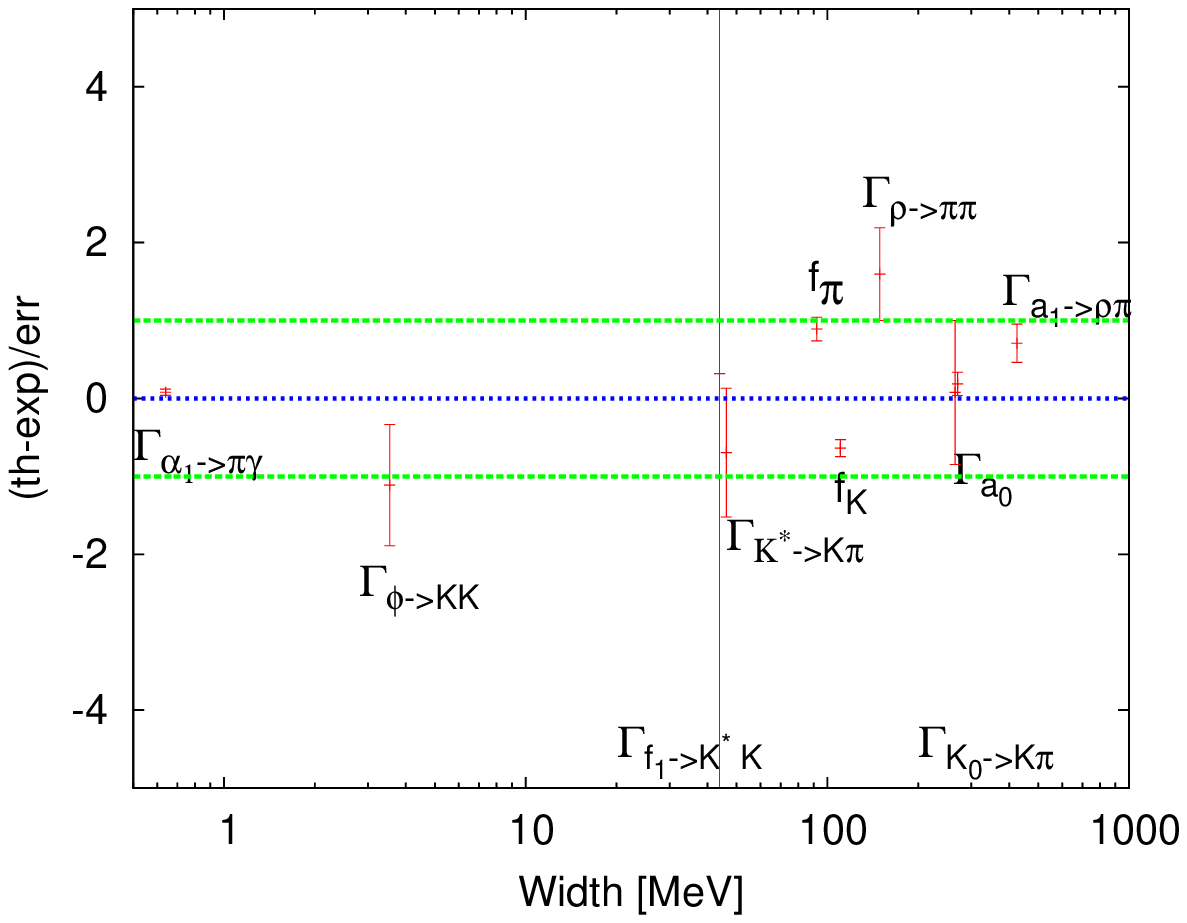}}
\end{tabular}
\end{center}
\caption{Comparison of theory and experiment for observables from Table II.}%
\label{Figure}%
\end{figure}

In Fig.\ \ref{Figure} we present the results of Table \ref{Comparison1} in
a slightly different way: as the difference of the theoretical and
experimental values divided by experimental error; the error bars correspond
to the theoretical error values from our fit.

In Table~\ref{Tab:param} we finally present the values of the
parameters of the model together with their errors as obtained from our best
fit. We remark
that, since the quantities entering our fit are
not affected by interactions with the glueball, this result holds also
when including the latter.

\begin{table}[th]
\centering
\begin{tabular}
[c]{|c|c|}\hline
Parameter & Value\\\hline
$C_{1}$ [GeV$^{2}$] & $-0.9183 \pm0.0006$\\\hline
$C_{2}$ [GeV$^{2}$] & $0.4135 \pm0.0147$\\\hline
$c_{1}$ [GeV$^{-2}$] & $450.5420 \pm7.0339$\\\hline
$\delta_{S}$ [GeV$^{2}$] & $0.1511 \pm0.0038$\\\hline
$g_{1}$ & $5.8433 \pm0.0176$\\\hline
$g_{2}$ & $3.0250 \pm0.2329$\\\hline
$\phi_{N}$ [GeV] & $0.1646 \pm0.0001$\\\hline
$\phi_{S}$ [GeV] & $0.1262 \pm0.0001$\\\hline
$h_{2}$ & $9.8796 \pm0.6627$\\\hline
$h_{3}$ & $4.8667 \pm0.0864$\\\hline
$\lambda_{2}$ & $68.2972 \pm0.0435$\\\hline
\end{tabular}
\caption{Parameters and their errors.}%
\label{Tab:param}%
\end{table}

Before moving to the consequences of the fit, we discuss two important
aspects of our study:

\bigskip

{\it (i)} \emph{Mass ordering in the scalar sector:} As evident from
Table~\ref{Comparison1}, our fit yields $m_{a_{0}}<m_{K_{0}^{\star}}$ whereas
the experimental mass ordering is opposite. The reason for our result is the
pattern of explicit symmetry breaking implemented in our model, which renders
$\bar{q}q$ states with a strange quark approximately 100 MeV 
heavier than non-strange states. However, other mechanisms, 
as for instance mixing of the currently present $\bar{q}q$ states
with light tetraquark states, may occur that change this mass ordering
\cite{Black,Mixing,Giacosamixing}: namely, in this mixing scenario a pure
quarkonium and a pure tetraquark in the isovector sector mix to form the
resonances $a_{0}(1450)$ and $a_{0}(980)$, while a similar
mixing in the isodoublet sector leads to the resonances $K_{0}^{\ast}%
(1450)$ and $K_{0}^{\ast}(800)$. The fact that the mass
ordering $m_{a_{0}}>m_{K_{0}^{\star}}$ is realized in nature can be understood
from a larger mixing angle, and therefore a larger level repulsion, in the
isovector sector. A detailed study of this scenario necessitates the inclusion
of a full nonet of light states, see also the discussion in the
Conclusions.

An additional point can be raised about the ratio of the full decay
widths of the $a_{0}(1450)$ and $K_{0}^{\star}(1430)$ resonances. From
$SU(3)_{V}$ symmetry arguments for $\bar{q}q$ states one expects the full
decay widths of these resonances to scale as \cite{Black} 
\begin{equation}
\frac{\Gamma_{a_{0}(1450)}}{\Gamma_{K_{0}^{\star}(1430)}}=1.51
\end{equation}
whereas experimental data suggest \cite{PDG} 
\begin{equation}
\frac{\Gamma_{a_{0}(1450)}}{\Gamma_{K_{0}^{\star}(1430)}}\sim0.9\text{.}%
\end{equation}
In Ref.\ \cite{Black}, this problem was also solved by introducing tetraquark
fields at the level of an effective Lagrangian and by studying
tetraquark-quarkonium mixing in the $I=1$ and $I=1/2$
channels. Contrarily, in our considerations so far only $\bar{q}q$ states were
taken into account. Thus one
might expect that our fit results should be closer to the $SU(3)$ limit rather
than to the experimental ratio. However, the opposite is the case: as evident
from Table II, our fit results for $a_{0}(1450)$ and $K_{0}^{\star}(1430)$
reproduce the experimental ratio $\sim0.9$. Our analysis shows that the reason
is the inclusion of the chiral-anomaly term and, in particular, of
(axial-)vector mesons. Artificially decoupling (axial-)vector states (i.e.,
setting $g_{1}=h_{2}=h_{3}=0$) and removing the chiral-anomaly term from our
Lagrangian (i.e., setting $c_{1}=0$) we obtain unphysically large decay widths
$\Gamma_{a_{0}(1450)}\sim14$ GeV, $\Gamma_{K_{0}^{\star}(1430)}$ $\sim10$ GeV
-- and also $\Gamma_{a_{0}(1450)}/\Gamma_{K_{0}^{\star}(1430)}$ $\sim1.4$,
very close to the value obtained from $SU(3)$ symmetry. The ratio would be
$\sim1.3$ if only (axial-)vectors were decoupled. Hence, just as in our
previous publication \cite{Paper1}, we emphasize again the crucial importance
of the (axial-)vector degrees of freedom for scalar-meson phenomenology.
However, there still remains an open question how additional light scalar fields
will influence the results in the $a_{0}$-$K_{0}^{\star}$ sector.\newline

{\it (ii)} \emph{Unitarity corrections:} Our results for masses
and decay widths are valid at tree level. Most particles in our model are rather
narrow, and thus justify -- at least in a first approximation -- this procedure.
This is also in agreement with large-$N_{c}$ considerations,
according to which the role of the contribution of mesonic loops is
suppressed. However, the axial-vector state $a_{1}(1230)$ and the
scalar states $a_{0}(1450)$ and $K_{0}^{\star}(1430),$ as
well as the scalar-isoscalar state $f_{0}(1370)$ to be discussed in
the next subsection, have a large width (of the order of $300$ MeV or
more). Then, the role of loops and unitarity corrections
for both the masses and the decay widths needs to be discussed. Concerning the mass
shifts due to mesonic loops, we notice that an indirect indication that their
contribution is not too large is the fact that the resonances $a_{1}
(1260)$ and $f_{1}(1285)$ have a similar mass, although the
former has a much larger decay width than the latter. In a theoretical study
of this system one should determine the pole position of the resonance: the
mass is then usually denoted as the real part of the pole. As shown recently
in Ref.\ \cite{wolkanowski}, the value of the real part of the pole does not
differ too much from the bare (tree-level) mass, not even for large
coupling constants. Concerning the influence of loops on the value of
decay widths, we note that Ref.\ \cite{FG} has shown that the
corrections are small as long as the ratio $\Gamma/m$ is small. For
instance, for the case of $f_{0}(1370)$, the average decay width
reported by the Particle Data Group is $350$ MeV whereas the
average resonance mass is $1350$ MeV. This implies $\Gamma
/m\sim0.26$. The ratio is almost as small as in the case of the 
$\rho$ meson where $\Gamma/m\sim150/750=0.2$ -- and it is
known from chiral perturbation theory that such resonances obtain only small
corrections upon unitarizations [see, e.g., Ref.\ \cite{P1}].

We thus conclude that the role of loops should not modify the 
picture presented here. It should, however, be stressed that the inclusion of loops
is a necessary step for the future: on the one hand we can numerically
evaluate the role of mesonic loops and thus quantitatively verify our
statements, on the other hand, even if large variations for the states above 1
GeV do not occur as we expect, an interesting question is how poles on the
complex plane emerge. Namely, these states could arise as tetraquark states as
mentioned above, but could also represent dynamically generated
states, see alsoe.g.\
Ref.\ \cite{pennington}. [For a detailed discussion of this point including the
interrelations of tetraquark, molecular, and dynamically generated states we
refer to Ref. \cite{dynrec}.]

\subsection{Consequences of the fit}

\label{consequences}

\subsubsection{Scalar-isoscalar mesons}

We now turn to the scalar-isoscalar mesons. We shall discuss four different aspects:

\begin{enumerate}
\item[(a)] \textit{Results in the large-$N_{c}$ limit ($\lambda_{1}=h_{1}%
=0$):}\\[0.1cm]From the fit in the previous section we cannot immediately
obtain the masses of the scalar-isoscalar states of the model because their
masses do not depend solely on the combination $C_{1}=m_{0}^{2}+\lambda
_{1}\left(  \phi_{N}^{2}+\phi_{S}^{2}\right)  $, but separately on the
parameters $m_{0}^{2}$ and $\lambda_{1}$. Similarly, the decay rates of the
scalar-isoscalar mesons do not depend solely on the combination $C_{2}%
=m_{1}^{2}+\frac{h_{1}}{2}\left(  \phi_{N}^{2}+\phi_{S}^{2}\right)  $, but
separately on the parameters $m_{1}$ and $h_{1}$.

Interestingly, the parameters $\lambda_{1}$ and $h_{1}$ are large-$N_{c}$
suppressed. Setting them to zero we obtain a prediction for the masses and
decay widths of the scalar-isoscalar states in the large-$N_{c}$ limit.
Moreover, when $\lambda_{1}=h_{1}=0,$ there is also no mixing between $\bar
{n}n$ and $\bar{s}s$ states: $f_{0}^{L}\equiv\sigma_{N}=\bar{n}n\equiv(\bar
{u}u+\bar{d}d)/\sqrt{2}$ and $f_{0}^{H}\equiv\sigma_{S}\equiv\bar{s}s.$ Their
masses read:
\begin{equation}
m_{f_{0}^{L}}=1362.7~\text{MeV, } \; m_{f_{0}^{H}}=1531.7~\text{MeV}\;.
\end{equation}
These masses clearly lie above 1 GeV. Comparing these values to the
experimental values for the three isoscalar states above 1 GeV \cite{PDG}
\begin{equation}
m_{f_{0}(1370)}=(1350\pm150)~\text{MeV},\; m_{f_{0}(1500)} = (1505
\pm75)~\text{MeV},\; m_{f_{0}(1710)} = (1720 \pm86)~\text{MeV}\;,
\label{above}%
\end{equation}
we see that the mass of $f_{0}^{L}$ is well compatible with that of the
resonance $f_{0}(1370).$ [Note that $m_{f_{0}(1370)}$ in Eq.\ (\ref{above})
emulates the PDG mass interval (1200--1500) MeV.] The mass of $f_{0}^{H}$
appears to be close to that of $f_{0}(1500)$. Remember, though, that we
artificially enlarged the experimental error to $5\%$; the actual error is
only 6 MeV, and then the mass of $f_{0}^{H}$ would be (just) outside the
experimental error. Considering the decays of $f_{0}^{H}$ we shall provide
evidence that an assignment of $f_{0}^{H}$ to $f_{0}(1710)$ is also possible
(and even rather likely).

For $\lambda_{1}=h_{1}=0$ the decay rates of $f_{0}^{L,H}$ into $\pi\pi$ and
$KK$ read
\begin{align}
\Gamma_{f_{0}^{L}\rightarrow\pi\pi}  &  = 520~\text{MeV, } \; \Gamma
_{f_{0}^{L}\rightarrow KK}= 129~\text{MeV,}\\
\Gamma_{f_{0}^{H}\rightarrow\pi\pi}  &  = 0~\text{MeV, } \; \Gamma_{f_{0}%
^{H}\rightarrow KK} = 422~\text{MeV.}%
\end{align}
For the experimental values we quote the PDG values \cite{PDG}:
\begin{align}
\Gamma_{f_{0}(1370)\rightarrow\pi\pi}  &  =(250\pm100)\text{ MeV, }\;
\Gamma_{f_{0}(1370)\rightarrow KK}\lesssim\Gamma_{f_{0}(1370)\rightarrow\pi
\pi}\;,\nonumber\\
\Gamma_{f_{0}(1500)\rightarrow\pi\pi}  &  =(38\pm5)\text{ MeV, }\;
\Gamma_{f_{0}(1500)\rightarrow KK}=(9.4\pm2.3)\text{ MeV}\;,\nonumber\\
\Gamma_{f_{0}(1710)\rightarrow\pi\pi}  &  =(29.3\pm6.5)\text{ MeV, }\;
\Gamma_{f_{0}(1710)\rightarrow KK}=(71.4\pm29.1)\text{ MeV}\;. \label{SI2}%
\end{align}
Note that for the resonance $f_{0}(1370)$ no branching ratios into $\pi\pi$
and $KK$ are reported in Ref.\ \cite{PDG}. The value for $\Gamma
_{f_{0}(1370)\rightarrow\pi\pi}$ is our estimate from PDG and the review
\cite{buggf0}, whereas $\Gamma_{f_{0}(1370)\rightarrow KK}$ is our estimate
from results presented in Refs.\ \cite{PDG,f01370KK}.

Our results for $f_{0}^{L}$ are in agreement with the experimental decay
widths of $f_{0}(1370).$ Our theoretical value for $f_{0}^{H}\rightarrow KK$
turns out to be too large, while $f_{0}^{H}\rightarrow\pi\pi$ vanishes,
because $f_{0}^{H}\equiv\sigma_{S}$ is a pure $\bar{s}s$ state. Nevertheless,
our model predicts the existence of a scalar-isoscalar state which decays
predominantly into kaons; this is indeed the decay pattern shown by
$f_{0}(1710).$ For these reasons we suggest to identify our state $f_{0}^{H}$ as
(predominantly) $f_{0}(1710)$. This identification will be further tested when
$\lambda_{1} \neq0$, $h_{1} \neq0$ below.

It should be stressed that a quantitative study of the scalar-isoscalar system
cannot be performed at present because our model contains only two states,
while three resonances appear (this is also the reason why we did not include
the scalar-isoscalar states into the fit). As many studies confirm
\cite{glueball}, the mixing with the scalar glueball can be sizable: a
reliable analysis of the scalar-isoscalar states can only be performed when
taking the scalar glueball into account. The preliminary results in the
$N_{f}=2$ sector \cite{Stani} have indeed shown that the glueball and
quarkonia degrees of freedom interact strongly and that the decay patterns are
sizably influenced.

\item[(b)] \textit{Results for $\lambda_{1}\neq0$ , $h_{1}\neq0$:}\\[0.1cm]A
nonvanishing value of the parameter $\lambda_{1}$ induces a mixing of the pure
states $\sigma_{N}$ and $\sigma_{S}$, see Eq.\ (\ref{eq:sigNS}). Our fit
determines only the value of the linear combination $C_{1}$ rather than the
value of $\lambda_{1}$, see discussion before Eq.\ (\ref{11p}). Nonetheless, a
range of values for $\lambda_{1}$ can be estimated using the value of $C_{1}$
from the fit and the condition $m_{0}^{2}<0$. Consequently, we also obtain a
range of values for our isoscalar masses $m_{f_{0}^{L,H}}$ and isoscalar decay
widths. The masses vary in the following intervals: $415$~MeV $\leq
m_{f_{0}^{L}}\leq1460$~MeV and $1480$~MeV $\leq m_{f_{0}^{H}}\leq1981$~MeV.
Considering these mass values, $f_{0}^{L}$ may correspond to either
$f_{0}(500)$, $f_{0}(980)$, or $f_{0}(1370)$ and $f_{0}^{H}$ may correspond to
either $f_{0}(1500)$ or $f_{0}(1710)$ \cite{PDG}. Therefore, a mere
calculation of scalar masses does not allow us to assign the scalar states
$f_{0}^{L}$ and $f_{0}^{H}$ to physical resonances. In order to resolve this
ambiguity, we will calculate various decay widths of the states $f_{0}^{L}$
and $f_{0}^{H}$ and compare them to data \cite{PDG}.

The dependence of $\Gamma_{f_{0}^{L,H}\rightarrow\pi\pi}$ on $m_{f_{0}^{L,H}}$
is presented in Fig.\ \ref{Pions}.

\begin{figure}[th]
\begin{center}%
\begin{tabular}
[c]{cc}%
\resizebox{78mm}{!}{\includegraphics{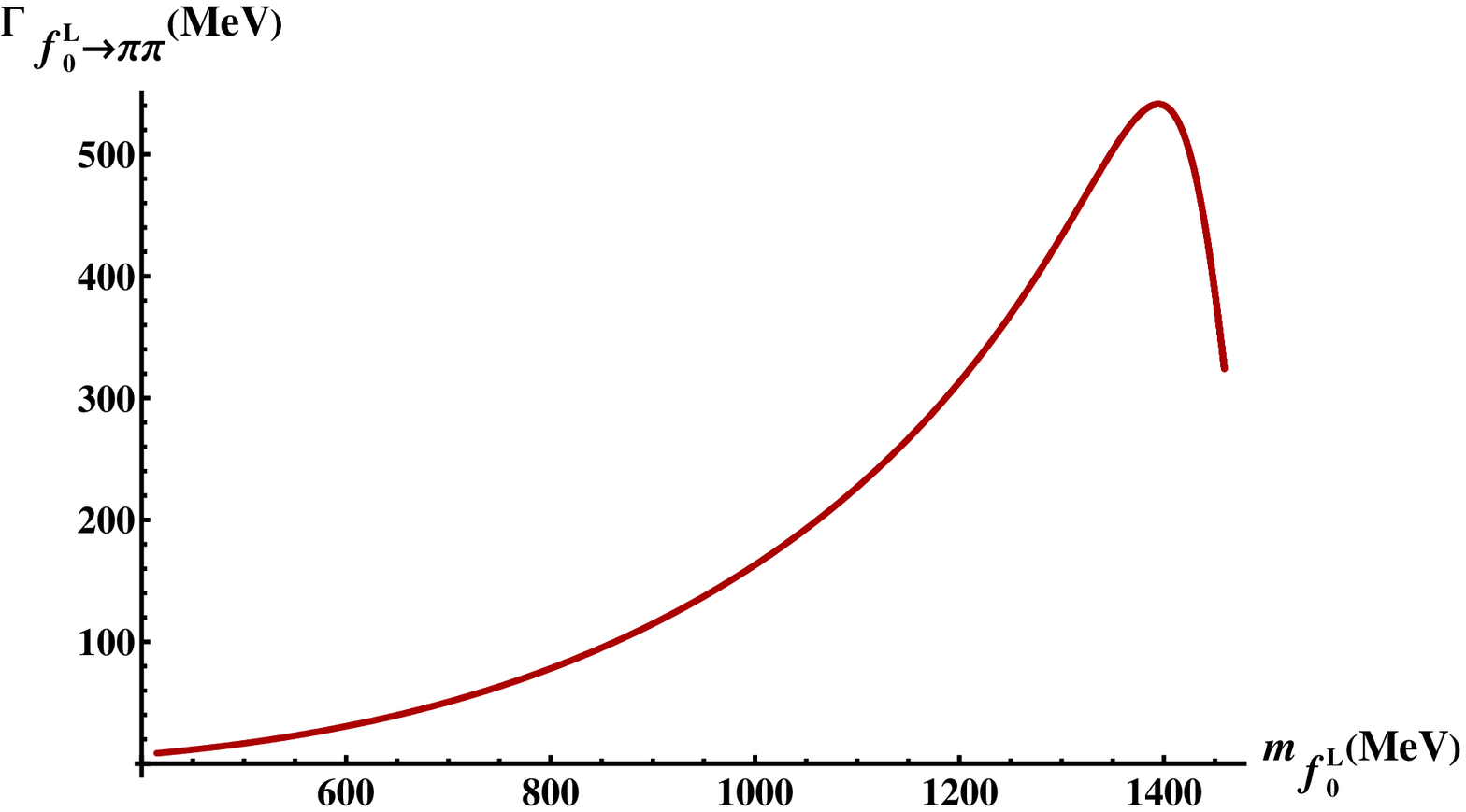}} &
\resizebox{78mm}{!}{\includegraphics{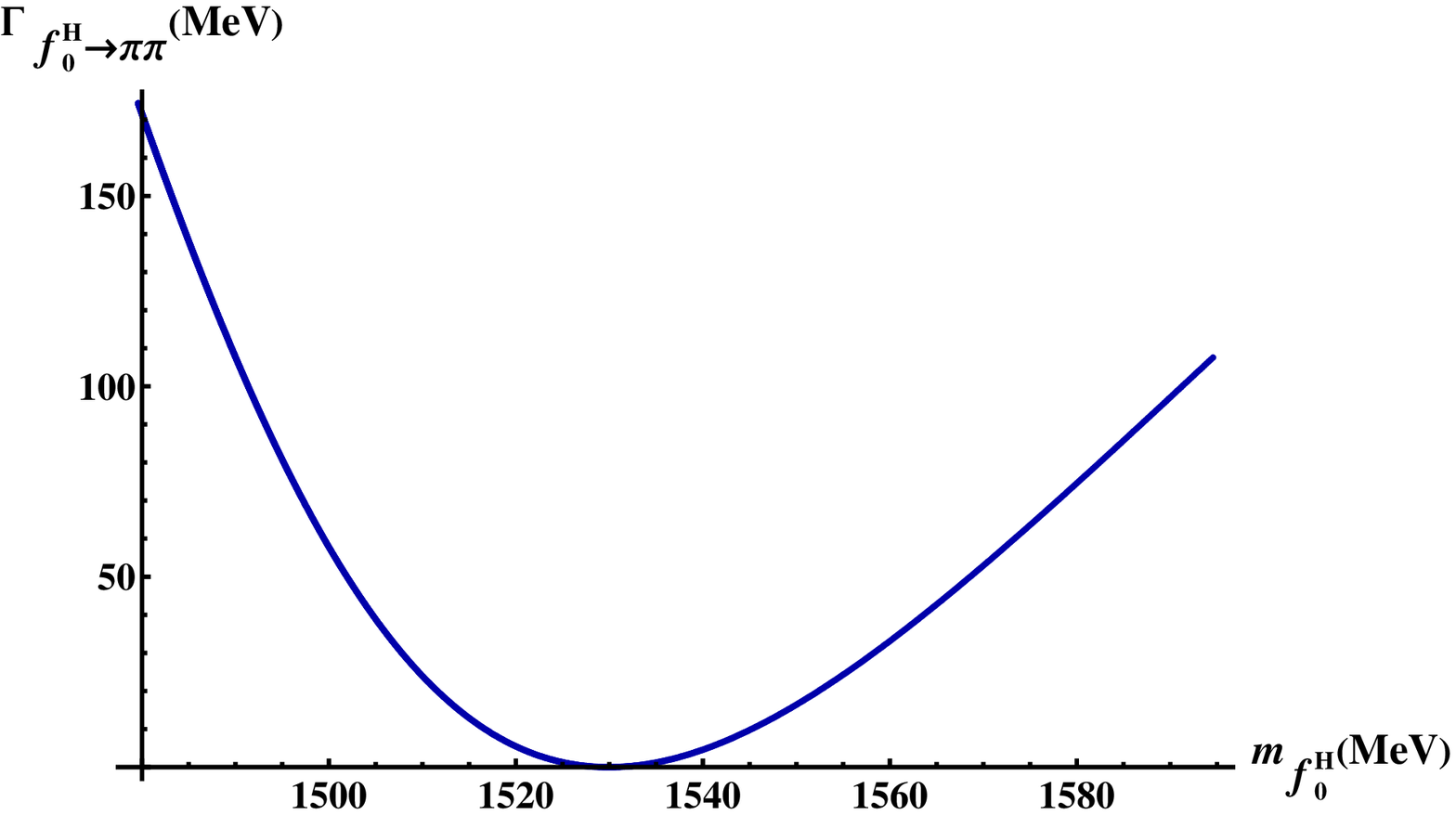}}
\end{tabular}
\end{center}
\caption{$\Gamma_{f_{0}^{L}\rightarrow\pi\pi}$ and $\Gamma_{f_{0}%
^{H}\rightarrow\pi\pi}$ as functions of $m_{f_{0}^{L}}$ and $m_{f_{0}^{H}}$,
respectively.}%
\label{Pions}%
\end{figure}

The decay width $\Gamma_{f_{0}^{L}\rightarrow\pi\pi}$ is consistent with the
experimental range $(250\pm100)$ MeV for the $f_{0}(1370)\rightarrow\pi\pi$
decay width, if 1000 MeV $\lesssim m_{f_{0}^{L}}\leq$ 1460 MeV. Other
assignments can be excluded: (\textit{i}) our $f_{0}^{L}$ state cannot be
assigned to $f_{0}(500)$ as $\Gamma_{f_{0}^{L}\rightarrow\pi\pi}\lesssim20$
MeV in the (new) PDG mass range $(400-550)$ MeV for $f_{0}(500)$. Therefore,
this strongly disfavors $f_{0}(500)$ as a $\bar{q}q$ state. (\textit{ii})
$\Gamma_{f_{0}^{L}\rightarrow\pi\pi}$ varies between approximately 140 MeV and
170 MeV in the mass interval of the $f_{0}(980)$ resonance, i.e., the interval
between 970 MeV and 1010 MeV. The PDG result for the full decay width of the
$f_{0}(980)$ resonance is between 40 MeV and 100 MeV, with the $\pi\pi$
channel being dominant \cite{PDG}, i.e., about a factor two smaller than our
theoretical values. Consequently, the assignment of $f_{0}^{L}$ to\ $f_{0}%
(980)$ is also disfavored by our analysis. Therefore, (although there could
still be mixing with other states) we assign our $f_{0}^{L}$ state to
$f_{0}(1370)$, thus supporting the interpretation of the latter state as a
quarkonium. Although experimental data is not conclusive, we mention that the
decay width $\Gamma_{f_{0}^{L}\rightarrow KK}$ shown in Fig.\ \ref{Kaons} is
consistent with the data.

In the mass interval $1500$ MeV $\lesssim f_{0}^{H} \lesssim1560$ MeV, our
results for the decay width $\Gamma_{f_{0}^{H}\rightarrow\pi\pi}$ are
consistent with experimental values for both $f_{0}(1500)$ and $f_{0}(1710)$,
see Eq.\ (\ref{SI2}). However, the $KK$ decay widths present us with a
different conclusion: as already indicated in the case $\lambda_{1} = h_{1}%
=0$, our $f_{0}^{H}$ state decays much more abundantly into kaons than into
pions, and the experimental data suggest only one physical resonance with the
same feature: $f_{0}(1710)$. The other resonance $f_{0}(1500)$ decays
preferably into pions. Thus (and although the mass value for $f_{0}^{H}$ is
too small in the range where the decay width into pions agrees with the
experimental value) we assign our $f_{0}^{H}$ state to $f_{0}(1710)$ and
inspect in the following whether this assignment is justified by other data.

\begin{figure}[th]
\begin{center}%
\begin{tabular}
[c]{cc}%
\resizebox{78mm}{!}{\includegraphics{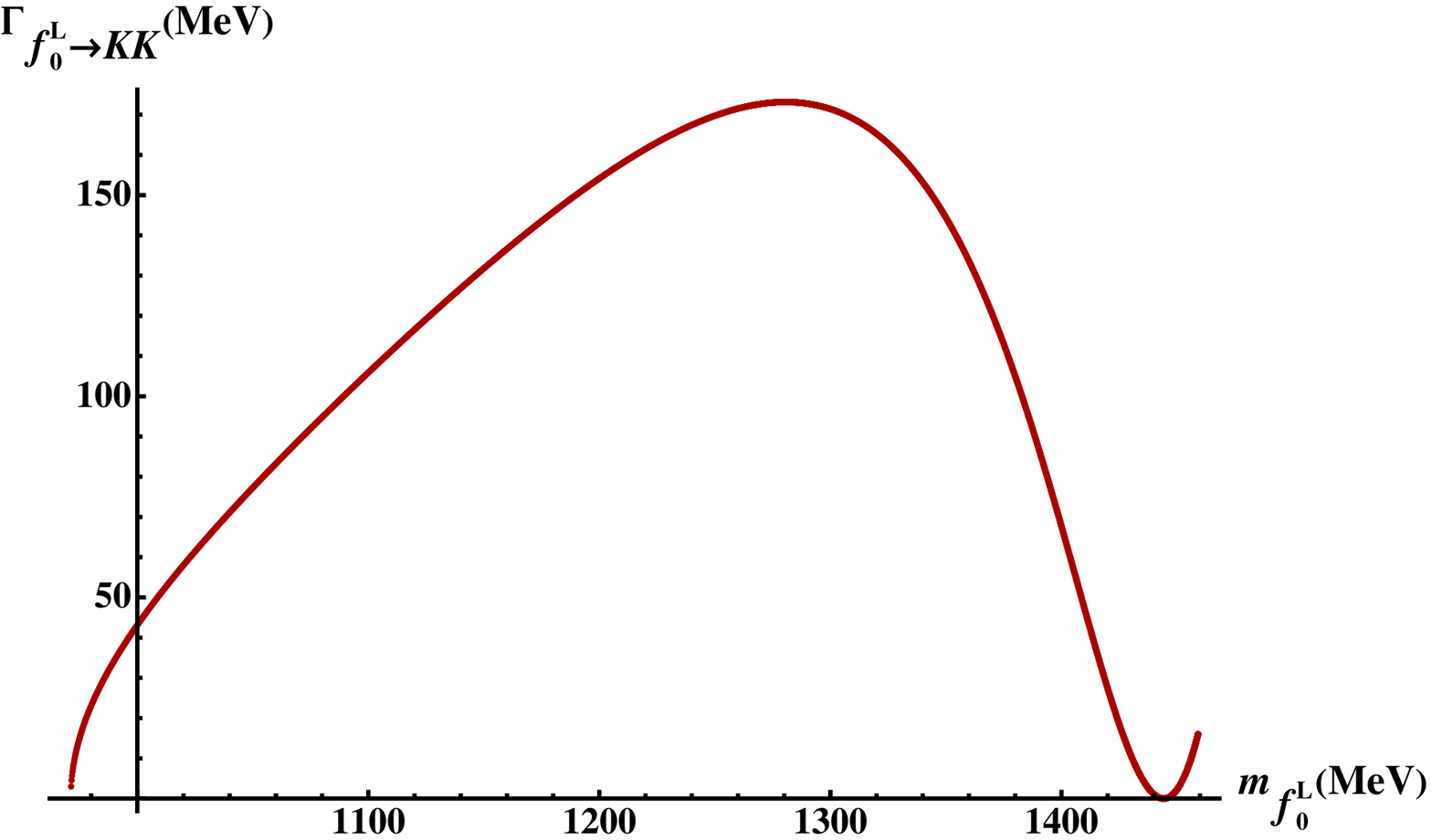}} &
\resizebox{78mm}{!}{\includegraphics{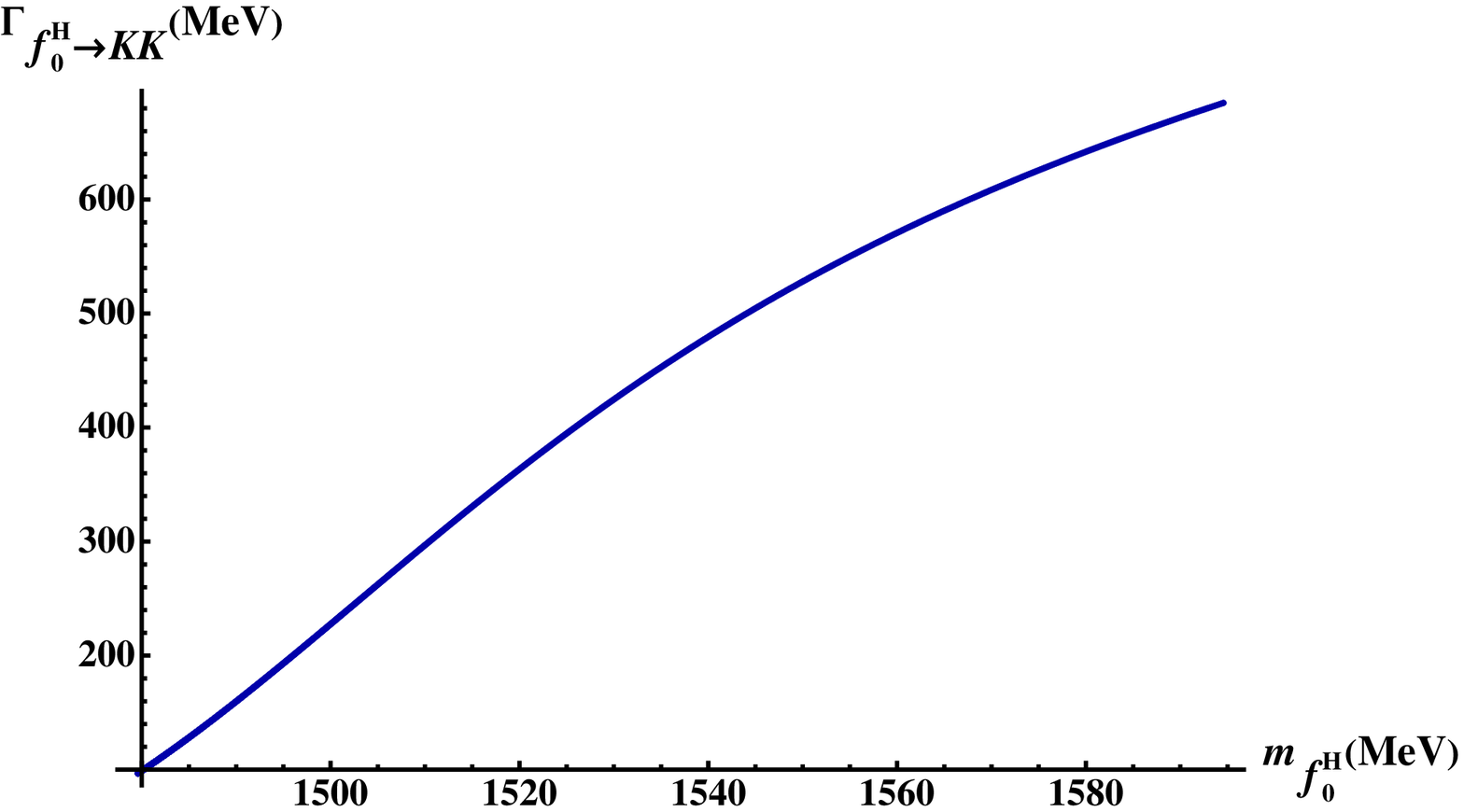}}
\end{tabular}
\end{center}
\caption{$\Gamma_{f_{0}^{L}\rightarrow KK}$ and $\Gamma_{f_{0}^{H}\rightarrow
KK}$ as functions of $m_{f_{0}^{L}}$ and $m_{f_{0}^{H}}$, respectively.}%
\label{Kaons}%
\end{figure}

We observe from Fig.\ \ref{Kaons} that $\Gamma_{f_{0}^{H}\rightarrow KK}$
rises rapidly and remains above the PDG result $\Gamma_{f_{0}(1710)\rightarrow
KK}=(71.4\pm29.1)$ MeV in the entire mass interval $m_{f_{0}^{H}}\gtrsim1500$
MeV. Although the absolute value of the decay width is rather large, several
ratios of decay widths can be described correctly by our fit, most notably the
$\pi\pi/KK$ decay ratio presented in Fig.\ \ref{SPPKK2}.

\begin{figure}[th]
\begin{center}%
\begin{tabular}
[c]{cc}%
\resizebox{78mm}{!}{\includegraphics{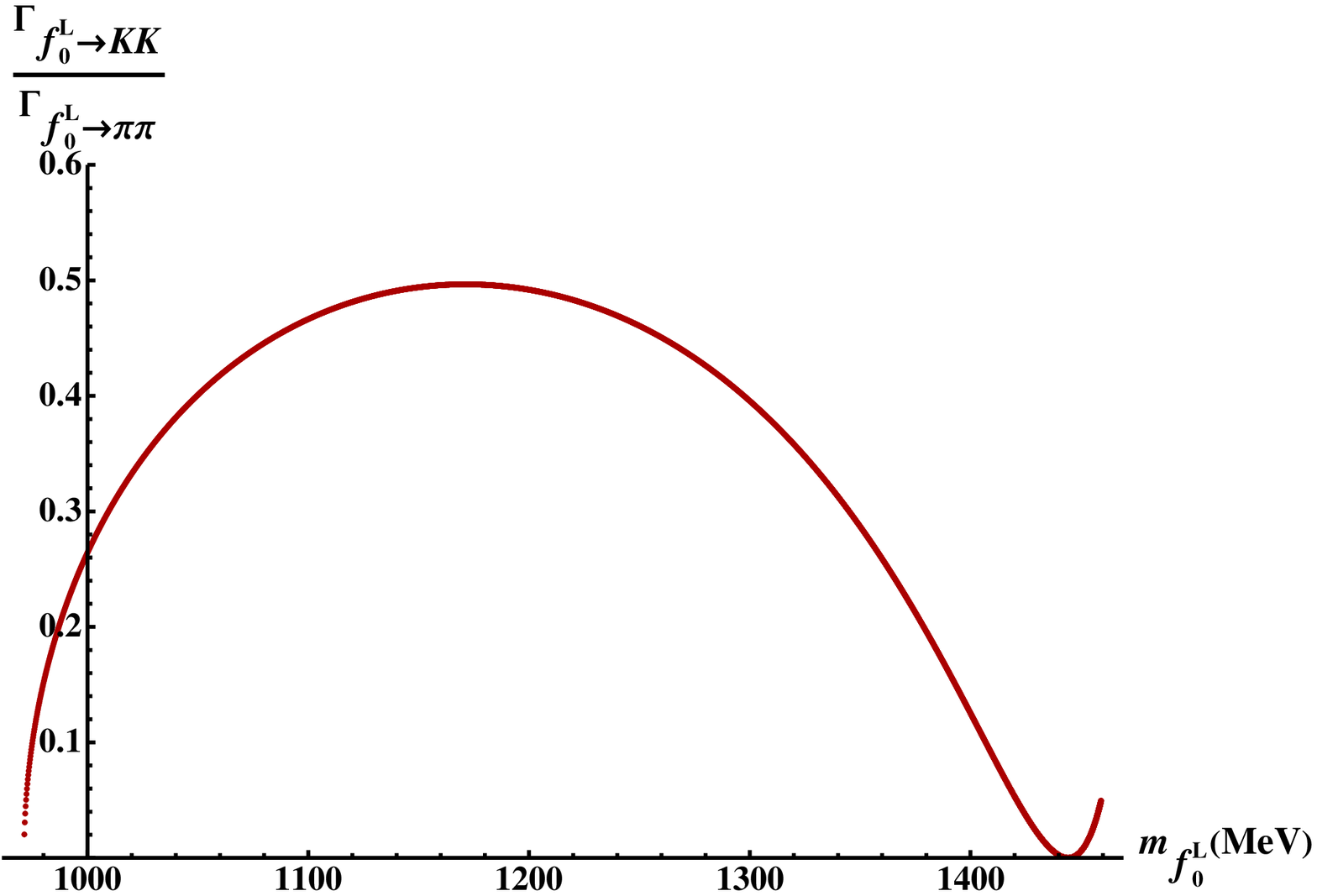}} &
\resizebox{78mm}{!}{\includegraphics{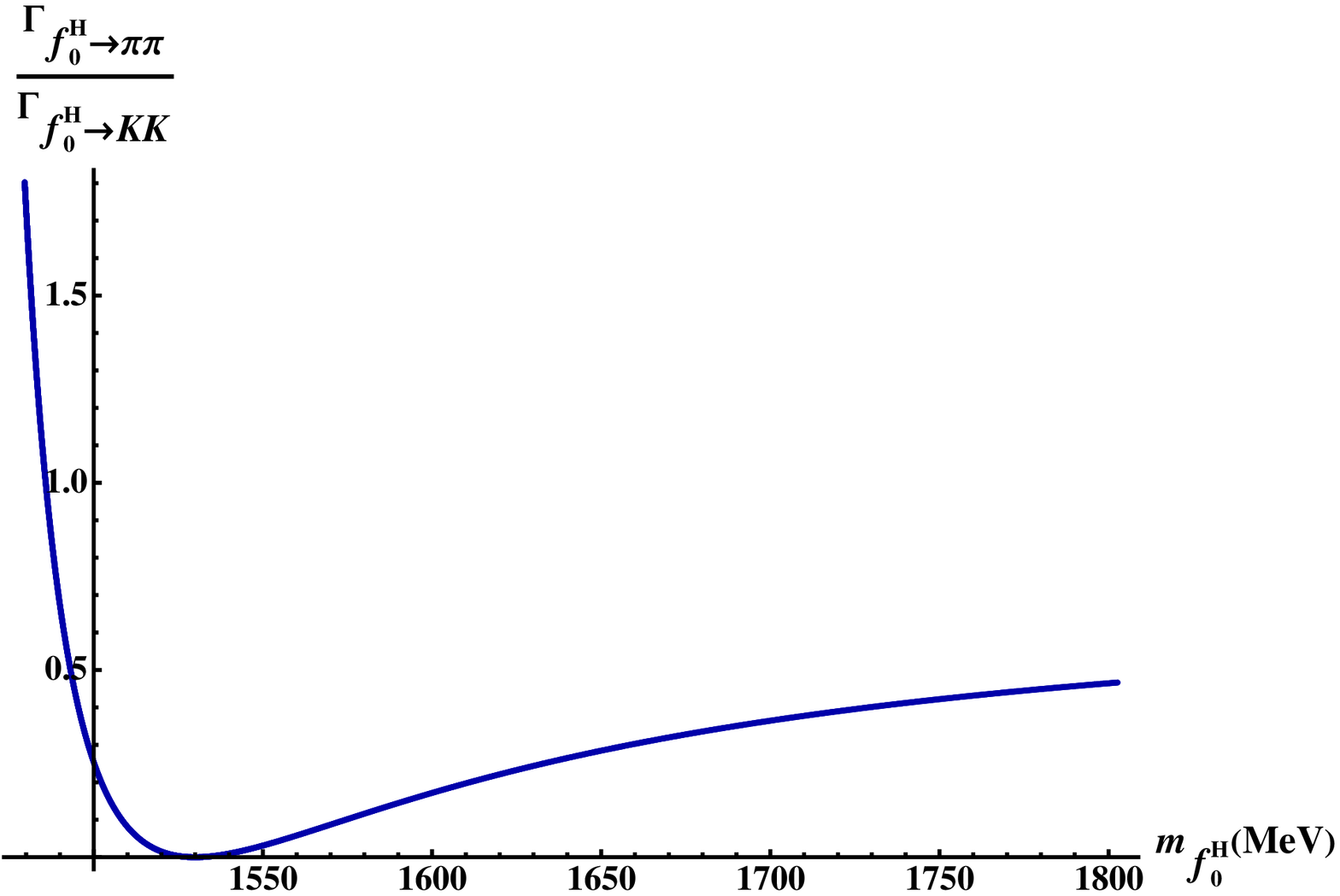}}
\end{tabular}
\end{center}
\caption{Left panel: ratio $\Gamma_{f_{0}^{L}\rightarrow KK}/\Gamma_{f_{0}%
^{L}\rightarrow\pi\pi}$ as function of $m_{f_{0}^{L}}$. Right panel: ratio
$\Gamma_{f_{0}^{H}\rightarrow\pi\pi}/\Gamma_{f_{0}^{H}\rightarrow KK}$ as
function of $m_{f_{0}^{H}}$.}%
\label{SPPKK2}%
\end{figure}

Let us first discuss results for $\Gamma_{f_{0}^{L}\rightarrow KK}%
/\Gamma_{f_{0}^{L}\rightarrow\pi\pi}$ (left panel of Fig.\ \ref{SPPKK2}). We
observe that the ratio varies between $0.49$ for $m_{f_{0}^{L}}=1152$ MeV and
$0$ for $m_{f_{0}^{L}}=1444$ MeV. Experimental data regarding this ratio for
$f_{0}(1370)$ are unfortunately inconclusive, $\Gamma_{f_{0}(1370)\rightarrow
KK}/\Gamma_{f_{0}(1370)\rightarrow\pi\pi}=0.08\pm0.08$ is quoted by the BESII
Collaboration \cite{Ablikim:2004} and $\Gamma_{f_{0}(1370)\rightarrow
KK}/\Gamma_{f_{0}(1370)\rightarrow\pi\pi}=$ $0.91\pm0.20$ is the value given
by the OBELIX Collaboration \cite{Bargiotti:2003}; for the mass interval shown
in Fig.\ \ref{SPPKK2}, our result is most consistent with the value of the
WA102 Collaboration \cite{Barberis:1999}, $\Gamma_{f_{0}(1370)\rightarrow
KK}/\Gamma_{f_{0}(1370)\rightarrow\pi\pi}=$ $0.46\pm0.15\pm0.11$. The
ambiguities in the experimental value of this ratio do not allow us to
constrain our parameters, although our $f_{0}^{L}$ state is compatible with
the $f_{0}(1370)$ data also for this particular case.

Let us now discuss results for $\Gamma_{f_{0}^{H}\rightarrow\pi\pi}%
/\Gamma_{f_{0}^{H}\rightarrow KK}$ (right panel of Fig.\ \ref{SPPKK2}). As
already mentioned, the analysis of the two-pion decay did not allow for a
definitive assignment of our $f_{0}^{H}$ state, as it could correspond either
to $f_{0}(1500)$ or to $f_{0}(1710)$. The PDG data suggest the ratio
$\Gamma_{f_{0}(1500)\rightarrow KK}/\Gamma_{f_{0}(1500)\rightarrow\pi\pi
}=0.246\pm0.026$ or $\Gamma_{f_{0}(1500)\rightarrow\pi\pi}/\Gamma
_{f_{0}(1500)\rightarrow KK}=4.065\pm0.430$, respectively, whereas, for
$f_{0}(1710)$, we use the WA102 ratio $\Gamma_{f_{0}(1710)\rightarrow\pi\pi
}/\Gamma_{f_{0}(1710)\rightarrow KK}=0.2\pm0.06$ \cite{Barberis:1999}. The
latter is only marginally consistent with the one preferred by the PDG
[$\Gamma_{f_{0}(1710)\rightarrow\pi\pi}^{\text{PDG}}/\Gamma_{f_{0}%
(1710)\rightarrow KK}^{\text{PDG}}=0.41_{-0.17}^{+0.11}$], originally
published by the BESII Collaboration \cite{f0(1710)-2006-BESII}, that suffers
from a large background in the $\pi^{+} \pi^{-}$ channel (approximately 50\%)
and is therefore omitted from our considerations. We observe from the right
panel of Fig.\ \ref{SPPKK2} that our ratio $\Gamma_{f_{0}^{H}\rightarrow\pi
\pi}/\Gamma_{f_{0}^{H}\rightarrow KK}$ never corresponds to the one
experimentally determined for $f_{0}(1500)$. Although the ratio shows a strong
increase near the left border of the mass interval, decreasing the mass beyond
this border would violate the constraint $m_{0}^{2}<0$. Thus, the experimental
value is out of reach. Conversely, our ratio describes exactly the value
$\Gamma_{f_{0}(1710)\rightarrow\pi\pi}/\Gamma_{f_{0}(1710)\rightarrow
KK}=0.2\pm0.06$ if $m_{f_{0}^{H}}=1502_{+3}^{-2}$ MeV. [We discard the second
possibility $m_{f_{0}^{H}}=1611_{-23}^{+27}$ MeV as the decay width
$\Gamma_{f_{0}^{H}\rightarrow KK}$ would exceed 600 MeV, see Fig.\ \ref{Kaons}%
, i.e., an order of magnitude larger than the experimental value, see
Eq.\ (\ref{SI2}).] This implies $\lambda_{1}=-4.1\mp0.7$ -- the parameter
$\lambda_{1}$ remains close to the large-$N_{c}$ limit and is much smaller
than $\lambda_{2}$, see Table~\ref{Tab:param}.

The contribution of the pure-strange state $\sigma_{S}$\ to $f_{0}^{H}$ is
then approximately 96\%, as can be calculated from Eqs.\
(\ref{eq:eigen_mass_angle}) -- (\ref{A16}). Thus our predominantly strange
state $f_{0}^{H}$ describes the $\pi\pi/KK$ ratio of $f_{0}(1710)$, and not
that of $f_{0}(1500)$, although the correct description requires that
$m_{f_{0}^{H}}$ corresponds to the mass value of $f_{0}(1500)$. The correct
description of the decay ratio is an indication that our assignment of
$f_{0}^{H}$ to $f_{0}(1710)$ is the correct one, whereas the fact that
$m_{f_{0}^{H}}$ is smaller than $m_{f_{0}(1710)}$ indicates the necessity to
include the coupling to the third isoscalar degree of freedom of our
model: the glueball. As
concluded from the $N_{f}=2$ version of our model \cite{Stani}, the
$f_{0}(1500)$ is predominantly a glueball state and thus considering the
glueball state is expected to induce a level repulsion in the masses. This may
shift the currently too small mass value of the predominantly strange
quarkonium from approximately $1.5$~GeV to $1.7$~GeV, where it is
experimentally found. Nonetheless, our results clearly demonstrate that scalar
quarkonia are found above, rather than below, $1$~GeV.

Until now we have only considered the case where one of our large-$N_{c}$
suppressed parameters ($\lambda_{1}$) is nonzero. We have also investigated
the influence of non-vanishing values for the other large-$N_{c}$ suppressed
parameter, $h_{1}$, on the decay widths $\Gamma_{f_{0}^{L,H}\rightarrow\pi\pi
}$ and $\Gamma_{f_{0}^{L,H}\rightarrow KK}$ [see Eqs.\ (\ref{aa}), (\ref{bb}),
(\ref{cc}), and (\ref{dd})]. Consistency with the large-$N_{c}$ deliberations
requires us to keep $h_{1}$ smaller than, or in the vicinity of, $h_{2}$ and
$h_{3}$, see Table~\ref{Tab:param}. We again observe that our ratio
$\Gamma_{f_{0}^{H}\rightarrow\pi\pi}/\Gamma_{f_{0}^{H}\rightarrow KK}$ never
corresponds to the one of the $f_{0}(1500)$ resonance whereas the ratio
$\Gamma_{f_{0}(1710)\rightarrow\pi\pi}/\Gamma_{f_{0}(1710)\rightarrow
KK}=0.2\pm0.06$ is correctly described if $h_{1}\sim-9$. In this case,
$m_{f_{0}^{H}}$ rises to approximately 1540 MeV and is thus outside the PDG
result $m_{f_{0}(1500)}=(1505\pm6)$ MeV but still too small when compared to
$m_{f_{0}(1710)}=(1720\pm6)$ MeV. Thus, it is still necessary to include a
glueball degree of freedom into our model. Nonetheless, the
qualitative correspondence of our predominantly non-strange quarkonium to
$f_{0}(1370)$ and of our predominantly strange quarkonium to $f_{0}(1710)$,
and also the conclusion that scalar $\bar{q}q$ states are located in the
energy region above 1 GeV, remain valid in the case $\lambda_{1}\neq0\neq
h_{1}$.

The assignment of $f_{0}^{H}$ to $f_{0}(1710)$ is further justified
considering decays into $\eta$ and $\eta^{\prime}$ mesons.

\item[(c)] \textit{$\eta\eta$ and $\eta\eta^{\prime}$ decay channels for the
scalar-isoscalar mesons:}\\[0.1cm]PDG data suggest the following values of
$\eta\eta$ decay widths for $f_{0}(1500)$ and $f_{0}(1710)$ \cite{PDG}
\begin{equation}
\Gamma_{f_{0}(1500)\rightarrow\eta\eta}=(5.56\pm1.34)\,\text{MeV}%
\;,\;\;\;\Gamma_{f_{0}(1710)\rightarrow\eta\eta}=(38.6\pm18.8)\,\text{MeV}\;.
\label{Gammaetaeta}%
\end{equation}
Our analysis always yields $\Gamma_{f_{0}^{H}\rightarrow\eta\eta}>20$ MeV;
there is consequently no value of $m_{f_{0}^{H}}$ at which $\Gamma_{f_{0}%
^{H}\rightarrow\eta\eta}$ would be compatible with $\Gamma_{f_{0}%
(1500)\rightarrow\eta\eta}$. However, $m_{f_{0}^{H}}=1502_{+3}^{-2}$ MeV
[obtained requiring $\Gamma_{f_{0}^{H}\rightarrow\pi\pi}/\Gamma_{f_{0}%
^{H}\rightarrow KK}=\Gamma_{f_{0}(1710)\rightarrow\pi\pi}/\Gamma
_{f_{0}(1710)\rightarrow KK}$, i.e., $\lambda_{1}=-4.1\mp0.7$ and $h_{1}=0$]
yields
\begin{equation}
\Gamma_{f_{0}^{H}\rightarrow\eta\eta}=49.6_{-3.3}^{+4.1}\,\text{MeV}\;,
\end{equation}
(where the errors arise from the uncertainty in $\lambda_{1}$ only), i.e., in
agreement with the experimental value for $f_0(1710)$ 
quoted in Eq.\ (\ref{Gammaetaeta}). The
same parameter set also yields
\begin{equation}
\Gamma_{f_{0}^{L}\rightarrow\eta\eta}\simeq33\,\text{MeV}\;;
\end{equation}
purely a prediction, as $\Gamma_{f_{0}(1370)\rightarrow\eta\eta}$ has not yet
been measured.

The choice $\lambda_{1}=-4.1\mp0.7,$ $h_{1}=0$ also yields
\begin{align}
\Gamma_{f_{0}^{L}\rightarrow\eta\eta}/\Gamma_{f_{0}^{L}\rightarrow KK}  &
=0.194_{-0.003}^{+0.002}\;\text{, }\;\;\Gamma_{f_{0}^{H}\rightarrow\eta\eta
}/\Gamma_{f_{0}^{H}\rightarrow KK}=0.204_{-0.002}^{+0.001}\;\text{,}\\
\Gamma_{f_{0}^{L}\rightarrow\eta\eta}/\Gamma_{f_{0}^{L}\rightarrow\pi\pi}  &
=0.087_{-0.006}^{+0.005}\;\text{, }\;\;\Gamma_{f_{0}^{H}\rightarrow\eta\eta
}/\Gamma_{f_{0}^{H}\rightarrow\pi\pi}=1.02_{+0.42}^{-0.23}\;,
\end{align}
whereas experimental data read%
\begin{align}
\Gamma_{f_{0}(1500)\rightarrow\eta\eta}/\Gamma_{f_{0}(1500)\rightarrow KK}  &
=0.59\pm0.12\;\;\text{ \cite{PDG} }\;,\\
\Gamma_{f_{0}(1500)\rightarrow\eta\eta}/\Gamma_{f_{0}(1500)\rightarrow\pi\pi}
&  =0.145\pm0.027\;\;\text{ \cite{PDG}}\;,\\
\Gamma_{f_{0}(1710)\rightarrow\eta\eta}/\Gamma_{f_{0}(1710)\rightarrow KK}  &
=0.48\pm0.15\;\;\text{ \cite{Barberis:2000}}\;,\\
\Gamma_{f_{0}(1710)\rightarrow\eta\eta}/\Gamma_{f_{0}(1710)\rightarrow KK}  &
=0.46_{-0.38}^{+0.70}\;\;\text{ \cite{Anisovich:2001} }\;,\\
\Gamma_{f_{0}(1710)\rightarrow\eta\eta}/\Gamma_{f_{0}(1710)\rightarrow\pi\pi}
&  =2.40\pm1.04\;\;\text{ \cite{Barberis:1999,Barberis:2000}}\;.
\end{align}
In all cases, the $f_{0}(1500)$ data are off by several standard deviations
from our theoretical results, while there is good agreement with the results
for $f_{0}(1710)$.

We have also considered decays involving the $\eta^{\prime}$ meson. Since the
threshold for $\eta\eta^{\prime}$ decays is at approximately 1.5 GeV, it
suffices to consider decays of $f_{0}^{H}$ only. The value $\lambda
_{1}=-4.1\mp0.7$ yields
\begin{align}
\Gamma_{f_{0}^{H}\rightarrow\eta\eta^{\prime}}  &  =12.7_{-1.4}^{+1.1}%
~\text{MeV}\;,\\
\Gamma_{f_{0}^{H}\rightarrow\eta\eta^{\prime}}/\Gamma_{f_{0}^{H}\rightarrow
\pi\pi}  &  =0.26_{+0.04}^{-0.03}~\text{,}\\
\Gamma_{f_{0}^{H}\rightarrow\eta\eta^{\prime}}/\Gamma_{f_{0}^{H}%
\rightarrow\eta\eta}  &  =0.26_{-0.05}^{+0.04}\;,\\
\Gamma_{f_{0}^{H}\rightarrow\eta\eta^{\prime}}/\Gamma_{f_{0}^{H}\rightarrow
KK}  &  =0.05\pm0.01~\text{,}%
\end{align}
whereas experimental data read
\begin{align}
\Gamma_{f_{0}(1500)\rightarrow\eta\eta^{\prime}}  &  =(2.1\pm1.0)~\text{MeV}%
\;,\\
\Gamma_{f_{0}(1500)\rightarrow\eta\eta^{\prime}}/\Gamma_{f_{0}%
(1500)\rightarrow\pi\pi}  &  =0.055\pm0.024~\text{,}\\
\Gamma_{f_{0}(1500)\rightarrow\eta\eta^{\prime}}/\Gamma_{f_{0}%
(1500)\rightarrow\eta\eta}  &  =0.38\pm0.16\;\text{.}%
\end{align}
The decay ratio $\Gamma_{f_{0}(1500)\rightarrow\eta\eta^{\prime}}%
/\Gamma_{f_{0}(1500)\rightarrow KK}$ is unknown; there are also no data for
the $\eta\eta^{\prime}$ decay channel of $f_{0}(1710)$. Still we observe that
neither $\Gamma_{f_{0}^{H}\rightarrow\eta\eta^{\prime}}$ nor $\Gamma
_{f_{0}^{H}\rightarrow\eta\eta^{\prime}}/\Gamma_{f_{0}^{H}\rightarrow\pi\pi}$
describe the corresponding experimental results for $f_{0}(1500)$. Indeed the
only piece of experimental data regarding $f_{0}(1500)$ that is described by
our fit results is the one for the $\eta\eta^{\prime}/\eta\eta$ decay ratio.
Nonetheless, all the other results regarding decay ratios obtained by our
analysis clearly demonstrate the correspondence of our predominantly strange
state $f_{0}^{H}$ to $f_{0}(1710)$; $f_{0}^{L}$ was found to correspond to
$f_{0}(1370)$ already in the discussion of the $\pi\pi$ decay channel.
Consequently, we conclude that $f_{0}(1370)$ and $f_{0}(1710)$ are favoured as
scalar $\bar{q}q$ states. However, we also stress again that the mass of our
$f_{0}^{H}$ remains too low when compared to $m_{f_{0}(1710)}$ due to a
missing scalar-glueball state expected to shift $m_{f_{0}^{H}}$ to
$m_{f_{0}(1710)}$ by level repulsion.
\end{enumerate}

\subsubsection{Mixing in the pseudoscalar-isoscalar sector}

Our Lagrangian (\ref{eq:Lagrangian}) implements the mixing of two pure
pseudoscalar isosinglet states, $\eta_{N}\equiv(\bar{u}u+\bar{d}d)/\sqrt{2}$
and $\eta_{S}\equiv\bar{s}s$. The mixing term is presented in
Eq.\ (\ref{eq:etaNS}). The mass terms for $\eta_{N}$ (\ref{eq:etaN}) and
$\eta_{S}$ (\ref{eq:etaS}) are determined by our fit parameters presented in
Table~\ref{Tab:param}. The same parameters also determine the $\eta$%
-$\eta^{\prime}$ mixing angle as [see Eqs.\ (\ref{eq:eigen_mass_angle}) --
(\ref{A16})]
\begin{equation}
\theta_{\eta}=-44.6^{\circ}\;.
\end{equation}
The result is close to maximal mixing, i.e., our result
  suggests a slightly larger mixing than those of Ref.\ \cite{etaM}.

\subsubsection{The axial-vector kaon state $K_{1}$}

The $K_{1}$ state has not been assigned to a physical resonance and included
into our fit, because the PDG listing suggest two distinct assignment
candidates: $K_{1}(1270)$ and $K_{1}(1400)$, expected to mix \cite{burakovsky}%
. Therefore, mass and decay widths are left as predictions of this work.

There are three decay widths of the $K_{1}$ state that can be calculated
within our model at tree level: $K_{1}$ $\rightarrow K^{\star}\pi$, $\rho K$,
and $\omega K$. They account for approximately $70\%$ of the full
$K_{1}(1270)$ decay width and almost $100\%$ of the full $K_{1}(1400)$ decay
width \cite{PDG}. Using the parameter values stated in Table~\ref{Tab:param}
it is possible to calculate the mass, Eq.\ (\ref{m_K_1}), as well as the decay
width [via the generic decay-width formula (\ref{GK1VP})] of our $K_{1}$
state. We obtain the following results: \
\begin{equation}
m_{K_{1}}=1282\,\text{ MeV}\;,\;\;\Gamma_{K_{1}\rightarrow K^{\star}\pi
}=205\,\text{MeV}\;,\;\;\Gamma_{K_{1}\rightarrow\rho K}=44~\text{MeV}\;,
\;\;\Gamma_{K_{1}\rightarrow\omega K}=15~\text{MeV}\;.
\end{equation}
The mass is within $2\sigma$ of $m_{K_{1}(1270)}=(1272\pm7)$ MeV and thus
rather close to the experimental result. However, the value of the full decay
width is $264$~MeV, while the PDG data read $\Gamma_{K_{1}(1270)}=(90\pm
20)$~MeV and $\Gamma_{K_{1}(1400)}=(174\pm13)$~MeV. Our result is therefore
approximately three times too large when compared to the data for
$K_{1}(1270)$ and approximately $50\%$ too large when compared to the data for
$K_{1}(1400)$; errors have been omitted from the calculation. These results
demonstrate the necessity to include a pseudovector $I(J^{PC})=1(1^{+-})$
nonet into our model and implement its mixing with the already present
axial-vector nonet. A possible mixing term between an axial-vector nonet
$A_{1}$ and a pseudovector nonet $B_{1}$ reads%
\begin{equation}
\mathrm{Tr}(\Delta\lbrack A_{1}^{\mu},B_{1\mu}])\;, \label{K11}%
\end{equation}
with $\Delta$ from Eq.\ (\ref{eq:expl_sym_br_delta}). Various studies have
indeed found the mixing to be non-negligible \cite{burakovsky}; see also
Ref.\ \cite{Cheng:2011}.

\subsubsection{Branching ratios of $a_{0}(1450)$}

As a consequence of our fit we can determine the branching ratio of the
resonance $a_{0}(1450)$ into $KK$, $\pi\eta,$ and $\pi\eta^{\prime}.$ We
obtain the following values [see Eqs.\ (\ref{B1}), (\ref{B2}), and
(\ref{B3})]:%
\begin{equation}
\Gamma_{a_{0}\rightarrow\eta\pi}=(115.4\pm6.2)~\text{MeV,}\;\Gamma
_{a_{0}\rightarrow\eta^{\prime}\pi}=(21.5\pm1.4)~\text{MeV,}\;\Gamma
_{a_{0}\rightarrow KK}=(128.8\pm3.9)~\text{MeV.}\;
\end{equation}
This leads to the following branching ratios:%
\begin{equation}
\frac{\Gamma_{a_{0}\rightarrow\eta^{\prime}\pi}}{\Gamma_{a_{0}\rightarrow
\eta\pi}}=0.19\pm0.02\;\text{,}\;\;\frac{\Gamma_{a_{0}\rightarrow KK}}%
{\Gamma_{a_{0}\rightarrow\eta\pi}}=1.12\pm0.07\;,
\end{equation}
which should be compared with the experimental results \cite{PDG}
\begin{equation}
\frac{\Gamma_{a_{0}(1450)\rightarrow\eta^{\prime}\pi}}{\Gamma_{a_{0}%
(1450)\rightarrow\eta\pi}}=0.35\pm0.16\;\text{,}\;\;\frac{\Gamma
_{a_{0}(1450)\rightarrow KK}}{\Gamma_{a_{0}(1450)\rightarrow\eta\pi}}%
=0.88\pm0.23\;\text{.}%
\end{equation}
Our results are, within errors, consistent with the data.

\bigskip

\subsubsection{Contributions to the mass of the $\rho$ meson}

The mass of the $\rho$ meson consists of three terms,
\begin{equation}
\label{rhomass}m_{\rho}^{2} = m_{1}^{2} + \frac{1}{2}(h_{1} + h_{2} +
h_{3})\phi_{N}^{2} + \frac{h_{1}}{2} \phi_{S}^{2}\;.
\end{equation}
The first term $m_{1}^{2}$ is generated by the condensation of the dilaton
field $G,$ $m_{1}^{2}\propto G_{0}^{2}.$ The second term is proportional to
the chiral condensate $\phi_{N}^{2}.$ In the large-$N_{c}$ limit the parameter
$h_{1}=0$ and we can determine the terms from the results of the fit:
$m_{1}=0.643$ GeV and $\sqrt{(h_{2}+h_{3})/2}\phi_{N}=0.447$~GeV. It turns out
that the glueball-driven term is dominant.

It is interesting to note that models based on QCD sum rules
\cite{Hatsuda-Lee} or Brown-Rho scaling \cite{Brown-Rho} predict that
$m_{\rho}^{k} \propto\phi_{N}$, where $k=2$ in the first and $k=3$ in the
latter case.
The order parameter for chiral symmetry breaking $\phi_{N}$
decreases as a function of nuclear matter density $n$: for small densities it
is well known [see, e.g.\ Ref.\ \cite{Hatsuda-Lee}] that
\[
\frac{\phi_{N}|_{n}}{\phi_{N}|_{\text{vacuum}}} \approx1 - 0.3 \frac{n}{n_{0}%
}\;,
\]
so these models predict a substantial decrease of $m_{\rho}$ already at
nuclear matter saturation density, $n_{0}$. In this respect, it is interesting
to evaluate the value of the $\rho$ mass at $n_{0}$ according to
Eq.\ (\ref{rhomass}). We assume that the glueball-driven term $m_{1}^{2}$ does
not vary. This can be motivated by considering that the glueball is massive
and a substantial decrease in its mass may occur only at much higher density.
Assuming that the linear dependence on the density of the condensate holds up
to saturation density, we obtain that $m_{\rho}$ decreases by about $70$~MeV
compared to $90-130$~MeV predicted by Refs.\ \cite{Hatsuda-Lee,Brown-Rho}.

Obviously, these considerations are only qualitative but could represent the
starting point of interesting studies of vector mesons in the medium, which is
an important aim of several experimental investigations \cite{CERES}. To this
end, one has to calculate the behavior of the dilaton and chiral condensates
in the medium within the same theoretical framework.

\section{CONCLUSIONS}

\label{sec:conclusion}

We have presented a linear sigma model with three quark flavors. The model
implements the symmetries of QCD, the discrete $CPT$ symmetry, the global
chiral $U(N_{f})_{L}\times U(N_{f})_{R}$ symmetry, and the breaking mechanisms
of the last: spontaneous (due to the chiral condensate), explicit (due to
non-vanishing quark masses) as well as at the quantum level [the $U(1)_{A}$
anomaly]. Moreover, it implements also dilatation symmetry and its
breaking (the so-called trace anomaly) in the YM sector of the theory.
In this way, besides explicit breaking of dilatation symmetries arising from
the nonzero current quark masses and the trace and axial anomalies in the
gauge sector, all other interaction terms in our Lagrangian carry mass
dimension equal to four. Furthermore, requiring analyticity in the fields makes the
number of allowed terms finite.

The model has been used to describe meson states up to energies of $\sim1.7$
GeV. This energy region exhibits numerous resonances, related by scattering
reactions and decays. For this reason, a realistic model of QCD degrees of
freedom in the mentioned energy region should describe as many of the
resonances as possible. Thus, we have constructed a linear sigma model that
contains scalar (two isoscalars, $f_{0}^{L}$ and $f_{0}^{H}$, as well as an
isotriplet, ${a}_{0}$, and two isodoublets, $K_{0}^{\star}$), pseudoscalar
($\pi$, $K$, $\eta$, $\eta^{\prime}$), vector ($\rho$, $\omega$, $K^{\star}$,
$\phi$), and axial-vector [$a_{1}$, $K_{1}$, $f_{1}(1285)$, $f_{1}(1420)$]
degrees of freedom. We have thus constructed a single model that contains
(pseudo)scalars and (axial-)vectors both in the non-strange and strange
channels. To our knowledge, this is the first time that such a comprehensive
approach has been presented.\newline

The model, dubbed extended Linear Sigma Model (eLSM), has allowed us to study
the overall phenomenology of mesons and, in particular, to explore the nature
of scalar and axial-vector resonances. In order to test our model we have
performed a global fit to $21$ experimental quantities involving both the
(pseudo)scalar and the (axial-)vector masses and decays. Due to mixing with
the scalar glueball (not included here explicitly, because its
coupling to the other mesons was neglected), we did not include the scalar-isoscalar
resonances in the fit. Similarly, we have omitted the axial-vector resonance
$K_{1},$ due to the fact that in reality a large mixing of two kaonic fields
from the $1^{++}$ and $1^{+-}$ nonets takes place.\newline

One of the central questions of our discussion has been the assignment of the
scalar states: to this end we have tested the possible scenarios for the
isotriplet and isodoublet scalar states by assigning our scalar fields
${a}_{0}$ to $a_{0}(980)$ or $a_{0}(1450)$ and $K_{0}^{\star}$ to
$K_{0}^{\star}(800)$ or $K_{0}^{\star}(1430)$. The outcome is univocal: the
global fit works well only if the states $a_{0}(1450)$ and $K_{0}^{\star
}(1430)$ are interpreted as quark-antiquark states. On the contrary, the other
combinations deliver unacceptably large values of $\chi^{2}$, see
Table~\ref{Tab:khi}. We thus conclude that the scalar $I=1$ and $I=1/2$ states
lie above 1 GeV and have to be identified with the resonances $a_{0}(1450)$
and $K_{0}^{\star}(1430).$ Moreover, the overall phenomenology described by
the fit is very good, see Table~\ref{Comparison1}. It is then possible to
properly describe many different mesonic masses and decays within a unified
treatment based on the symmetries of QCD. It should be stressed that the
inclusion of the (axial-)vector mesons has a crucial impact on our results and
represents the most important new ingredient of our approach. The good
agreement with data also shows that the axial-vector mesons can be
interpreted, just as their vector chiral partners, as quark-antiquark
states.\newline

We have then studied the consequences of our fit. We have primarily
concentrated on the scalar-isoscalar sector which was not included in the fit.
In the limit $N_{c}\rightarrow\infty$ it is possible to make clear predictions
for the two states $f_{0}^{L}$ and $f_{0}^{H}$. Their masses lie above 1 GeV
and their decay patterns have led us to identify $f_{0}^{L}$ with
(predominantly) $f_{0}(1370)$ and $f_{0}^{H}$ with (predominantly)
$f_{0}(1710).$ The theoretical decay rates of $f_{0}^{L}$ are in agreement
with experiment; $f_{0}^{H}$ decays only into kaons, but turns out to be too
wide. At a qualitative level, this large-$N_{c}$ result clearly shows that
also the scalar-isoscalar quark-antiquark states lie above 1 GeV. The overall
situation in the scalar-isoscalar sector can be slightly improved when
adding large-$N_{c}$ suppressed terms. Some of
these terms correspond to interactions between mesons and a
glueball state. These were neglected here, but are
necessary for a full quantitative study. Then
starting from $\sigma_{N}%
=\sqrt{1/2}(\bar{u}u+\bar{d}d)$, $G=gg$, and $\sigma
_{S}=\bar{s}s$ one aims to describe properly $f_{0}(1370),$
$f_{0}(1500),$ and $f_{0}(1710)$.

Additionally, we have studied other consequences of the fit, such as
predictions for the $\eta\eta$ channel of scalar-isoscalar states, the $\eta
$-$\eta^{\prime}$ mixing, the prediction for the axial-vector kaonic state,
and for the ${a}_{0}$ branching ratios. Finally, we have also discussed the
origin of the mass of the $\rho$ meson and, at a qualitative level, its
possible value at nuclear matter density.\newline

There is, however, one question that remains open. Interpreting
resonances above 1 GeV as $\bar{q}q$ states leads to questions about the
nature of $f_{0}(500)$, $a_{0}(980),$ $f_{0}(980)$, and $K_{0}^{\star}(800)$.
Their presence is necessary for the correct description of $\pi\pi$ scattering
lengths [see, e.g.\ Ref.\ \cite{Paper1}]. There are two possibilities:
(\textit{i}) they can arise as (quasi-)molecular states of $\pi\pi$ or $KK$,
respectively, or (\textit{ii}) as tetraquark states
\cite{varietq,Giacosamixing}. The question is whether the attraction is large
enough so that these states are bound, or whether the attraction is not
sufficient so that they are (unstable) resonances in the continuum. This
question and deciding whether possibility (\textit{i}) or (\textit{ii}) is
realized in nature represent interesting starting points for future studies,
along the lines of the Bethe-Salpeter approach of Ref.\ \cite{Fischer}, the
lattice approach of Ref.\ \cite{Wagner}, or even holographic approaches such
as those presented in Ref.\ \cite{Sakai}. In either case, one may include the
low-lying isoscalar states as interpolating fields in our Lagrangian, such as
has been done in Ref.\ \cite{Mixing}.\newline

The present model can be also studied in the baryonic sector, see
Ref.\ \cite{gallas} for the two-flavor case. The very same ideas of chiral
symmetry and dilatation invariance can be applied to the baryonic sector as
well. The extension of the model to three flavors in the baryonic sector would
surely represent an interesting problem: a multitude of data on decays and
masses is available to make a precise test of our approach.\newline

Additionally, restoration of chiral symmetry at nonzero temperature and
density is one of the fundamental questions of modern hadron physics, see,
e.g.\ Refs.\ \cite{RS,susagiu}, in which the two-flavor version of this model
has been studied at nonzero density, or Ref.\ \cite{Andreas} for alternative
approaches to the exploration of nonzero chemical potential. Linear sigma
models are appropriate to study chiral symmetry restoration because they
contain from the onset not only (pseudo)scalar and (axial-)vector mesons but
also their chiral partners; mass degeneration of chiral partners represents a
signal for the chiral transition. Therefore, we also plan to apply the model
to study chiral symmetry restoration at nonzero temperatures and densities.

\section*{Acknowledgments}

Gy.\ Wolf and P.\ Kov{\'a}cs thank the Institute for Theoretical Physics of
Goethe University for its hospitality, where part of this work was done. They
were partially supported by the Hungarian OTKA funds T71989 and T101438. The
work of D.\ Parganlija and F.\ Giacosa was supported by the Foundation of the
Polytechnical Society Frankfurt. This work was financially supported by the
Helmholtz International Center for FAIR within the framework of the LOEWE
program (Landesoffensive zur Entwicklung Wissenschaftlich-\"{O}konomischer
Exzellenz) launched by the State of Hesse. D.H.\ Rischke thanks R.\ Longacre
and R.\ Pisarski for enlightening discussions.

\appendix

\section{Tree-level masses}

\label{App:mass}

After spontaneous symmetry breaking (see Sec.~\ref{ssec:SSB_and_mass}), from
the quadratic terms of the Lagrangian the (squared) mass matrices for the
different fields are read off as
\begin{align}
(m_{S}^{2})_{ij}  &  =S_{ip}(m_{0}^{2}\delta_{pq}+4F_{pqlm}\phi_{l}\phi
_{m})S_{qj},\\
(m_{P}^{2})_{ij}  &  =S_{ip}(m_{0}^{2}\delta_{pq}-9c_{1}G_{plm}\phi_{l}%
\phi_{m}G_{ql^{\prime}m^{\prime}}\phi_{l^{\prime}}\phi_{m^{\prime}}%
+4H_{pq,lm}\phi_{l}\phi_{m})S_{qj},\\
(m_{V}^{2})_{ij}  &  =S_{ip}(m_{1}^{2}\delta_{pq}+J_{pqlm}\phi_{l}\phi
_{m})S_{qj},\\
(m_{A}^{2})_{ij}  &  =S_{ip}(m_{1}^{2}\delta_{pq}+J_{pqlm}^{\prime}\phi
_{l}\phi_{m})S_{qj},
\end{align}
where $S_{qj}$ is the $(9\times9)$ transformation matrix to the non strange --
strange base,
\begin{equation}
S = \frac{1}{\sqrt{3}}\left(
\begin{array}
[c]{cccc}%
\sqrt{2} & 0 & \dots & 1\\
0 & 1 & \dots & 0\\
\vdots & \vdots & \ddots & \vdots\\
1 & 0 & \dots & -\sqrt{2}%
\end{array}
\right)  ,
\end{equation}
and the $F,H,G,J,J^{\prime}$ coefficient tensors are
\begin{align}
F_{ijkl}  &  =\frac{\lambda_{1}}{4}\left(  \delta_{ij}\delta_{kl}+\delta
_{ik}\delta_{jl}+\delta_{il}\delta_{jk}\right)  +\frac{\lambda_{2}}{8}\left(
d_{ijm}d_{klm}+d_{ikm}d_{jlm}+d_{ilm}d_{jkm}\right)  ,\\
H_{ij,kl}  &  =\frac{\lambda_{1}}{4}\delta_{ij}\delta_{kl}+\frac{\lambda_{2}%
}{8}\left(  d_{ijm}d_{klm}+f_{ikm}f_{jlm}+f_{ilm}f_{jkm}\right)  ,\\
G_{ijk}  &  =\frac{1}{6}\left[  d_{ijk}+\frac{9}{2}d_{000}\delta_{i0}%
\delta_{j0}\delta_{k0}-\frac{3}{2}\left(  \delta_{i0}d_{jk0}+\delta
_{j0}d_{ik0}+\delta_{k0}d_{ij0}\right)  \right]  ,\\
J_{ijkl}  &  =g_{1}^{2}f_{ikm}f_{jlm}+\frac{h_{1}}{2}\delta_{ij}\delta
_{kl}+\frac{h_{2}}{2}d_{ijm}d_{klm}+\frac{h_{3}}{4}\left(  d_{ikm}%
d_{jlm}+d_{ilm}d_{jkm}-f_{ikm}f_{jlm}-f_{ilm}f_{jkm}\right)  ,\\
J_{ijkl}^{\prime}  &  =g_{1}^{2}d_{ikm}d_{jlm}+\frac{h_{1}}{2}\delta
_{ij}\delta_{kl}+\frac{h_{2}}{2}d_{ijm}d_{klm}-\frac{h_{3}}{4}\left(
d_{ikm}d_{jlm}+d_{ilm}d_{jkm}-f_{ikm}f_{jlm}-f_{ilm}f_{jkm}\right)  .
\end{align}
As can be seen from Eqs.\ \eqref{eq:etaN}--\eqref{eq:etaNS}, and
Eqs.\ \eqref{eq:sigN}--\eqref{eq:sigNS}, there is a mixing in the pseudoscalar
and scalar $N-S$ sectors, which can be resolved by the following
two-dimensional orthogonal transformations
\begin{equation}
O_{\eta/\sigma} = \left(
\begin{array}
[c]{cc}%
\cos\theta_{\eta/\sigma} & \sin\theta_{\eta/\sigma}\\
-\sin\theta_{\eta/\sigma} & \cos\theta_{\eta/\sigma}%
\end{array}
\right)  \text{,} \label{eq:orthog}%
\end{equation}
where $\theta_{\eta/\sigma}$ are the pseudoscalar and scalar mixing angles.
The (squared) mass eigenvalues can be written with the help of the mixing
angles as
\begin{align}
m_{\eta_{1}/\sigma_{1}}^{2}  &  =m_{\eta_{N}/\sigma_{N}}^{2}\cos^{2}%
\theta_{\eta/\sigma}+m_{\eta_{NS}/\sigma_{NS}}^{2}\sin2\theta_{\eta/\sigma
}+m_{\eta_{S}/\sigma_{S}}^{2}\sin^{2}\theta_{\eta/\sigma}%
,\label{eq:eigen_mass_angle}\\
m_{\eta_{2}/\sigma_{2}}^{2}  &  =m_{\eta_{N}/\sigma_{N}}^{2}\sin^{2}%
\theta_{\eta/\sigma}-m_{\eta_{NS}/\sigma_{NS}}^{2}\sin2\theta_{\eta/\sigma
}+m_{\eta_{S}/\sigma_{S}}^{2}\cos^{2}\theta_{\eta/\sigma},{\nonumber}%
\end{align}
where, if we require that $m_{\eta_{2}/\sigma_{2}}^{2}>m_{\eta_{1}/\sigma_{1}%
}^{2}$, the mixing angles are given by
\begin{align}
\sin\theta_{\eta/\sigma}  &  =-\mathrm{sign}(m_{\eta_{NS}/\sigma_{NS}}%
^{2})\frac{1}{\sqrt{2}}\sqrt{1-\frac{m_{\eta_{S}/\sigma_{S}}^{2}-m_{\eta
_{N}/\sigma_{N}}^{2}}{\sqrt{(m_{\eta_{N}/\sigma_{N}}^{2}-m_{\eta_{S}%
/\sigma_{S}}^{2})^{2}+4m_{\eta_{NS}/\sigma_{NS}}^{4}}}}\;,\\
\cos\theta_{\eta/\sigma}  &  =\frac{1}{\sqrt{2}}\sqrt{1+\frac{m_{\eta
_{S}/\sigma_{S}}^{2}-m_{\eta_{N}/\sigma_{N}}^{2}}{\sqrt{(m_{\eta_{N}%
/\sigma_{N}}^{2}-m_{\eta_{S}/\sigma_{S}}^{2})^{2}+4m_{\eta_{NS}/\sigma_{NS}%
}^{4}}}}\;.
\end{align}
It can be seen that, if $m_{\eta_{NS}/\sigma_{NS}}^{2}>0$, then $-\pi
/2<\theta_{\eta/\sigma}<0$ and, if $m_{\eta_{NS}/\sigma_{NS}}^{2}<0$, then
$0<\theta_{\eta/\sigma}<\pi/2$. Substituting these expression into
Eq.~\eqref{eq:eigen_mass_angle} it is found that
\begin{align}
m_{\eta_{1}/\sigma_{1}}^{2}  &  =\frac{1}{2}\left[  m_{\eta_{N}/\sigma_{N}%
}^{2}+m_{\eta_{S}/\sigma_{S}}^{2}-\sqrt{(m_{\eta_{N}/\sigma_{N}}^{2}%
-m_{\eta_{S}/\sigma_{S}}^{2})^{2}+4m_{\eta_{NS}/\sigma_{NS}}^{4}}\right]  ,\\
m_{\eta_{2}/\sigma_{2}}^{2}  &  =\frac{1}{2}\left[  m_{\eta_{N}/\sigma_{N}%
}^{2}+m_{\eta_{S}/\sigma_{S}}^{2}+\sqrt{(m_{\eta_{N}/\sigma_{N}}^{2}%
-m_{\eta_{S}/\sigma_{S}}^{2})^{2}+4m_{\eta_{NS}/\sigma_{NS}}^{4}}\right]  ,
\label{A16}%
\end{align}
from which it is obvious that the condition $m_{\eta_{2}/\sigma_{2}}%
^{2}>m_{\eta_{1}/\sigma_{1}}^{2}$ is fulfilled. In this way we know that
$m_{\eta_{2}}$ must be identified as $m_{\eta^{\prime}}$, and similarly for
$\sigma$.

\section{Decay widths}

\label{App:decay}

In this section we show the explicit form of some of the most relevant decay
widths calculated from our model at tree-level using Eqs.\ (\ref{Gamma}). The
formulas below are organized according to the type of the decaying particle.

\subsection{Scalar-meson decay widths}

At first we are considering the decays of the scalar isotriplet ${a}_{0}$,
scalar kaon $K_{0}^{\star}$, and scalar isosinglets $f_{0}^{H/L}$. The
${a}_{0}$ state has three relevant decay channels, namely into $\eta\pi$,
$\eta^{\prime}\pi$, and $KK$, the first two of which are strongly connected
due to the mixing between $\eta$ and $\eta^{\prime}$. The $a_{0}%
\rightarrow\eta\pi$ and $a_{0}\rightarrow\eta^{\prime}\pi$ decay widths
\footnote{Obviously, the neutral and the charged $a_{0}$'s have the same
formulas for the decay widths.}, considering that in both cases $\mathcal{I}%
=1$, read
\begin{align}
\Gamma_{a_{0}\rightarrow\eta\pi}  &  =\frac{1}{8m_{a_{0}}\pi}\left[
\frac{(m_{a_{0}}^{2}-m_{\eta}^{2}-m_{\pi}^{2})^{2}-4m_{\eta}^{2}m_{\pi}^{2}%
}{4m_{a_{0}}^{4}}\right]  ^{1/2}|\mathcal{M}_{a_{0}\rightarrow\eta\pi}%
|^{2},\label{B1}\\
\Gamma_{a_{0}\rightarrow\eta^{\prime}\pi}  &  =\frac{1}{8m_{a_{0}}\pi}\left[
\frac{(m_{a_{0}}^{2}-m_{\eta^{\prime}}^{2}-m_{\pi}^{2})^{2}-4m_{\eta^{\prime}%
}^{2}m_{\pi}^{2}}{4m_{a_{0}}^{4}}\right]  ^{1/2}|\mathcal{M}_{a_{0}%
\rightarrow\eta^{\prime}\pi}|^{2}, \label{B2}%
\end{align}
with the following transition matrix elements,
\begin{align}
\mathcal{M}_{a_{0}\rightarrow\eta\pi}  &  =\cos\theta_{\pi}\mathcal{M}%
_{a_{0}\rightarrow\eta_{N}\pi}(m_{\eta})+\sin\theta_{\pi}\mathcal{M}%
_{a_{0}\rightarrow\eta_{S}\pi}(m_{\eta})\;,\\
\mathcal{M}_{a_{0}\rightarrow\eta^{\prime}\pi}  &  =\cos\theta_{\pi
}\mathcal{M}_{a_{0}\rightarrow\eta_{S}\pi}(m_{\eta^{\prime}})-\sin\theta_{\pi
}\mathcal{M}_{a_{0}\rightarrow\eta_{N}\pi}(m_{\eta^{\prime}})\;,
\end{align}
where
\begin{align}
\mathcal{M}_{a_{0}\rightarrow\eta_{N}\pi}(m)  &  =A_{a_{0}\eta_{N}\pi
}-B_{a_{0}\eta_{N}\pi}\frac{m_{a_{0}}^{2}-m^{2}-m_{\pi}^{2}}{2}+C_{a_{0}%
\eta_{N}\pi}m_{a_{0}}^{2},\\
\mathcal{M}_{a_{0}\rightarrow\eta_{S}\pi}(m)  &  =A_{a_{0}\eta_{S}\pi},
\end{align}
and
\begin{align}
A_{a_{0}\eta_{N}\pi}  &  =Z_{\pi}^{2}(c_{1}\phi_{S}^{2}-\lambda_{2})\phi
_{N},\\
B_{a_{0}\eta_{N}\pi}  &  =-2\frac{g_{1}^{2}\phi_{N}}{m_{a_{1}}^{2}}\left[
1-\frac{1}{2}\frac{Z_{\pi}^{2}\phi_{N}^{2}}{m_{a_{1}}^{2}}(h_{2}%
-h_{3})\right]  ,\\
C_{a_{0}\eta_{N}\pi}  &  =g_{1}Z_{\pi}^{2}w_{a_{1}},\\
A_{a_{0}\eta_{S}\pi}  &  =\frac{1}{2}c_{1}Z_{\pi}Z_{\eta_{S}}\phi_{N}^{2}%
\phi_{S}.
\end{align}
The $a_{0}\rightarrow KK$ decay width includes two subchannels, $K^{0}\bar
{K}^{0}$ and $K^{-}K^{+}$, which results in the isospin factor $\mathcal{I}%
=2$, and accordingly the decay width is found to be%
\begin{equation}
\Gamma_{a_{0}\rightarrow KK}=\frac{1}{8m_{a_{0}}\pi}\sqrt{1-\left(
\frac{2m_{K}}{m_{a_{0}}}\right)  ^{2}}\left\vert A_{a_{0}KK}-\frac{1}%
{2}B_{a_{0}KK}(m_{a_{0}}^{2}-2m_{K}^{2})+C_{a_{0}KK}m_{a_{0}}^{2}\right\vert
^{2}\;, \label{B3}%
\end{equation}
where
\begin{align}
A_{a_{0}KK}  &  =\lambda_{2}Z_{K}^{2}\left(  \phi_{N}-\frac{\phi_{S}}{\sqrt
{2}}\right)  ,\\
B_{a_{0}KK}  &  =Z_{K}^{2}w_{K_{1}}\left\{  g_{1}-\frac{1}{2}w_{K_{1}}\left(
(g_{1}^{2}+h_{2})\phi_{N}+\sqrt{2}(g_{1}^{2}-h_{3})\phi_{S}\right)  \right\}
,\\
C_{a_{0}KK}  &  =-\frac{g_{1}}{2}Z_{K}^{2}w_{K_{1}}.
\end{align}
Now turning to the scalar kaon there is only one relevant decay channel,
$K_{0}^{\star}(\text{or }K_{0})\rightarrow K\pi$, for which the decay width
reads
\begin{equation}%
\begin{split}
\Gamma_{K_{0}\rightarrow K\pi}=\frac{3}{8\pi m_{K_{0}}}\left[  \frac
{(m_{K_{0}}^{2}-m_{\pi}^{2}-m_{K}^{2})^{2}-4m_{\pi}^{2}m_{K}^{2}}{4m_{K_{0}%
}^{4}}\right]  ^{1/2}  &  \left[  \frac{}{}A_{K_{0}K\pi}+(C_{K_{0}K\pi
}+D_{K_{0}K\pi}-B_{K_{0}K\pi}) \right. \\
&  \times\left.  \frac{m_{K_{0}}^{2}-m_{K}^{2}-m_{\pi}^{2}}{2}+C_{K_{0}K\pi
}m_{K}^{2}+D_{K_{0}K\pi}m_{\pi}^{2}\right]  \;,
\end{split}
\end{equation}
where
\begin{align}
A_{K_{0}K\pi}  &  =\frac{Z_{\pi}Z_{K}Z_{K_{0}}}{\sqrt{2}}\lambda_{2}\phi
_{S},\\
B_{K_{0}K\pi}  &  =\frac{Z_{\pi}Z_{K}Z_{K_{0}}}{4}w_{a_{1}}w_{K_{1}}\left[
2g_{1}\frac{w_{a_{1}}+w_{K_{1}}}{w_{a_{1}}w_{K_{1}}}+(2h_{3}-h_{2}-3g_{1}%
^{2})\phi_{N}-\sqrt{2}(g_{1}^{2}+h_{2})\phi_{S})\right]  ,\\
C_{K_{0}K\pi}  &  =\frac{Z_{\pi}Z_{K}Z_{K_{0}}}{2}[-g_{1}(iw_{K^{\star}%
}+w_{K_{1}})+\sqrt{2}iw_{K^{\star}}w_{K_{1}}(g_{1}^{2}-h_{3})\phi_{S}],\\
D_{K_{0}K\pi}  &  =\frac{Z_{\pi}Z_{K}Z_{K_{0}}}{4} \left \{ 2g_{1}(iw_{K^{\star}%
}-w_{a_{1}})+iw_{K^{\star}}w_{a_{1}} \left [(2h_{3}-h_{2}-3g_{1}^{2})\phi_{N}%
+\sqrt{2}(g_{1}^{2}+h_{2})\phi_{S} \right ] \right \}.
\end{align}
Finally, for the $f_{0}^{{L/H}}$ particles there are two relevant decay
channels: the two-pion and the two-kaon channels. It is important to note that
due to the particle mixing between $f_{0}^{{L}}$ and $f_{0}^{{H}}$, the matrix
elements are linear combinations that depend on the scalar mixing angle
$\theta_{\sigma}$ (see Appendix \ref{App:mass}), as can be seen explicitly
below. The decay widths in the $\pi\pi$ channel are
\begin{align}
\Gamma_{f_{0}^{{L}}\rightarrow\pi\pi}  &  =\frac{3}{32\pi m_{f_{0}^{\text{L}}%
}}\sqrt{1-\left(  \frac{2m_{\pi}}{m_{f_{0}^{\text{L}}}}\right)  ^{2}%
}\left\vert \mathcal{M}_{f_{0}^{{L}}\rightarrow\pi\pi}\right\vert ^{2}\;,\\
\Gamma_{f_{0}^{{H}}\rightarrow\pi\pi}  &  =\frac{3}{32\pi m_{f_{0}^{H}}}%
\sqrt{1-\left(  \frac{2m_{\pi}}{m_{f_{0}^{H}}}\right)  ^{2}}\left\vert
\mathcal{M}_{f_{0}^{{H}}\rightarrow\pi\pi}\right\vert ^{2}\;,
\end{align}
where an isospin factor of $3/2$ was considered\footnote{There are two
subchannels, namely the $\pi^{+}\pi^{-}$ and the $\pi^{0}\pi^{0}$, which would
mean $\mathcal{I}=2$. However, since the two $\pi^{0}$ are indistinguishable,
the solid-angle integral (there is no angular dependence at tree-level) ends
up as $2\pi$ instead of $4\pi$, which means a factor of $1/2$ in case of
$\pi^{0}\pi^{0}$ compared to $\pi^{+}\pi^{-}$, thus $\mathcal{I}=1/2+1$.} and
the matrix elements are
\begin{align}
\mathcal{M}_{f_{0}^{{L}}\rightarrow\pi\pi}  &  =-\sin\theta_{\sigma
}\mathcal{M}_{f_{0}\pi}^{{H}}(m_{f_{0}^{{L}}})+\cos\theta_{\sigma}%
\mathcal{M}_{f_{0}\pi}^{{L}}(m_{f_{0}^{{L}}}),\\
\mathcal{M}_{f_{0}^{{H}}\rightarrow\pi\pi}  &  =\cos\theta_{\sigma}%
\mathcal{M}_{f_{0}\pi}^{{H}}(m_{f_{0}^{{H}}})+\sin\theta_{\sigma}%
\mathcal{M}_{f_{0}\pi}^{{L}}(m_{f_{0}^{{H}}}),\\
\mathcal{M}_{f_{0}\pi}^{{L}}(m)  &  =2Z_{\pi}^{2}\phi_{N}\left\{  \frac
{g_{1}^{2}}{2}\frac{m^{2}}{m_{a_{1}}^{2}}\left[  1+\left(  1-\frac{2m_{\pi
}^{2}}{m^{2}}\right)  \frac{m_{1}^{2}+h_{1}\phi_{S}^{2}/2+2\delta_{N}%
}{m_{a_{1}}^{2}}\right]  -\left(  \lambda_{1}+\frac{\lambda_{2}}{2}\right)
\right\}  ,\label{aa}\\
\mathcal{M}_{f_{0}\pi}^{{H}}(m)  &  =2Z_{\pi}^{2}\phi_{S}\left\{  -\frac
{g_{1}^{2}}{4}\frac{m^{2}}{m_{a_{1}}^{2}}\left(  1-\frac{2m_{\pi}^{2}}{m^{2}%
}\right)  \frac{h_{1}\phi_{N}^{2}}{m_{a_{1}}^{2}}-\lambda_{1}\right\}  .
\label{bb}%
\end{align}
In the $KK$ channel, where $\mathcal{I}=2$, the decay widths read
\begin{align}
\Gamma_{f_{0}^{H}\rightarrow KK}  &  =\frac{1}{8\pi m_{f_{0}^{H}}}%
\sqrt{1-\left(  \frac{2m_{K}}{m_{f_{0}^{H}}}\right)  ^{2}}\left\vert
\mathcal{M}_{f_{0}^{H}\rightarrow KK}\right\vert ^{2},\\
\Gamma_{f_{0}^{L}\rightarrow KK}  &  =\frac{1}{8\pi m_{f_{0}^{L}}}%
\sqrt{1-\left(  \frac{2m_{K}}{m_{f_{0}^{L}}}\right)  ^{2}}\left\vert
\mathcal{M}_{f_{0}^{L}\rightarrow KK}\right\vert ^{2},
\end{align}
where the matrix elements, using the notations $H_{N}\equiv\left(  g_{1}%
^{2}+2h_{1}+h_{2}\right)  /4 $ and $H_{S}\equiv\left(  g_{1}^{2}+h_{1}%
+h_{2}\right)  /2 $, are
\begin{align}
\mathcal{M}_{f_{0}^{L}\rightarrow KK}  &  =-\sin\theta_{\sigma}\mathcal{M}%
_{f_{0}K}^{H}(m_{f_{0}^{L}})+\cos\theta_{\sigma}\mathcal{M}_{f_{0}K}%
^{L}(m_{f_{0}^{L}}),\\
\mathcal{M}_{f_{0}^{H}\rightarrow KK}  &  =\cos\theta_{\sigma}\mathcal{M}%
_{f_{0}K}^{H}(m_{f_{0}^{H}})+\sin\theta_{\sigma}\mathcal{M}_{f_{0}K}%
^{L}(m_{f_{0}^{H}}),\\
\mathcal{M}_{f_{0}K}^{L}(m)  &  =-Z_{K}^{2}\left[  (2\lambda_{1}+\lambda
_{2})\phi_{N}-\frac{\lambda_{2}}{\sqrt{2}}\phi_{S}+g_{1}w_{K_{1}}(m_{K}%
^{2}-m^{2})+w_{K_{1}}^{2}\left(  2H_{N}\phi_{N}-\frac{h_{3}-g_{1}^{2}}%
{\sqrt{2}}\phi_{S}\right)  \frac{m^{2}-2m_{K}^{2}}{2}\right]  ,\label{cc}\\
\mathcal{M}_{f_{0}K}^{H}(m)  &  =-Z_{K}^{2}\left[  2(\lambda_{1}+\lambda
_{2})\phi_{S}-\frac{\lambda_{2}}{\sqrt{2}}\phi_{N}+\sqrt{2}g_{1}w_{K_{1}%
}(m_{K}^{2}-m^{2})+w_{K_{1}}^{2}\left(  2H_{S}\phi_{S}-\frac{h_{3}-g_{1}^{2}%
}{\sqrt{2}}\phi_{N}\right)  \frac{m^{2}-2m_{K}^{2}}{2}\right]  . \label{dd}%
\end{align}

\subsection{Vector-meson decay widths}

In the case of the vector mesons we are considering the decays of the $\rho$
meson, the $K^{\star}$ vector kaon, and the $\phi$ meson. All these particles
have only one relevant decay channel. The first is the $\rho\rightarrow\pi\pi$
decay which has the following quite simple decay width formula
\begin{equation}
\Gamma_{\rho\rightarrow\pi\pi}=\frac{m_{\rho}^{5}}{48\pi m_{a_{1}}^{4}}\left[
1-\left(  \frac{2m_{\pi}}{m_{\rho}}\right)  ^{2}\right]  ^{3/2}\left[
g_{1}Z_{\pi}^{2}-\frac{g_{2}}{2}\left(  Z_{\pi}^{2}-1\right)  \right]  ^{2}.
\end{equation}
The next one is the $K^{\star}\rightarrow K\pi$ decay, in case of which the
decay width reads
\begin{equation}
\Gamma_{K^{\star}\rightarrow K\pi}=\frac{m_{K^{\star}}}{8\pi}\left[
\frac{(m_{K^{\star}}^{2}-m_{\pi}^{2}-m_{K}^{2})^{2}-4m_{\pi}^{2}m_{K}^{2}%
}{4m_{K^{\star}}^{4}}\right]  ^{3/2}(A_{K^{\star}K\pi}-B_{K^{\star}K\pi
}+C_{K^{\star}K\pi})^{2}\;,
\end{equation}
where the constants are defined as
\begin{align}
A_{K^{\star}K\pi}  &  =\frac{1}{2}Z_{\pi}Z_{K}\left[  -g_{1}+\sqrt{2}w_{K_{1}%
}(g_{1}^{2}-h_{3})\phi_{S}\right]  \;, {\nonumber}\\
B_{K^{\star}K\pi}  &  =\frac{1}{4}Z_{\pi}Z_{K}\left[  2g_{1}+w_{a_{1}}%
(-3g_{1}^{2}-h_{2}+2h_{3})\phi_{N}+\sqrt{2}w_{a_{1}}(g_{1}^{2}+h_{2})\phi
_{S}\right]  \;, {\nonumber}\\
C_{K^{\star}K\pi}  &  =\frac{1}{2}Z_{\pi}Z_{K}w_{a_{1}}w_{K_{1}}%
g_{2}m_{K^{\star}}^{2}\;.
\end{align}
Finally, the $\phi\rightarrow KK$ decay width reads
\begin{equation}
\Gamma_{\phi\rightarrow KK}=\frac{m_{\phi}^{5}}{192\pi m_{K_{1}}^{4}}\left[
1-\left(  \frac{2m_{K}}{m_{\phi}}\right)  ^{2}\right]  ^{3/2}\left[
2g_{1}Z_{K}^{2}\left(  1+\frac{\delta_{N}-\delta_{S}}{m_{\phi}^{2}}\right)
-g_{2}(Z_{K}^{2}-1)\right]  ^{2}.
\end{equation}

\subsection{Axial-vector-meson decay widths}

Turning to axial-vector mesons, the considered decays are the $a_{1}$ decay
with two relevant channels and the $f_{1S}$ decay with one relevant channel.
Since in case of the $a_{1}\rightarrow\rho\pi$ and $f_{1S}\rightarrow
KK^{\star}$ decays, the decaying as well as one of the resulting particle are
(axial-)vector mesons, the matrix elements have a more complicated form than
in the previous cases, as can be seen below. The first considered decay width
is the one of the $a_{1}\rightarrow\pi\gamma$ process, which takes the
following simple form
\begin{equation}
\Gamma_{a_{1}\rightarrow\pi\gamma}=\frac{e^{2}g_{1}^{2}\phi_{N}^{2}}{96\pi
m_{a_{1}}}Z_{\pi}^{2}\left[  1-\left(  \frac{m_{\pi}}{m_{a_{1}}}\right)
^{2}\right]  ^{3}\;. \label{eq:a1pigamma}%
\end{equation}
In case of the $a_{1}\rightarrow\rho\pi$ decay width one has to consider two
channels ($\rho^{+}\pi^{-},\rho^{-}\pi^{+}$), thus $\mathcal{I}=2$ and the
decay width is found to be
\begin{equation}
\Gamma_{a_{1}\rightarrow\rho\pi}=\frac{1}{12m_{a_{1}}\pi}\left[
\frac{(m_{a_{1}}^{2}-m_{\rho}^{2}-m_{\pi}^{2})^{2}-4m_{\rho}^{2}m_{\pi}^{2}%
}{4m_{a_{1}}^{4}}\right]  ^{1/2}\left[  |V_{\mu\nu}|^{2}-\frac{|V_{\mu\nu
}k_{\rho}^{\nu}|^{2}}{m_{\rho}^{2}}-\frac{|V_{\mu\nu}k_{a_{1}}^{\mu}|^{2}%
}{m_{a_{1}}^{2}}+\frac{|V_{\mu\nu}k_{a_{1}}^{\mu}k_{\rho}^{\nu}|^{2}}{m_{\rho
}^{2}m_{a_{1}}^{2}}\right]  \; , \label{eq:decay_a1_rho_pi}%
\end{equation}
where $V_{\mu\nu}$ is the vertex following from the relevant part of the
Lagrangian,
\begin{equation}
V_{\mu\nu}=iZ_{\pi}\phi_{N}\left\{  (g_{1}^{2}-h_{3})\,g_{\mu\nu}+\frac
{g_{1}g_{2}}{m_{a_{1}}^{2}}[k_{\pi\mu}k_{a_{1}\nu}+k_{\rho\mu}k_{\pi\nu
}-k_{\pi}\cdot(k_{\rho}+k_{a_{1}})g_{\mu\nu}]\right\}  \label{eq:V_mu_nu}%
\end{equation}
and $k_{a_{1}}^{\mu}=(m_{a_{1}},\mathbf{0})$, $k_{\rho}^{\mu}=(E_{\rho
},\mathbf{k})$ and $k_{\pi}^{\mu}=(E_{\pi},-\mathbf{k})$ are the four-momenta
of $a_{1}$, $\rho$, and $\pi$ in the rest frame of $a_{1}$, respectively.
Using the following kinematic relations
\begin{align}
&  k_{\pi}\cdot k_{\rho}=\frac{m_{a_{1}}^{2}-m_{\rho}^{2}-m_{\pi}^{2}}%
{2},{\nonumber}\\
&  k_{a_{1}}\cdot k_{\pi}=m_{a_{1}}E_{\pi}=\frac{m_{a_{1}}^{2}+m_{\pi}%
^{2}-m_{\rho}^{2}}{2},\label{eq:kin_rel}\\
&  k_{a_{1}}\cdot k_{\rho}=m_{a_{1}}E_{\rho}=\frac{m_{a_{1}}^{2}+m_{\rho}%
^{2}-m_{\pi}^{2}}{2},{\nonumber}\\
&  \mathbf{k}^{2}=\frac{(m_{a_{1}}^{2}-m_{\pi}^{2}-m_{\rho}^{2})^{2}-4m_{\pi
}^{2}m_{\rho}^{2}}{4m_{a_{1}}^{2}},{\nonumber}%
\end{align}
the terms in Eq.\ \eqref{eq:decay_a1_rho_pi} are given by
\begin{align}
&  |V_{\mu\nu}|^{2}=Z_{\pi}^{2}\phi_{N}^{2}\left\{  4(g_{1}^{2}-h_{3}%
)^{2}+\frac{g_{1}^{2}g_{2}^{2}}{m_{a_{1}}^{4}}\left[  \frac{5}{2}(m_{a_{1}%
}^{2}-m_{\rho}^{2})^{2}+\frac{1}{2}m_{\pi}^{2}(2m_{a_{1}}^{2}+2m_{\rho}%
^{2}-m_{\pi}^{2})\right]  -6\frac{g_{1}g_{2}(g_{1}^{2}-h_{3})}{m_{a_{1}}^{2}%
}(m_{a_{1}}^{2}-m_{\rho}^{2})\right\}  ,{\nonumber}\\
&  \frac{|V_{\mu\nu}k_{\rho}^{\nu}|^{2}}{m_{\rho}^{2}}=Z_{\pi}^{2}\phi_{N}%
^{2}\left\{  (g_{1}^{2}-h_{3})^{2}-\frac{g_{1}^{2}g_{2}^{2}}{m_{\rho}^{2}%
}(E_{\rho}^{2}-m_{\rho}^{2})+2\frac{g_{1}g_{2}(g_{1}^{2}-h_{3})}{m_{\rho}^{2}%
}(E_{\pi}^{2}-m_{\pi}^{2})\right\}  ,\label{eq:V_prod}\\
&  \frac{|V_{\mu\nu}k_{a_{1}}^{\nu}|^{2}}{m_{a_{1}}^{2}}=Z_{\pi}^{2}\phi
_{N}^{2}\left\{  (g_{1}^{2}-h_{3})^{2}-\frac{g_{1}^{2}g_{2}^{2}}{m_{a_{1}}%
^{4}}m_{\rho}^{2}\mathbf{k}^{2}-2\frac{g_{1}g_{2}(g_{1}^{2}-h_{3})}{m_{a_{1}%
}^{2}}(E_{\rho}^{2}-m_{\rho}^{2})\right\}  ,{\nonumber}\\
&  \frac{|V_{\mu\nu}k_{a_{1}}^{\mu}k_{\rho}^{\nu}|^{2}}{m_{\rho}^{2}m_{a_{1}%
}^{2}}=Z_{\pi}^{2}\phi_{N}^{2}\frac{(g_{1}^{2}-h_{3})^{2}}{m_{\rho}^{2}%
}E_{\rho}^{2}.{\nonumber}%
\end{align}
Analogously to the previous case, the width of the $f_{1S}\rightarrow
KK^{\star}$ decay which includes four sub-channels ($K^{0}\bar{K}^{\star0}$,
$\bar{K}^{0}K^{\star0}$, $K^{-}K^{\star+}$, $K^{+}K^{\star-}$), resulting in
$\mathcal{I}=4$, becomes
\begin{equation}
\Gamma_{f_{1S}\rightarrow KK^{\star}}=\frac{1}{6m_{f_{1S}}\pi}\left[
\frac{(m_{f_{1S}}^{2}-m_{K^{\star}}^{2}-m_{\pi}^{2})^{2}-4m_{K^{\star}}%
^{2}m_{K}^{2}}{4m_{f_{1S}}^{4}}\right]  ^{1/2}\left[  |\tilde{V}_{\mu\nu}%
|^{2}-\frac{|\tilde{V}_{\mu\nu}k_{K^{\star}}^{\nu}|^{2}}{m_{K^{\star}}^{2}%
}-\frac{|\tilde{V}_{\mu\nu}k_{f_{1S}}^{\mu}|^{2}}{m_{f_{1S}}^{2}}%
+\frac{|\tilde{V}_{\mu\nu}k_{f_{1S}}^{\mu}k_{K^{\star}}^{\nu}|^{2}%
}{m_{K^{\star}}^{2}m_{f_{1S}}^{2}}\right]  , \label{eq:decay_f1S_K_Kstar}%
\end{equation}
where $\tilde{V}_{\mu\nu}$ has the same Lorentz structure as
Eq.~\eqref{eq:V_mu_nu} and the only difference is in the constants in front of
the different terms. More explicitly,
\begin{align}
\tilde{V}_{\mu\nu}  &  =iZ_{K}\left\{  A_{fKK}\,g_{\mu\nu}+B_{fKK}[k_{K\mu
}k_{f_{1S}\nu}+k_{K^{\star}\mu}k_{K\nu}-k_{K}\cdot(k_{K^{\star}}+k_{f_{1S}%
})g_{\mu\nu}]\right\}  ,\\
&  A_{fKK}=\frac{1}{4}\left[  g_{1}^{2}(\sqrt{2}\phi_{N}-6\phi_{S})+\sqrt
{2}h_{2}(\phi_{N}-\sqrt{2}\phi_{S})+4h_{3}\phi_{S}\right]  ,\\
&  B_{fKK}=-\frac{1}{\sqrt{2}}g_{2}w_{K_{1}}.
\end{align}
The kinematic relations are the same as in Eq.~\eqref{eq:kin_rel} with the
following substitutions: $a_{1}\rightarrow f_{1S},\pi\rightarrow
K,\rho\rightarrow K^{\star}$, while the expressions analogous to
Eq.~\eqref{eq:V_prod} are,
\begin{align}
&  |\tilde{V}_{\mu\nu}|^{2}=Z_{K}^{2}\left\{  4A_{fKK}^{2}+B_{fKK}^{2}\left[
\frac{5}{2}(m_{f_{1S}}^{2}-m_{K^{\star}}^{2})^{2}+\frac{1}{2}m_{K}%
^{2}(2m_{f_{1S}}^{2}+2m_{K^{\star}}^{2}-m_{K}^{2})\right]  -6A_{fKK}%
B_{fKK}(m_{f_{1S}}^{2}-m_{K^{\star}}^{2})\right\}  ,{\nonumber}\\
&  \frac{|\tilde{V}_{\mu\nu}k_{K^{\star}}^{\nu}|^{2}}{m_{K^{\star}}^{2}}%
=Z_{K}^{2}\left\{  A_{fKK}^{2}-B_{fKK}^{2}(E_{K^{\star}}^{2}-m_{K^{\star}}%
^{2})+2A_{fKK}B_{fKK}(E_{K}^{2}-m_{K}^{2})\right\}  ,\label{eq:V_prod2}\\
&  \frac{|\tilde{V}_{\mu\nu}k_{f_{1S}}^{\nu}|^{2}}{m_{f_{1S}}^{2}}=Z_{K}%
^{2}\left\{  A_{fKK}^{2}-B_{fKK}^{2}m_{K^{\star}}^{2}\mathbf{k}^{2}%
-2A_{fKK}B_{fKK}(E_{K^{\star}}^{2}-m_{K^{\star}}^{2})\right\}  ,{\nonumber}\\
&  \frac{|\tilde{V}_{\mu\nu}k_{f_{1S}}^{\mu}k_{K^{\star}}^{\nu}|^{2}%
}{m_{K^{\star}}^{2}m_{f_{1S}}^{2}}=Z_{K}^{2}\frac{A_{fKK}^{2}}{m_{K^{\star}%
}^{2}}E_{K^{\star}}^{2}.{\nonumber}%
\end{align}

Analogously, the width for a generic decay of the form $K_{1}\rightarrow V{P}%
$, where ${V}$ denotes a vector and ${P}$ a pseudoscalar state, reads%
\begin{equation}
\Gamma_{K_{1}\rightarrow V{P}}=\mathcal{I}\frac{k(m_{K_{1}},m_{V},m_{{P}}%
)}{8\pi m_{K_{1}}^{2}}\frac{1}{3}\left[  \left\vert \tilde{V}_{K_{1}V{P}}%
^{\mu\nu}\right\vert ^{2}-\frac{\left\vert \tilde{V}_{K_{1}V{P}}^{\mu\nu
}p_{K_{1}\mu}\right\vert ^{2}}{m_{K_{1}}^{2}}-\frac{\left\vert \tilde
{V}_{K_{1}V{P}}^{\mu\nu}p_{V\nu}\right\vert ^{2}}{m_{V}^{2}}+\frac{\left\vert
\tilde{V}_{K_{1}V{P}}^{\mu\nu}p_{K_{1}\mu}p_{V\nu}\right\vert ^{2}}{m_{V}%
^{2}m_{K_{1}}^{2}}\right]  \;, \label{GK1VP}%
\end{equation}
where $\mathcal{I}=3$ for the $K_{1} \rightarrow\rho K$ and $K_{1} \rightarrow
K^{\star}\pi$ decays and $\mathcal{I}=1$ for the $K_{1} \rightarrow\omega_{N}
K$ decay,
\begin{equation}
k(m_{a},m_{b},m_{c})=\frac{1}{2m_{a}}\sqrt{m_{a}^{4}-2m_{a}^{2}\,(m_{b}%
^{2}+m_{c}^{2})+(m_{b}^{2}-m_{c}^{2})^{2}}\theta(m_{a}-m_{b}-m_{c}) \;,
\label{kabc}%
\end{equation}
(the theta function ensures that the decay width vanishes below threshold)
and
\begin{equation}
\tilde{V}_{K_{1}V{P}}^{\mu\nu}=i\left\{  A_{K_{1}}g^{\mu\nu}+B_{K_{1}}%
[p_{V}^{\mu}p_{{P}}^{\nu}+p_{{P}}^{\mu}p_{K_{1}}^{\nu}-(p_{{P}}\cdot
p_{V})g^{\mu\nu}-(p_{K_{1}}\cdot p_{{P}})g^{\mu\nu}]\right\}  \text{.}
\label{hK1VP}%
\end{equation}
The values of $A_{K_{1}}$, $B_{K_{1}}$, and $C_{K_{1}}$ depend on the process
considered: $A_{K_{1}}=\{A_{K_{1}K^{\star}\pi},A_{K_{1}\rho K},A_{K_{1}%
\omega_{N}K}\}$ and $B_{K_{1}}=\{B_{K_{1}K^{\star}\pi},B_{K_{1}\rho
K},B_{K_{1}\omega_{N}K}\}$. The coefficients read
\begin{align}
A_{K_{1}K^{\star}\pi}  &  =\frac{i}{\sqrt{2}}Z_{\pi}(h_{3}-g_{1}^{2})\phi
_{S}\;,\label{AK1KstarK}\\
B_{K_{1}K^{\star}\pi}  &  =-\frac{i}{2}Z_{\pi}g_{2}w_{a_{1}} \label{BK1KstarK}%
\\
A_{K_{1}\rho K}  &  =\frac{i}{4}Z_{K}\left[  g_{1}^{2}(\phi_{N}+\sqrt{2}%
\phi_{S})-h_{2}(\phi_{N}-\sqrt{2}\phi_{S})-2h_{3}\phi_{N}\right]
\;,\label{AK1rK}\\
B_{K_{1}\rho K}  &  =\frac{i}{2}Z_{K}g_{2}w_{K_{1}}\;,\label{BK1rK}\\
A_{K_{1}\omega_{N}K}  &  =-\frac{i}{4}Z_{K}[g_{1}^{2}(\phi_{N}+\sqrt{2}%
\phi_{S})-h_{2}(\phi_{N}-\sqrt{2}\phi_{S})-2h_{3}\phi_{N}]\;,\label{AK1oK}\\
B_{K_{1}\omega_{N}K}  &  =-\frac{i}{2}Z_{K}g_{2}w_{K_{1}}\;\text{.}
\label{BK1oK}%
\end{align}


\begin{thebibliography}{99}                                                                                               %


\bibitem {PDG}J.~Beringer \textit{et al}. (Particle Data Group),
Phys.\ Rev.\ D\textbf{86}, 010001 (2012).



\bibitem {amslerrev}C.~Amsler and N.~A.~Tornqvist,
Phys.\ Rept.\ \textbf{389}, 61 (2004);
E.~Klempt and A.~Zaitsev,
Phys.\ Rept.\ \textbf{454}, 1 (2007) [arXiv:0708.4016 [hep-ph]].


\bibitem {f0(1790)}A.~V.~Anisovich \textit{et al.},
Phys.\ Lett.\ B \textbf{449}, 154 (1999); M.~Ablikim \textit{et al.} [BES
Collaboration],
Phys.\ Lett.\ B \textbf{603}, 138 (2004) [arXiv:hep-ex/0409007];
M.~Ablikim \textit{et al.} [BES Collaboration],
Phys.\ Lett.\ B \textbf{607}, 243 (2005) [hep-ex/0411001];
M.~Ablikim \textit{et al.} [BES Collaboration],
Phys.\ Rev.\ D \textbf{72}, 092002 (2005) [arXiv:hep-ex/0508050];
D.~V.~Bugg,
arXiv:hep-ph/0603018;
A.~V.~Anisovich, D.~V.~Bugg, V.~A.~Nikonov, A.~V.~Sarantsev and
V.~V.~Sarantsev,
Phys.\ Rev.\ D \textbf{85}, 014001 (2012) [arXiv:1110.4333 [hep-ex]].


\bibitem {KappaY}A.~V.~Anisovich and A.~V.~Sarantsev,
Phys.\ Lett.\ B \textbf{413}, 137 (1997) [hep-ph/9705401];
R.~Delbourgo and M.~D.~Scadron,
Int.\ J.\ Mod.\ Phys.\ A \textbf{13}, 657 (1998) [hep-ph/9807504];
D.~Black, A.~H.~Fariborz, S.~Moussa, S.~Nasri and J.~Schechter,
Phys.\ Rev.\ D \textbf{64}, 014031 (2001) [hep-ph/0012278];
M.~D.~Scadron, F.~Kleefeld, G.~Rupp and E.~van Beveren,
Nucl.\ Phys.\ A \textbf{724}, 391 (2003) [hep-ph/0211275];
D.~V.~Bugg,
Phys.\ Lett.\ B \textbf{572}, 1 (2003) [Erratum-ibid.\ B \textbf{595}, 556
(2004)].


\bibitem {KappaN}S.~N.~Cherry and M.~R.~Pennington,
Nucl.\ Phys.\ A \textbf{688}, 823 (2001) [hep-ph/0005208];
J.~M.~Link \textit{et al.} [FOCUS Collaboration],
Phys.\ Lett.\ B \textbf{621}, 72 (2005) [hep-ex/0503043].


\bibitem {Leutwyler}
I.~Caprini, G.~Colangelo and H.~Leutwyler,
Phys.\ Rev.\ Lett.\ \textbf{96}, 132001 (2006) [hep-ph/0512364];
F.~J.~Yndurain, R.~Garcia-Martin and J.~R.~Pelaez,
Phys.\ Rev.\ D \textbf{76}, 074034 (2007) [hep-ph/0701025];
H.~Leutwyler,
AIP Conf.\ Proc.\ \textbf{1030}, 46 (2008) [arXiv:0804.3182 [hep-ph]];
R.~Kaminski, R.~Garcia-Martin, P.~Grynkiewicz and J.~R.~Pelaez,
Nucl.\ Phys.\ Proc.\ Suppl.\ \textbf{186}, 318 (2009) [arXiv:0811.4510
[hep-ph]];
R.~Garcia-Martin, R.~Kaminski, J.~R.~Pelaez and J.~Ruiz de Elvira,
Phys.\ Rev.\ Lett.\ \textbf{107}, 072001 (2011) [arXiv:1107.1635 [hep-ph]].


\bibitem {buggf0}D.~V.~Bugg,
Eur.\ Phys.\ J.\ C \textbf{52}, 55 (2007) [arXiv:0706.1341 [hep-ex]].


\bibitem {av}M.~F.~M.~Lutz and E.~E.~Kolomeitsev,
Nucl.\ Phys.\ A \textbf{730}, 392 (2004) [arXiv:nucl-th/0307039];
M.~Wagner and S.~Leupold,
Phys.\ Lett.\ B \textbf{670}, 22 (2008) [arXiv:0708.2223 [hep-ph]];
M.~Wagner and S.~Leupold,
Phys.\ Rev.\ D \textbf{78}, 053001 (2008) [arXiv:0801.0814 [hep-ph]];
S.~Leupold and M.~Wagner,
arXiv:0807.2389 [nucl-th].




\bibitem {Achim}A.~Heinz, S.~Struber, F.~Giacosa and D.~H.~Rischke,
Phys.\ Rev.\ D \textbf{79}, 037502 (2009) [arXiv:0805.1134 [hep-ph]];
A.~Heinz, S.~Struber, F.~Giacosa and D.~H.~Rischke,
Acta Phys.\ Polon.\ Supp.\ \textbf{3}, 925 (2010) [arXiv:1006.5393 [hep-ph]].






\bibitem {RS}S.~Struber and D.~H.~Rischke,
Phys.\ Rev.\ D \textbf{77}, 085004 (2008) [arXiv:0708.2389 [hep-th]].


\bibitem {SSB}C.~Vafa and E.~Witten,
Nucl.\ Phys.\ B \textbf{234}, 173 (1984);
L.~Giusti and S.~Necco,
JHEP \textbf{0704}, 090 (2007) [arXiv:hep-lat/0702013].




\bibitem {Hooft}G.~'t Hooft,
Phys.\ Rept.\ \textbf{142}, 357 (1986).


\bibitem {gellmanlevy}

J.~S.~Schwinger,
Annals Phys.\ \textbf{2}, 407 (1957);
M.~Gell-Mann and M.~Levy,
Nuovo Cim.\ \textbf{16}, 705 (1960);
S.~Weinberg,
Phys.\ Rev.\ Lett.\ \textbf{18}, 188 (1967).


\bibitem {weinberg}

J.~S.~Schwinger,
Phys.\ Lett.\ B \textbf{24}, 473 (1967);
S.~Weinberg,
Phys.\ Rev.\ \textbf{166}, 1568 (1968).


\bibitem {Paper1}D.~Parganlija, F.~Giacosa and D.~H.~Rischke,
Phys.\ Rev.\ D \textbf{82}, 054024 (2010) [arXiv:1003.4934 [hep-ph]].


\bibitem {gallas}S.~Gallas, F.~Giacosa and D.~H.~Rischke,
Phys.\ Rev.\ \textbf{D82}, 014004 (2010) [arXiv:0907.5084 [hep-ph]].


\bibitem {geffen}S.~Gasiorowicz and D.~A.~Geffen,
Rev.\ Mod.\ Phys.\ \textbf{41}, 531 (1969).






\bibitem {urban}P.~Ko and S.~Rudaz,
Phys.\ Rev.\ D \textbf{50}, 6877 (1994).

\bibitem {UBW}M.~Urban, M.~Buballa and J.~Wambach,
Nucl.\ Phys.\ \textbf{A697}, 338-371 (2002) [hep-ph/0102260].


\bibitem {Lenaghan:2000ey}J.~T.~Lenaghan, D.~H.~Rischke and
J.~Schaffner-Bielich,
Phys.\ Rev.\ D \textbf{62}, 085008 (2000) [nucl-th/0004006];


\bibitem {Szep}P.~Kovacs and Z.~Szep,
Phys.\ Rev.\ D \textbf{75}, 025015 (2007) [hep-ph/0611208];
P.~Kovacs and Z.~Szep,
Phys.\ Rev.\ D \textbf{77}, 065016 (2008) [arXiv:0710.1563 [hep-ph]].


\bibitem {chpt}J.~Gasser and H.~Leutwyler,
Annals Phys.\ \textbf{158}, 142 (1984);
see also S.~Scherer,
Adv.\ Nucl.\ Phys.\ \textbf{27}, 277 (2003) [arXiv:hep-ph/0210398]
and refs.\ therein.


\bibitem {chptvm}M.~Bando, T.~Kugo and K.~Yamawaki,
Phys.\ Rept.\ \textbf{164}, 217 (1988); G.~Ecker, J.~Gasser, A.~Pich and E.~de
Rafael,
Nucl.\ Phys.\ B \textbf{321}, 311 (1989);
E.~E.~Jenkins, A.~V.~Manohar and M.~B.~Wise,
Phys.\ Rev.\ Lett.\ \textbf{75}, 2272 (1995) [arXiv:hep-ph/9506356].
C.~Terschlusen and S.~Leupold,
Prog.\ Part.\ Nucl.\ Phys.\ \textbf{67}, 401 (2012) [arXiv:1111.4907
[hep-ph]].

\bibitem {Stani}S.~Janowski, D.~Parganlija, F.~Giacosa and D.~H.~Rischke,
Phys.\ Rev.\ D \textbf{84}, 054007 (2011) [arXiv:1103.3238 [hep-ph]].




\bibitem {References}D.~Parganlija, F.~Giacosa and D.~H.~Rischke,
AIP Conf.\ Proc.\ \textbf{1030}, 160 (2008) [arXiv:0804.3949 [hep-ph]];
D.~Parganlija, F.~Giacosa and D.~H.~Rischke,
PoS CONFINEMENT \textbf{8}, 070 (2008) [arXiv:0812.2183 [hep-ph]];
D.~Parganlija, F.~Giacosa and D.~H.~Rischke,
arXiv:0911.3996 [nucl-th];
D.~Parganlija, F.~Giacosa and D.~H.~Rischke,
Acta Phys.\ Polon.\ Supp.\ \textbf{3}, 963 (2010) [arXiv:1004.4817 [hep-ph]].


\bibitem {References2}D.~Parganlija, F.~Giacosa, D.~H.~Rischke, P.~Kovacs and
G.~Wolf,
Int.\ J.\ Mod.\ Phys.\ A \textbf{26}, 607 (2011) [arXiv:1009.2250 [hep-ph]];
D.~Parganlija, F.~Giacosa, P.~Kovacs and G.~Wolf,
AIP Conf.\ Proc.\ \textbf{1343}, 328 (2011) [arXiv:1011.6104 [hep-ph]];
P.~Kovacs, G.~Wolf, F.~Giacosa and D.~Parganlija,
EPJ Web Conf.\ \textbf{13}, 02006 (2011) [arXiv:1102.4732 [hep-ph]];
D.~Parganlija,
Acta Phys.\ Polon.\ Supp.\ \textbf{4}, 727 (2011) [arXiv:1105.3647 [hep-ph]];
D.~Parganlija,
arXiv:1109.4331 [hep-ph].


\bibitem {Morningstar}C.~Morningstar and M.~J.~Peardon,
AIP Conf.\ Proc.\ \textbf{688}, 220 (2004) [arXiv:nucl-th/0309068];
M.~Loan, X.~Q.~Luo and Z.~H.~Luo,
Int.\ J.\ Mod.\ Phys.\ A \textbf{21}, 2905 (2006) [arXiv:hep-lat/0503038];
E.~B.~Gregory, A.~C.~Irving, C.~C.~McNeile, S.~Miller and Z.~Sroczynski,
PoS \textbf{LAT2005}, 027 (2006) [arXiv:hep-lat/0510066];
Y.~Chen \textit{et al.},
Phys.\ Rev.\ D \textbf{73}, 014516 (2006) [arXiv:hep-lat/0510074];
C.~M.~Richards \textit{et al.} [UKQCD Collaboration],
Phys.\ Rev.\ D \textbf{82}, 034501 (2010) [arXiv:1005.2473 [hep-lat]].


\bibitem {glueball}
C.~Amsler and F.~E.~Close,
Phys.\ Rev.\ D \textbf{53}, 295 (1996) [arXiv:hep-ph/9507326];
W.~J.~Lee and D.~Weingarten,
Phys.\ Rev.\ D \textbf{61}, 014015 (2000) [arXiv:hep-lat/9910008];
F.~E.~Close and A.~Kirk,
Eur.\ Phys.\ J.\ C \textbf{21}, 531 (2001) [arXiv:hep-ph/0103173];
F.~Giacosa, T.~Gutsche, V.~E.~Lyubovitskij and A.~Faessler,
Phys.\ Rev.\ D \textbf{72}, 094006 (2005) [arXiv:hep-ph/0509247];
F.~Giacosa, T.~Gutsche, V.~E.~Lyubovitskij and A.~Faessler,
Phys.\ Lett.\ B \textbf{622}, 277 (2005) [arXiv:hep-ph/0504033];
F.~Giacosa, T.~Gutsche and A.~Faessler,
Phys. Rev. C \textbf{71}, 025202 (2005) [arXiv:hep-ph/0408085];
H.~Y.~Cheng, C.~K.~Chua and K.~F.~Liu,
Phys.\ Rev.\ D \textbf{74}, 094005 (2006) [arXiv:hep-ph/0607206];
L.~Bonanno and A.~Drago,
Phys.\ Rev.\ C \textbf{79}, 045801 (2009) [arXiv:0805.4188 [nucl-th]];
V.~Mathieu, N.~Kochelev and V.~Vento,
Int.\ J.\ Mod.\ Phys.\ E \textbf{18}, 1 (2009) [arXiv:0810.4453 [hep-ph]].




\bibitem {Doktorarbeit}D.~Parganlija, ``Quarkonium Phenomenology in Vacuum''
(PhD Thesis), urn:nbn:de:hebis:30:3-249891 UR, Institute for Theoretical
Physics of Frankfurt University (2012) [arXiv:1208.0204 [hep-ph]].


\bibitem {Meissner}U.~G.~Meissner,
Phys.\ Rept.\ \textbf{161}, 213 (1988).



\bibitem {dynrec}F.~Giacosa,
Phys.\ Rev.\ D \textbf{80}, 074028 (2009) [arXiv:0903.4481 [hep-ph]].





\bibitem {schechter}
C.~Rosenzweig, A.~Salomone and J.~Schechter,
Phys.\ Rev.\ D \textbf{24}, 2545 (1981);
A.~Salomone, J.~Schechter and T.~Tudron,
Phys.\ Rev.\ D \textbf{23}, 1143 (1981);
C.~Rosenzweig, A.~Salomone and J.~Schechter,
Nucl.\ Phys.\ B \textbf{206}, 12 (1982) [Erratum-ibid.\ B \textbf{207}, 546
(1982)];
A.~A.~Migdal and M.~A.~Shifman,
Phys.\ Lett.\ B \textbf{114}, 445 (1982);
H.~Gomm and J.~Schechter,
Phys.\ Lett.\ B \textbf{158}, 449 (1985);
R.~Gomm, P.~Jain, R.~Johnson and J.~Schechter,
Phys.\ Rev.\ D \textbf{33}, 801 (1986).


\bibitem {wittennc}E.~Witten,
Nucl.\ Phys.\ B \textbf{160}, 57 (1979);
S.~R.~Coleman, ``1/N,'' Published in Erice Subnuclear 1979:0011;
R.~F.~Lebed,
Czech.\ J.\ Phys.\ \textbf{49}, 1273 (1999) [arXiv:nucl-th/9810080].


\bibitem {Asner:1999}D.~M.~Asner \textit{et al.} [CLEO Collaboration],
Phys.\ Rev.\ D \textbf{61}, 012002 (2000) [hep-ex/9902022].



\bibitem {Bromberg}C.~Bromberg, J.~Dickey, G.~Fox, R.~Gomez, W.~Kropac,
J.~Pine, S.~Stampke and H.~Haggerty \textit{et al.},
Phys.\ Rev.\ D \textbf{22}, 1513 (1980).




\bibitem {Dionisi}C.~Dionisi \textit{et al.} [CERN-College de
France-Madrid-Stockholm Collaboration],
Nucl.\ Phys.\ B \textbf{169}, 1 (1980).


\bibitem {Giacosa:2009qh}F.~Giacosa and G.~Pagliara,
Nucl.\ Phys.\ A \textbf{833}, 138 (2010) [arXiv:0905.3706 [hep-ph]].


\bibitem {harada}M.~Harada and K.~Yamawaki,
Phys.\ Rept.\ \textbf{381}, 1 (2003) [arXiv:hep-ph/0302103].




\bibitem {burakovsky}L.~Burakovsky and J.~T.~Goldman,
Phys.\ Rev.\ D \textbf{57}, 2879 (1998) [hep-ph/9703271].




\bibitem {Hiller}A.~A.~Osipov, B.~Hiller, A.~H.~Blin and J.~da Providencia,
Annals Phys.\ \textbf{322}, 2021 (2007) [hep-ph/0607066].




\bibitem {Black}D.~Black, A.~H.~Fariborz and J.~Schechter,
Phys.\ Rev.\ D \textbf{61}, 074001 (2000) [hep-ph/9907516];

\bibitem {Mixing}
D.~Black, A.~H.~Fariborz, F.~Sannino and J.~Schechter,
Phys.\ Rev.\ D \textbf{59}, 074026 (1999) [hep-ph/9808415];
A.~H.~Fariborz, R.~Jora and J.~Schechter,
Phys.\ Rev.\ D \textbf{72}, 034001 (2005) [hep-ph/0506170];
A.~H.~Fariborz, R.~Jora and J.~Schechter,
Phys.\ Rev.\ D \textbf{76}, 014011 (2007) [hep-ph/0612200];
A.~H.~Fariborz, R.~Jora and J.~Schechter,
Phys.\ Rev.\ D \textbf{77}, 034006 (2008) [arXiv:0707.0843 [hep-ph]];
A.~H.~Fariborz, R.~Jora and J.~Schechter,
Phys.\ Rev.\ D \textbf{79}, 074014 (2009) [arXiv:0902.2825 [hep-ph]];
A.~H.~Fariborz, R.~Jora, J.~Schechter and M.~N.~Shahid,
Phys.\ Rev.\ D \textbf{83}, 034018 (2011) [arXiv:1012.4868 [hep-ph]];
A.~H.~Fariborz, R.~Jora, J.~Schechter and M.~N.~Shahid,
Phys.\ Rev.\ D \textbf{84}, 113004 (2011) [arXiv:1106.4538 [hep-ph]];
T.~K.~Mukherjee, M.~Huang and Q.~-S.~Yan,
arXiv:1203.5717 [hep-ph].



\bibitem {Giacosamixing}F.~Giacosa,
Phys.\ Rev.\ D \textbf{75}, 054007 (2007) [arXiv:hep-ph/0611388].



\bibitem {wolkanowski}F.~Giacosa and T.~Wolkanowski,
arXiv:1209.2332 [hep-ph].




\bibitem {FG}F.~Giacosa and G.~Pagliara,
Phys.\ Rev.\ C \textbf{76}, 065204 (2007) [arXiv:0707.3594 [hep-ph]].



\bibitem {P1}J.~R.~Pelaez,
Mod.\ Phys.\ Lett.\ A \textbf{19}, 2879 (2004) [hep-ph/0411107].




\bibitem {pennington}E.~van Beveren, T.~A.~Rijken, K.~Metzger, C.~Dullemond,
G.~Rupp and J.~E.~Ribeiro,
Z.\ Phys.\ C \textbf{30} (1986) 615 [arXiv:0710.4067 [hep-ph]];
N.~A.~Tornqvist,
Z.\ Phys.\ C \textbf{68} (1995) 647 [hep-ph/9504372];
M.~Boglione and M.~R.~Pennington,
Phys.\ Rev.\ D \textbf{65} (2002) 114010 [hep-ph/0203149];
E.~van Beveren, D.~V.~Bugg, F.~Kleefeld and G.~Rupp,
Phys.\ Lett.\ B \textbf{641} (2006) 265 [hep-ph/0606022]; J.~A.~Oller, E.~Oset
and J.~R.~Pelaez,
Phys.\ Rev.\ D \textbf{59} (1999) 074001 [Erratum-ibid.\ D \textbf{60}
(1999\ ERRAT,D75,099903.2007) 099906] [arXiv:hep-ph/9804209].

\bibitem {f01370KK}V.~A.~Polychronakos \textit{et al.},
Phys.\ Rev.\ D \textbf{19}, 1317 (1979);
A.~B.~Wicklund, D.~S.~Ayres, D.~H.~Cohen, R.~Diebold and A.~J.~Pawlicki,
Phys.\ Rev.\ Lett.\ \textbf{45}, 1469 (1980);
A.~Etkin \textit{et al.},
Phys.\ Rev.\ D \textbf{25}, 1786 (1982);
B.~V.~Bolonkin \textit{et al.},
Yad.\ Fiz.\ \textbf{46}, 799 (1987) [Nucl.\ Phys.\ B \textbf{309}, 426
(1988)];
G.~D.~Tikhomirov, I.~A.~Erofeev, O.~N.~Erofeeva and V.~N.~Luzin,
Phys.\ Atom.\ Nucl.\ \textbf{66}, 828 (2003) [Yad.\ Fiz.\ \textbf{66}, 860
(2003)]; V.~V.~Vladimirsky \textit{et al.},
Phys.\ Atom.\ Nucl.\ \textbf{69}, 493 (2006) [Yad.\ Fiz.\ \textbf{69}, 515
(2006)].


\bibitem {Ablikim:2004}M.~Ablikim \textit{et al.} [BES Collaboration],
Phys.\ Lett.\ B \textbf{607}, 243 (2005) [arXiv:hep-ex/0411001].


\bibitem {Bargiotti:2003}M.~Bargiotti \textit{et al.} [OBELIX Collaboration],
Eur.\ Phys.\ J.\ C \textbf{26}, 371 (2003).


\bibitem {Barberis:1999}D.~Barberis \textit{et al.} [WA102 Collaboration],
Phys.\ Lett.\ B \textbf{462}, 462 (1999) [arXiv:hep-ex/9907055].


\bibitem {f0(1710)-2006-BESII}M.~Ablikim \textit{et al.},
Phys.\ Lett.\ B \textbf{642}, 441 (2006) [arXiv:hep-ex/0603048].









\bibitem {Barberis:2000}D.~Barberis \textit{et al.} [WA102 Collaboration],
Phys.\ Lett.\ B \textbf{479}, 59 (2000) [arXiv:hep-ex/0003033].

\bibitem {Anisovich:2001}V.~V.~Anisovich, V.~A.~Nikonov and A.~V.~Sarantsev,
Phys.\ Atom.\ Nucl.\ \textbf{65}, 1545 (2002) [Yad.\ Fiz.\ \textbf{65}, 1583
(2002)] [arXiv:hep-ph/0102338].



\bibitem{etaM}
  M.~S.~Bhagwat, L.~Chang, Y.~-X.~Liu, C.~D.~Roberts and P.~C.~Tandy,
  Phys.\ Rev.\ C {\bf 76}, 045203 (2007)
  [arXiv:0708.1118 [nucl-th]];
  F.~Ambrosino, A.~Antonelli, M.~Antonelli, F.~Archilli, P.~Beltrame, G.~Bencivenni, S.~Bertolucci and C.~Bini {\it et al.},
  JHEP {\bf 0907}, 105 (2009)
  [arXiv:0906.3819 [hep-ph]];
  G.~Amelino-Camelia, F.~Archilli, D.~Babusci, D.~Badoni, G.~Bencivenni, J.~Bernabeu, R.~A.~Bertlmann and D.~R.~Boito {\it et al.},
  Eur.\ Phys.\ J.\ C {\bf 68}, 619 (2010)
  [arXiv:1003.3868 [hep-ex]];
  M.~C.~Chang, Y.~C.~Duh, J.~Y.~Lin, I.~Adachi, K.~Adamczyk, H.~Aihara, D.~M.~Asner and T.~Aushev {\it et al.},
  Phys.\ Rev.\ D {\bf 85}, 091102 (2012)
  [arXiv:1203.3399 [hep-ex]].





\bibitem {Cheng:2011}
R.~K.~Carnegie, R.~J.~Cashmore, W.~M.~Dunwoodie, T.~A.~Lasinski and
D.~W.~G.~Leith,
Phys.\ Lett.\ B \textbf{68}, 287 (1977);
J.~L.~Rosner,
Comments Nucl.\ Part.\ Phys.\ \textbf{16}, 109 (1986);
N.~Isgur and M.~B.~Wise,
Phys.\ Lett.\ B \textbf{232}, 113 (1989);
H.~G.~Blundell, S.~Godfrey and B.~Phelps,
Phys.\ Rev.\ D \textbf{53}, 3712 (1996) [arXiv:hep-ph/9510245];
F.~E.~Close and A.~Kirk,
Z.\ Phys.\ C \textbf{76}, 469 (1997) [arXiv:hep-ph/9706543];
D.~M.~Li, H.~Yu and Q.~X.~Shen,
Chin.\ Phys.\ Lett.\ \textbf{17}, 558 (2000) [arXiv:hep-ph/0001011];
D.~M.~Asner \textit{et al.} [CLEO Collaboration],
Phys.\ Rev.\ D \textbf{62}, 072006 (2000) [arXiv:hep-ex/0004002];
W.~S.~Carvalho, A.~S.~de Castro and A.~C.~B.~Antunes,
J.\ Phys.\ A \textbf{35}, 7585 (2002) [arXiv:hep-ph/0207372];
H.~Y.~Cheng,
Phys.\ Rev.\ D \textbf{67}, 094007 (2003) [arXiv:hep-ph/0301198];
T.~Barnes, N.~Black and P.~R.~Page,
Phys.\ Rev.\ D \textbf{68}, 054014 (2003) [arXiv:nucl-th/0208072];
D.~M.~B.~Li, B.~Ma, Y.~X.~Li, Q.~K.~Yao and H.~Yu,
Eur.\ Phys.\ J.\ C \textbf{37}, 323 (2004) [arXiv:hep-ph/0408214];
J.~Vijande, F.~Fernandez and A.~Valcarce,
J.\ Phys.\ G \textbf{31}, 481 (2005) [arXiv:hep-ph/0411299].
D.~M.~Li, B.~Ma and H.~Yu,
Eur.\ Phys.\ J.\ A \textbf{26}, 141 (2005) [arXiv:hep-ph/0509215];
D.~M.~Li and Z.~Li,
Eur.\ Phys.\ J.\ A \textbf{28}, 369 (2006) [arXiv:hep-ph/0606297];
H.~Hatanaka and K.~C.~Yang,
Phys.\ Rev.\ D \textbf{77}, 094023 (2008) [Erratum-ibid.\ D \textbf{78},
059902 (2008)] [arXiv:0804.3198 [hep-ph]];
H.~Y.~Cheng and C.~K.~Chua,
Phys.\ Rev.\ D \textbf{81}, 114006 (2010) [Erratum-ibid.\ D \textbf{82},
059904 (2010)] [arXiv:0909.4627 [hep-ph]]; K.~-C.~Yang,
Phys.\ Rev.\ D \textbf{84}, 034035 (2011) [arXiv:1011.6113 [hep-ph]];
H.~-Y.~Cheng,
Phys.\ Lett.\ B \textbf{707}, 116 (2012) [arXiv:1110.2249 [hep-ph]].


\bibitem {Hatsuda-Lee}T. Hatsuda, S.H. Lee, Phys. Rev. C\textbf{46}, R34 (1992).

\bibitem {Brown-Rho}G. Brown, M. Rho, Phys. Rev. Lett. \textbf{66}, 2720 (1991).

\bibitem {CERES}P. Salabura, Acta Phys. Pol. B\textbf{27}, 421 (1996); G.
Agakichiev \emph{et al.} Phys. Lett. B\textbf{422}, 405 (1998).


\bibitem {varietq}
R.~L.~Jaffe,
Phys.\ Rev.\ D \textbf{15}, 267 (1977);
R.~L.~Jaffe,
Phys.\ Rev.\ D \textbf{15}, 281 (1977);
L.~Maiani, F.~Piccinini, A.~D.~Polosa and V.~Riquer,
Phys.\ Rev.\ Lett.\ \textbf{93} (2004) 212002 [arXiv:hep-ph/0407017];
M.~Napsuciale and S.~Rodriguez,
Phys.\ Rev.\ D \textbf{70}, 094043 (2004) [arXiv:hep-ph/0407037]; F.~Giacosa,
Phys.\ Rev.\ D \textbf{74}, 014028 (2006) [arXiv:hep-ph/0605191].






\bibitem {Fischer}W.~Heupel, G.~Eichmann and C.~S.~Fischer,
arXiv:1206.5129 [hep-ph].




\bibitem {Wagner}C.~Michael \textit{et al.} [ETM Collaboration],
JHEP \textbf{1008}, 009 (2010) [arXiv:1004.4235 [hep-lat]];
R.~Baron, P.~.Boucaud, J.~Carbonell, A.~Deuzeman, V.~Drach, F.~Farchioni,
V.~Gimenez and G.~Herdoiza \textit{et al.},
JHEP \textbf{1006}, 111 (2010) [arXiv:1004.5284 [hep-lat]];
R.~Baron \textit{et al.} [European Twisted Mass Collaboration],
Comput.\ Phys.\ Commun.\ \textbf{182}, 299 (2011) [arXiv:1005.2042
[hep-lat]];
M.~Wagner [ETM and Y Collaborations],
PoS LATTICE \textbf{2010}, 162 (2010) [arXiv:1008.1538 [hep-lat]];
R.~Baron \textit{et al.} [ETM Collaboration],
arXiv:1009.2074 [hep-lat].
R.~Baron \textit{et al.} [ETM Collaboration],
PoS LATTICE \textbf{2010}, 123 (2010) [arXiv:1101.0518 [hep-lat]];
M.~Wagner [ETM Collaboration],
Acta Phys.\ Polon.\ Supp.\ \textbf{4}, 747 (2011) [arXiv:1103.5147
[hep-lat]].




\bibitem {Sakai}T.~Sakai and S.~Sugimoto,
Prog.\ Theor.\ Phys.\ \textbf{113}, 843 (2005) [hep-th/0412141];
T.~Sakai and S.~Sugimoto,
Prog.\ Theor.\ Phys.\ \textbf{114}, 1083 (2005) [hep-th/0507073];
D.~Li, M.~Huang and Q.~-S.~Yan,
arXiv:1206.2824 [hep-th].




\bibitem {susagiu}S.~Gallas, F.~Giacosa and G.~Pagliara,
Nucl.\ Phys.\ \textbf{A872}, 13-24 (2011) [arXiv:1105.5003 [hep-ph]].

\bibitem {Andreas}
A.~Schmitt, S.~Stetina and M.~Tachibana,
Phys.\ Rev.\ D \textbf{83}, 045008 (2011) [arXiv:1010.4243 [hep-ph]];
F.~Preis, A.~Rebhan and A.~Schmitt,
JHEP \textbf{1103}, 033 (2011) [arXiv:1012.4785 [hep-th]];
F.~Preis, A.~Rebhan and A.~Schmitt,
J.\ Phys.\ G G \textbf{39}, 054006 (2012) [arXiv:1109.6904 [hep-th]].



\end{thebibliography}
\end{document}